\documentclass[aps,prb,twocolumn,floats]{revtex4}
\usepackage{color,ulem,verbatim,float,wrapfig}
\usepackage{amssymb,amsbsy,amsmath,mathrsfs}
\usepackage{epsfig,subfigure}
\usepackage{graphicx}
\usepackage{epstopdf}
\usepackage{times}
\usepackage{color}
\usepackage{subfigure}

\usepackage{setspace}
\usepackage{bm} 
\usepackage{caption}
\newcommand\bea{\begin{eqnarray}}
\newcommand\eea{\end{eqnarray}}
\newcommand\beq{\begin{equation}}
\newcommand\eeq{\end{equation}}
\newcommand\bib{\bibitem}

\newcommand{\noi}{\noindent}
\newcommand{\non}{\nonumber}
\newcommand{\al}{\alpha}
\newcommand{\de}{\delta}

\newcommand{\ep}{\epsilon}
\newcommand{\ka}{\kappa}

\newcommand{\ta}{\theta}
\newcommand{\om}{\omega}
\newcommand{\dg}{\dagger}

\newcommand{\la}{\langle}
\newcommand{\ra}{\rangle}

\begin{document}

\title{Floquet engineering of edge states in the presence of staggered
potential and interactions}
\author{Samudra Sur$^1$ and Diptiman Sen$^{1,2}$}
\affiliation{$^1$Center for High Energy Physics, Indian Institute of Science,
Bengaluru 560012, India \\
$^2$Department of Physics, Indian Institute of Science, Bengaluru 560012,
India}

\begin{abstract}
We study the effects of a periodically driven electric field applied to 
a variety of tight-binding models in one dimension. We first consider a 
non-interacting system with or without a staggered on-site potential, and 
we find that periodic driving can generate states localized completely 
or partially near the ends of a finite-sized system. Depending on the system 
parameters, such states have Floquet eigenvalues lying either outside or 
inside the continuum of eigenvalues of the bulk states. In the former 
case we find that these states are completely localized at the ends and are 
true edge states, while in the latter case, the states are not
completely localized at the ends although the localization can be made
almost perfect by tuning the driving parameters.
We then consider a system of two bosonic particles which 
have an on-site Hubbard interaction and show that a periodically driven 
electric field can generate two-particle states which are localized at the 
ends of the system. We show that many of these effects can be understood using 
a Floquet perturbation theory which is valid in the limit of a large staggered 
potential or large interaction strength. Some of these effects can also
be understood qualitatively by considering time-independent Hamiltonians
which have a potential at the sites at the edges; Hamiltonians of these kinds
effectively appear in a Floquet-Magnus analysis of the driven problem.
Finally, we discuss how the edge states produced by periodic driving of a 
non-interacting system of fermions can be detected by measuring the 
differential conductance of the system. 
\end{abstract}


\maketitle

\section{Introduction}
\label{sec1}

Quantum systems whose Hamiltonians are periodically driven in time have 
been extensively studied in recent years. There has been tremendous progress, 
both theoretically~\cite{gross,kaya,das,russo,nag,sharma,laza,dutta,frank} and 
experimentally~\cite{bloch,kita,tarr,recht,langen,jotzu,eckardt,mciver}, in 
generating novel many-body phases of matter by using periodic driving and 
understanding various properties of such systems. Two particularly 
interesting phenomena which can occur are dynamical
localization~\cite{nag,dunlap,agarwala1,agarwala2} and the generation of 
states localized at the boundaries of the 
system~\cite{oka,kitagawa10,lindner11,jiang11,gu,trif12,gomez12,dora12,
cayssol13,liu13,tong13,rudner13,katan13,lindner13,kundu13,kundu14,schmidt13,
reynoso13,wu13,manisha13,perez,reichl14,carp,xiong,manisha17,zhou,deb,saha}.
A periodically driven system can be studied by calculating the Floquet 
operator $U_T$ which evolves the system over one time 
period $T$ of the driving. We can use $U_T$ to study the behavior of the 
system at stroboscopic intervals, i.e., at integer multiples of $T$. In the 
phenomenon of dynamical localization, particles appear to be stationary
if the system is viewed stroboscopically, namely, certain states with one
or more particles located at some particular places are eigenstates of $U_T$. 
In the generation of boundary states, $U_T$ has some eigenstates which are 
localized near the boundaries of the system. 

The states of a periodically driven system are labeled by the eigenvalues of 
$U_T$. 
It is possible for states localized near the boundaries 
to have Floquet eigenvalues which lie within the continuum 
of eigenvalues of the bulk states; these are called `Floquet bound states in 
a continuum'~\cite{longhi,santander,zhu}. When we numerically 
find states which appear to be candidates for such bound states, we have to 
study their wave functions carefully to 
decide if they are true bound states (with normalizable wave functions) or 
if they merely have large values in some restricted regions of space but are 
not normalizable (for an infinite system size) because their wave functions 
do not go to zero fast enough outside those regions. [We note that in 
time-independent systems, it is possible for true bound states to appear in 
the continuum of energies of the bulk states. However, in such cases there are
symmetries which do not allow such states to hybridize with the bulk 
states~\cite{schult}. In the absence of any symmetries, the hybridization with
bulk states prevents the existence of true bound states in the continuum].

It is known that interactions between particles can lead to the formation of 
multiparticle bound states at the edges of a system when there is no 
driving~\cite{pinto}, while interactions along with periodic driving
can produce multiparticle bound states inside the bulk of a 
system~\cite{agarwala2}. This naturally leads to the question of whether
interactions and driving can produce such bound states at the edges 
of a system rather than in the bulk.

Finally, it is important to find ways of detecting edge states when they 
appear in a system. For instance, when a one-dimensional topological 
superconductor is generated by periodic driving of one of the system 
parameters, it is known that Majorana end modes can be generated and these
can give rise to peaks in the differential conductance at certain values
of the voltage bias applied across the system~\cite{kundu13}.

Keeping all the above considerations in mind, our paper is planned as 
follows. In Sec.~\ref{sec2}, we briefly introduce 
the periodically driven tight-binding models that we will study in 
detail. These include a non-interacting system and a 
Bose-Hubbard model with two particles. In both cases, the phase of the 
nearest-neighbor hopping will be taken to vary sinusoidally with time with 
a frequency $\om$ and an amplitude $a$; this describes the effect of a 
periodically varying electric field through the Peierls prescription. 
In Sec.~\ref{sec3}, we look at a non-interacting system with a 
single particle, with and without an on-site staggered potential $v$. We 
numerically calculate the Floquet operator $U_T$ which evolves the system 
through one time period $T = 2 \pi/\om$, and find its eigenvalues and 
eigenstates. In both cases, we hold the magnitude $g$ of the nearest-neighbor 
hopping fixed and study the ranges of the parameters $\om, ~a$
and $v$ for which one or more Floquet eigenstates appear near each edge of a 
long but finite system. 
When the staggered potential $v$ is much larger than $g$, we study the problem
analytically using a Floquet perturbation theory. The results obtained from
this are compared with those obtained numerically. We then study the time 
evolution of a state which is not a Floquet eigenstate and is initially 
localized at the edge of the system. 

In Sec.~\ref{sec4}, we study the Bose-Hubbard model with an on-site
interaction strength $u$. We consider a system with two particles and 
study the effect of periodic driving of the hopping phase to find the range 
of parameters $\om, ~a$ and $u$ in which there are Floquet eigenstates with 
the particles localized near the edges of the system. In the limit that the 
interaction strength $u$ is much larger than $g$, we again develop a 
Floquet perturbation theory to find when such bound states occur and see how
well this matches the numerical results. We also study the time
evolution of a state which initially has both particles at the edge.

In Sec.~\ref{sec5}, we study how the edge states can be detected using
transport measurements. To this end, we consider a tight-binding model of 
non-interacting fermions in which there are semi-infinite leads on the left 
and right which are weakly coupled to a finite length wire in the middle. 
In the wire, the hopping phase is periodically driven as in 
Sec.~\ref{sec2}. We find that when the differential conductance across the 
system has peaks when the chemical potential of the leads is equal to the 
quasienergies of the edge states of the isolated wire.

We present some additional material in the Appendices. In Appendix A we 
provide a brief introduction to Floquet theory and the calculation of Floquet 
eigenstates and eigenvalues. In Appendix B we use the Floquet-Magnus expansion 
to derive the effective Hamiltonian to first order in $1/\om$, where $\om$ 
is the driving frequency. This shows that an important effect of periodic 
driving in a finite system is to generate a potential at the sites at the two 
ends. In Appendix C we therefore study some time-independent models to 
understand qualitatively the role of such an edge potential in producing edge 
states. The first model is a non-interacting system with a staggered potential 
$v$ while the second model is the Bose-Hubbard model with an interaction 
strength $u$ and two particles. In both cases, we include a potential $A$ at 
the end sites. We study the conditions under which an edge state (consisting 
of one particle in the first model and two particles in the second model) 
appears near the ends.

Our main results are as follows. We find that a tight-binding model in 
one dimension can host one or more states at each end when the phase of the 
hopping amplitude is periodically driven in time. The range of driving
parameters where such edge states appear increases significantly when a 
staggered potential or an on-site Bose-Hubbard interaction is present. 
The edge states can be detected by measuring the differential conductance 
across a periodically driven wire. 

\section{Introduction to our periodically driven systems}
\label{sec2}

In this section we will briefly introduce the models that we will study to see 
if periodic driving of a finite-sized system can give rise to states which are 
localized at its ends. We will study two lattice models in one dimension, one 
without interactions and one with interactions, and look for edge states in 
each case. In this paper we will set the lattice spacing equal to unity and 
work in units where $\hbar = 1$ (unless mentioned explicitly).

\noi (i) We will first consider a tight-binding model, with possibly a 
staggered on-site potential, which is driven by an oscillating electric field. 
\begin{eqnarray}
H &=& - ~g ~\sum_{n=0}^{L-2} ~(e^{\frac{ia}{\om} \sin(\om t)} c_{n}^{\dg}
c_{n+1} ~+~ e^{-\frac{ia}{\om}\sin(\om t)}c_{n+1}^{\dg}c_{n}) \non \\
&& + ~v ~\sum_{n=0}^{L-1} ~(-1)^{n} ~c_{n}^{\dg}c_{n}. \label{ham1}
\end{eqnarray}
The time-dependent electric field appears through a vector potential in the 
phase of the nearest-neighbor hopping following the Peierls 
prescription~\cite{peierls} as follows. If the electric field is 
$\vec E = {\vec E}_0 \cos (\omega t)$, the vector potential will be given by 
$\vec A = - (c/\omega) \sin (\omega t) {\vec E}_0$, since ${\vec E} = -(1/c) 
\partial {\vec A}/\partial t$.
If $q$ is the charge of the particle, the phase of the hopping from a site at
${\vec r}_j$ to a site at ${\vec r}_i$ is given by $(q/\hbar c) {\vec A} \cdot
({\vec r}_i - {\vec r}_j) = - (q /\hbar \omega) \sin (\omega t) {\vec E}_0 
\cdot ({\vec r}_i - {\vec r}_j)$. The parameter $a$ in the phases in the first 
line of Eq.~\eqref{ham1} is therefore given by $a= -(q/\hbar) {\vec E}_0 \cdot 
({\vec r}_n - {\vec r}_{n+1})$. We have also allowed for a staggered on-site 
potential $v$ in the model, and we will study the model with and without $v$.

\noi (ii) We will then consider an interacting model of bosons, namely, the
Bose-Hubbard model which is again driven by a time-dependent electric field 
as described above.
\begin{eqnarray}
H &=& - ~g ~\sum_{n=0}^{L-2}(e^{\frac{ia}{\om}\sin(\om t)} c_{n}^{\dg}c_{n+1}
~+~ e^{-\frac{ia}{\om}\sin(\om t)}c_{n+1}^{\dg}c_{n}) \non \\
&& + ~\frac{u}{2} ~\sum_{n=0}^{L-1}\rho_{n} (\rho_{n}-1), \label{ham2} 
\end{eqnarray}
where $\rho_n = c_n^\dg c_n$. In this interacting model, we will study if
periodic driving can give rise to bound states of two bosons which are 
localized at one end of the system.

In all cases, we will calculate the Floquet operator $U_T$ which is a unitary
operator which time evolves the system from $t=0$ to $t=T$, where $T = 2\pi/
\om$ is the time period (see Appendix A). We will then study the eigenvalues 
and eigenstates of $U_T$. Note that it is sufficient to consider the case
$a \ge 0$ in Eqs.~\eqref{ham1} and \eqref{ham2}, since $a \to -a$ is 
equivalent to shifting the time by $T/2 = \pi/\om$, and the eigenvalues of 
$U_T$ do not change under time shifts (however, the eigenstates of $U_T$ 
change by a unitary transformation as discussed in Appendix A). We will 
sometimes use the fact that the Floquet eigenvalues are invariant under 
time shifts to choose values of the shift where $U_T$ has some special 
symmetries.

\section{Tight-binding model without interactions}
\label{sec3}

\subsection{Tight-binding model}
\label{sec3a}

We begin with a nearest-neighbor tight-binding model in one dimension.
Since we will only consider a system with one particle in this 
section, it does not matter if the particle is a fermion or a boson and
interactions between particles will not play any role. The 
time-independent (undriven) model has the Hamiltonian 
\beq H ~=~ - ~g ~\sum_{n=0}^{L-2} ~(c_{n}^{\dg}c_{n+1} ~+~ c_{n+1}^{\dg} 
c_{n}). \label{ham3} \eeq
When this model is driven by an oscillating electric field as discussed above,
the Hamiltonian is given by
\begin{eqnarray}
H &=& - ~g ~\sum_{n=0}^{L-2} ~(e^{\frac{ia}{\om} \sin(\om t)} c_{n}^{\dg}
c_{n+1} ~+~ e^{-\frac{ia}{\om}\sin(\om t)}c_{n+1}^{\dg}c_{n}). \non \\
&& \label{ham4} \end{eqnarray}
An undriven tight-binding model only admits extended states whose wave 
functions are given by plane waves on the lattice. We will find that 
periodic driving can generate edge states in an open-ended 
(finite length) system for certain values of the driving amplitude $a$. 

It is interesting to note that for an infinite chain in which 
$n$ goes from $-\infty$ to $\infty$ in Eq.~\eqref{ham4}, the effective
Hamiltonian $H_{eff}$ and therefore the energy-momentum dispersion 
can be found exactly (see Appendix B). The dispersion is found to be 
\beq E_k ~=~ - 2 g J_0 \left( \frac{a}{\om} \right) \cos k. \label{disp} \eeq
Interestingly, a flat band is generated if $J_0 (a/\om) = 0$ giving rise
to dynamical localization. 

Before presenting our numerical results, we discuss the concept of inverse 
participation ratio (IPR) which provides a measure of how well a wave function
is localized. Let $\psi_j(n)$ be the $j$-th Floquet eigenstate and $n$ runs 
over the lattice sites $0$ to $L-1$. We assume that this is normalized, so that 
$\sum_{n=0}^{L-1} | \psi_j(n) |^{2}=1 $. Then the IPR of the $j$-th 
eigenstate is defined as $I_j =\sum_{n=0}^{L-1} | \psi_j(n) |^{4}$.
If a state $\psi_j(n)$ is extended equally over all sites, then 
$| \psi_j(n) |^{2} =1/L$ for all $n$, which implies that
$I_j =\sum_{n=0}^{L-1} | \psi_j(n) |^{4}=1/L$.
But if $\psi_j(n)$ is localized over a distance $\xi$ (which is of the order 
of the decay length of the eigenstate, where the decay length remains 
constant as $L \to \infty$), then we have $| \psi_j(n) |^{2}\sim 
1/\xi$ in a region of length $\xi$ and $ \sim 0$ elsewhere; this
implies that $I_j \sim 1/\xi$ which remains finite as $L \to \infty$.
Thus, if $L$ is sufficiently large, a plot of $I_j$ versus $j$ will be able to 
distinguish between states which are localized (over a length scale $\xi 
\ll L$) and states which are extended. Once we find a state $\psi_j$ for which 
$I_j$ is significantly larger than $1/L$ (which is the value of the IPR for 
a completely extended state), we look at a plot of the probabilities 
$| \psi_j(n) |^{2}$ versus $m$ to see whether it is indeed an edge state. 
As discussed below, we will sometimes find that there are states
with large IPR but which are not true bound states at the edges; their
wave functions are much larger at the ends than in the bulk but the wave
functions do not go to zero in the bulk even when $L \to \infty$. As a 
result, the IPR for such states may be much larger than $1/L$ for system 
sizes $L$ of the order of 100 but the IPR would eventually become 
of order $1/L$ if $L$ was of the order of a million or more.

\begin{figure}[H]
\centering
\includegraphics[width=0.49\textwidth]{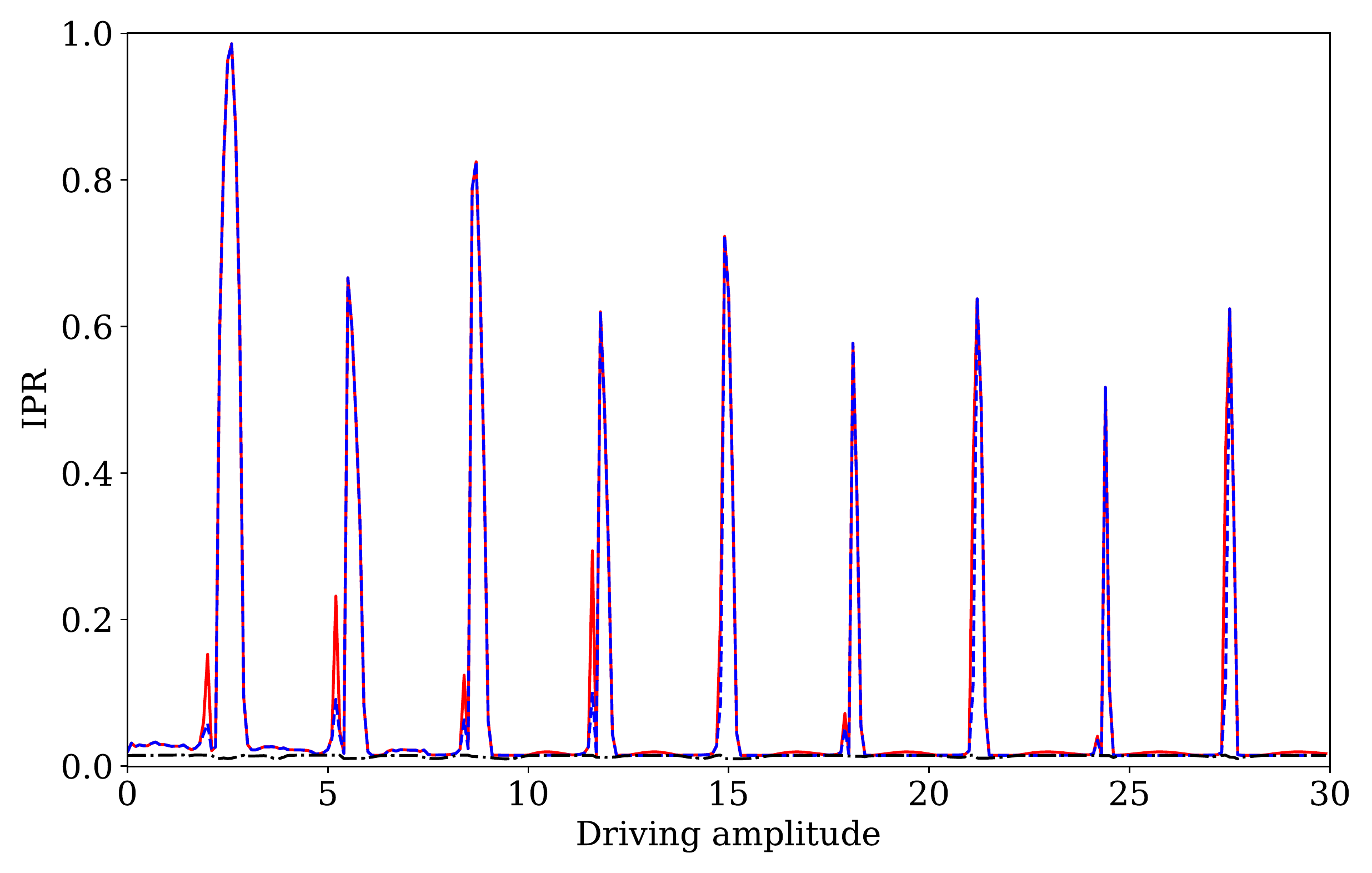}
\caption{Plots of the two maximum IPR values (red solid and blue dashed lines)
and minimum IPR value (black dash-dotted line) as a 
function of the driving amplitude $a$ in the range $[0,30]$. We see that 
states with maximum IPR $\gg 1/L$, called large-IPR states, appear in certain 
intervals of $a$. We have considered a 101-site system with $g=1$ and $\om=1$.} 
\label{ss1fig01} \end{figure}

We now present our numerical results. We have chosen the parameter values 
$g=1$ and $\om=1$ and the system size $L=101$. In Fig.~\ref{ss1fig01} we show 
plots of the largest and second largest values of the IPR as a function of $a$
in the range 0 to 30; these are shown by red and blue curves respectively.
We find that the maximum value of the IPR is very large in certain ranges of 
$a$. We will call these large-IPR states. We will discuss below if these 
states are truly localized at the edges of the system. 

\begin{figure}[H]
\centering
\subfigure[]{\includegraphics[width=0.49\textwidth]{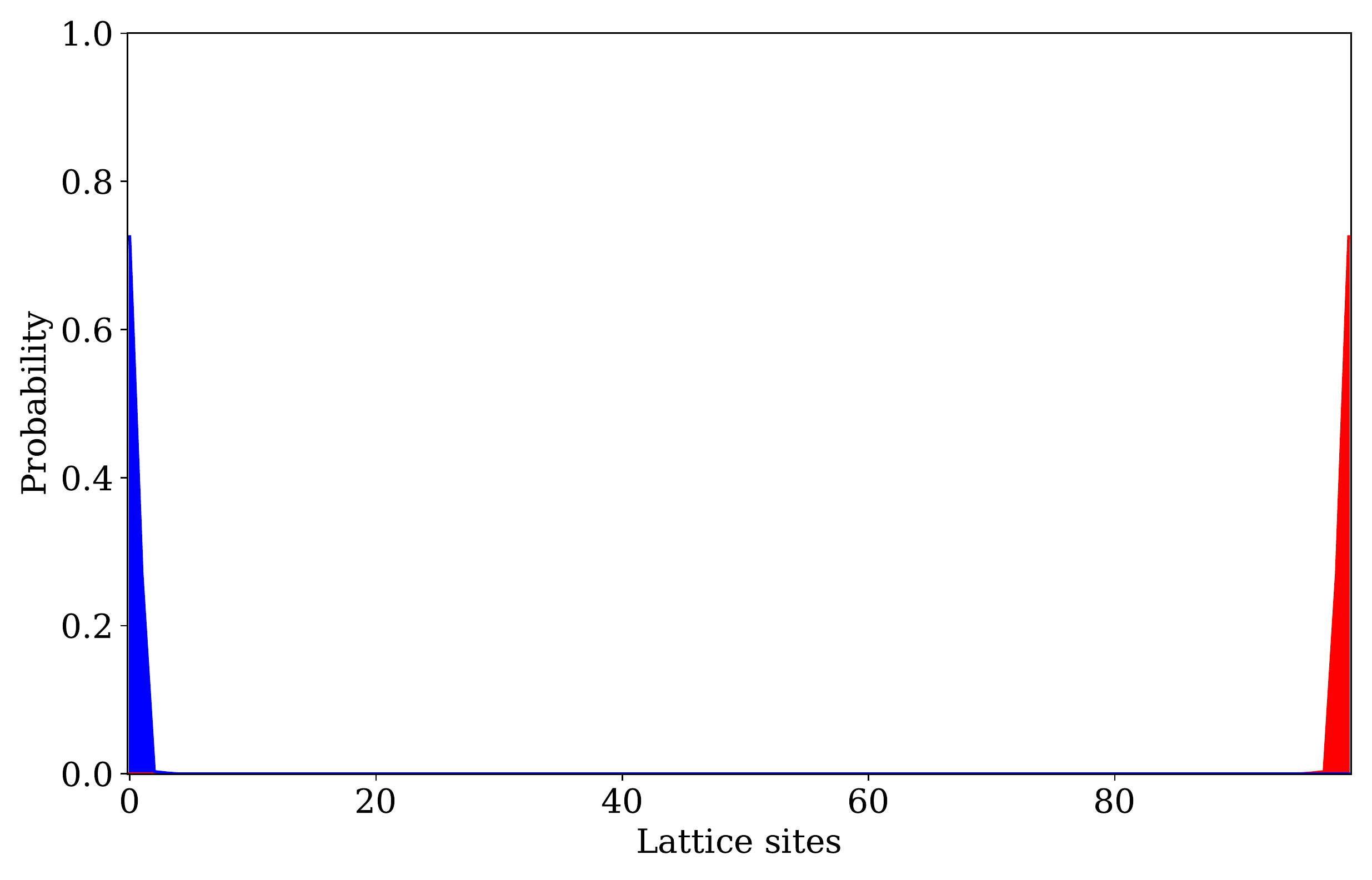}}
\subfigure[]{\includegraphics[width=0.49\textwidth]{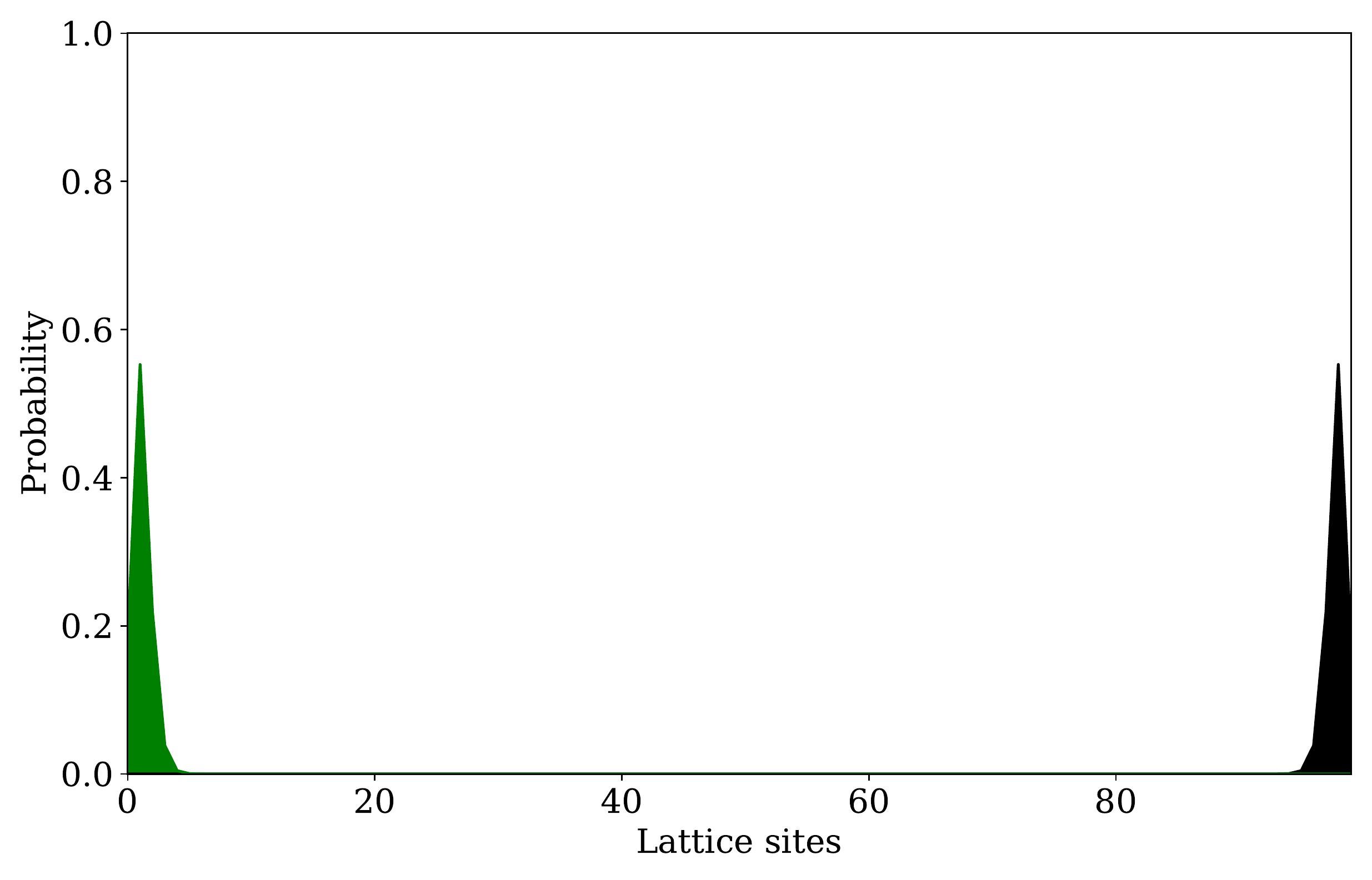}}
\caption{Four edge states for $a=5.6$ (type-1 region), for a 100-site system
with $g=1$ and $\om=1$.} \label{ss1fig02} \end{figure}

\begin{figure}[H]
\centering
\includegraphics[width=0.49\textwidth]{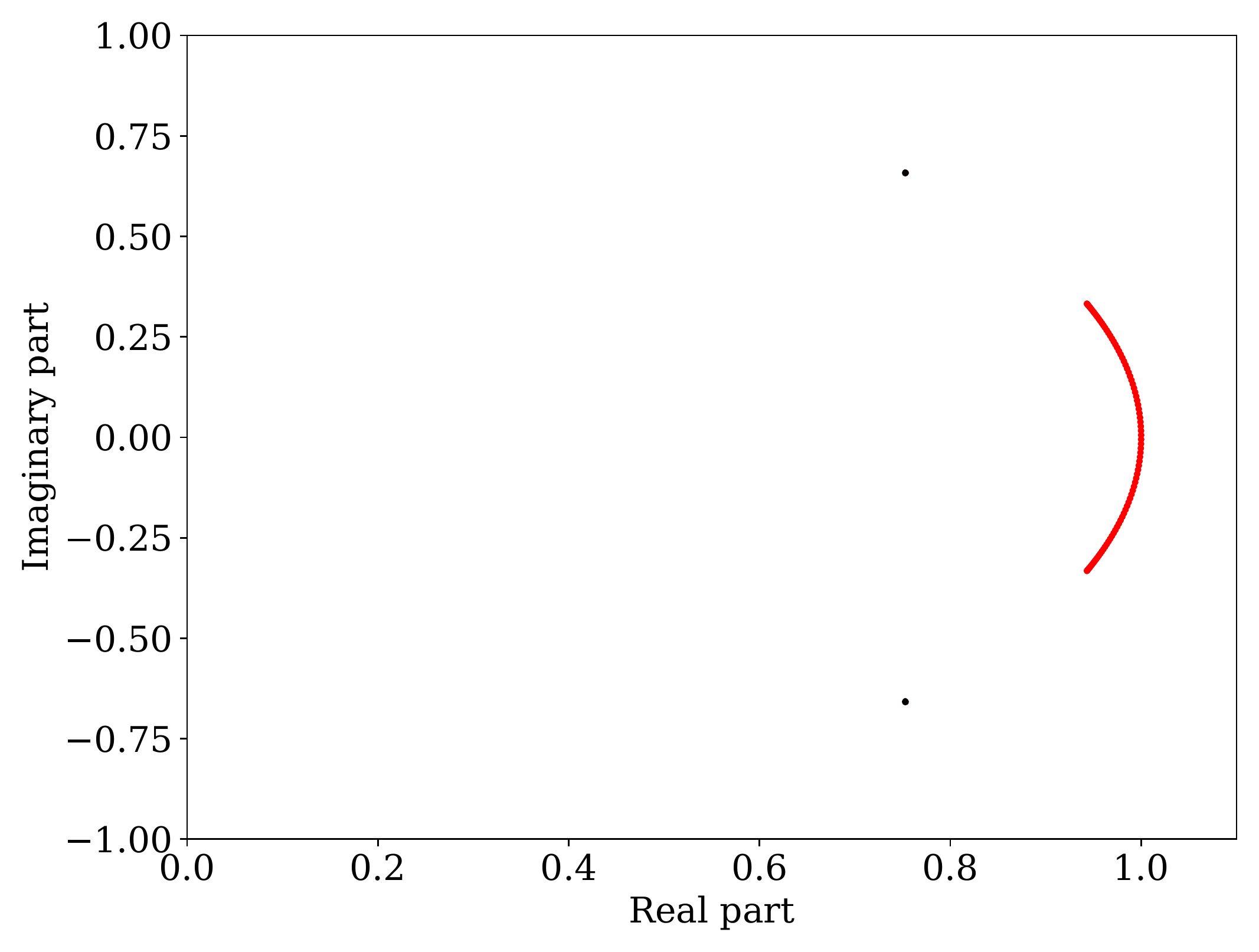}
\caption{Plot of the real and imaginary parts of the Floquet eigenvalues 
$a=5.6$ (type-1 region), for a 100-site system with $g=1$ and $\om=1$. 
There are four large-IPR states (shown in black) which are pairwise degenerate 
(hence we only see two black dots). These are well separated from the bulk 
states (shown by the red curve), and they correspond to bound states 
at the edges.} \label{ss1fig04} \end{figure}
\begin{figure}[H]
\centering
\includegraphics[width=0.49\textwidth]{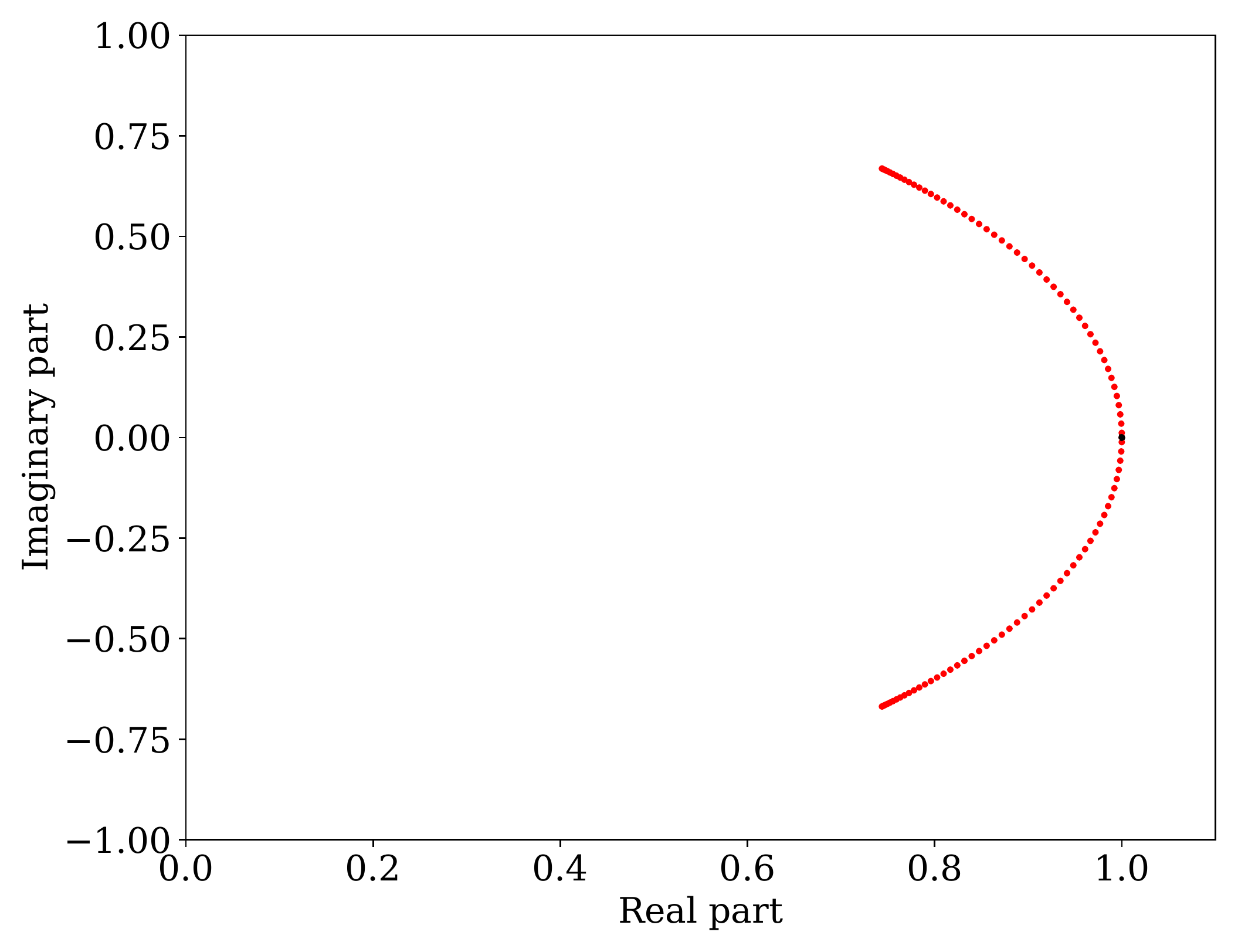}
\caption{Plot of the real and imaginary parts of the Floquet eigenvalues for 
$a=8.44$ (type-2 region), for a 100-site system with $g=1$ and $\om=1$.
There are two large-IPR states (shown in black) whose eigenvalues lie close to
1. These lie within the bulk states (shown by the red curve), and they are 
not true edge states.} \label{ss1fig05} \end{figure}

Interestingly we see that the peaks in the IPR occur in the vicinity of the 
zeros of the Bessel function $J_{0}(a/\om)$. An important thing to note is
that in most of the regions where large-IPR states exist, there are four such 
states (two at each end of the system). These are the regions where both red 
and blue curves have large values; we will call these type-1 regions. 
The probabilities $|\psi (n)|^2$ for the four states are shown in 
Figs.~\ref{ss1fig02} for a 100-site system with $g=1$, $\om = 1$ and $a=5.6$. 
The states shown in Figs.~\ref{ss1fig02} (a) and (b) have somewhat different 
profiles, one of them 
being closer to the end than the other. There also exist small regions where 
only the red curve have a large value in Fig.~\ref{ss1fig01}; we call these 
type-2 regions. These regions are given by the intervals $[1.97,2.07]$, 
$[5.17,5.27]$, $[8.40,8.47]$, etc. In these regions, we have two large-IPR
states, one at each end, if the number of sites $L$ is even, and one large-IPR
state (which has a mode localized at each end simultaneously) if $L$ is odd. 
For a 100-site system with $g=1$, $\om = 1$, and $a=8.44$, we find that
the probabilities for the edge states look very similar to the ones shown in
Fig.~\ref{ss1fig02} (a) and are therefore not shown here. The Floquet 
eigenvalues for $a=5.6$ (type-1) and $a=8.44$ (type-2) cases are shown in 
Figs.~\ref{ss1fig04} and \ref{ss1fig05}.

\begin{figure}[H]
\centering
\includegraphics[width=0.52\textwidth]{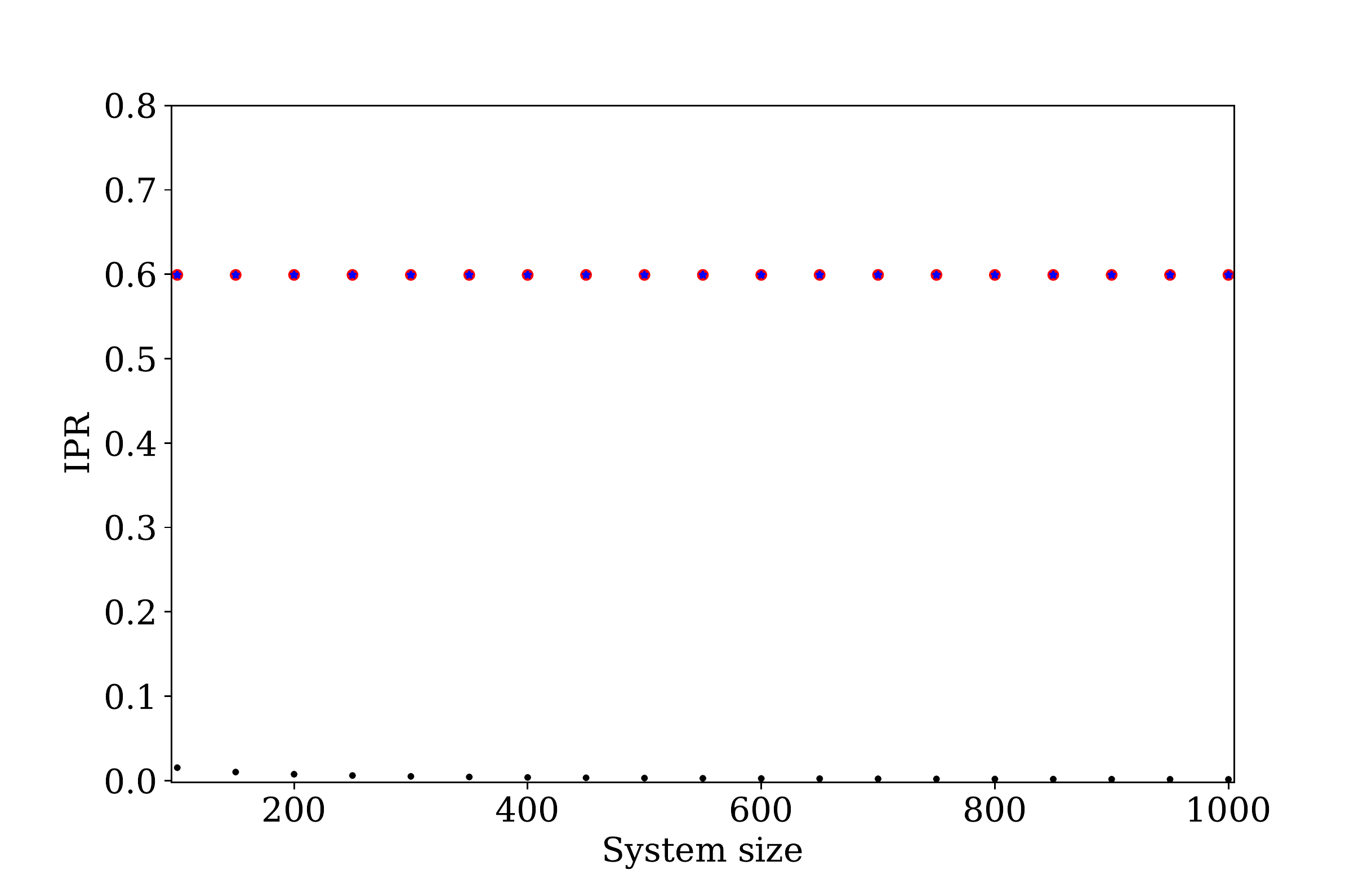}
\caption{Plot of the largest two IPRs and the minimum IPR versus system 
size from $L=100$ to 1000 (in steps of 50), for $g=1$, $\om=1$ and $a=5.6$ 
(type-1 region). The largest two IPRs correspond to the states at the two 
ends which are true edge states, hence the IPRs do not change with the system 
size. The minimum IPR corresponds to one of the bulk states.}
\label{ss1fig03} \end{figure}

We find that there is a significant difference between the large-IPR states 
of type-1 and type-2. The type-1 states are exponentially localized at the 
edges; their wave functions go to zero rapidly as we go away from the edges; 
hence they are true edge states. (For example, for $g=1$, $\om =1$ and 
$a=5.6$ which lies in the type-1 region, we find that for the edge states, the 
probability $|\psi (n)|^2$ in the middle of a 100-site system is only about 
$10^{-32}$ which is essentially zero). Hence their IPR remains large and
constant as the system size is increased. This is shown in 
Fig.~\ref{ss1fig03} for a system with $g=1$, $\om =1$, and $a=5.6$ (type-1 
region). The type-2 states have a large amplitude at the edges, but their 
wave functions approach some finite (although very small) values as we go away
from the edges. Thus the type-2 states are not perfectly localized at 
the edges; they have a small but finite weight deep inside the bulk, even
when the system size becomes very large. This difference in behavior is 
related to the following. We will see below that the Floquet 
eigenvalues of the type-1 edge states differ from those of the bulk states 
by a gap which remains finite as the system size $L \to \infty$.
In contrast, the Floquet eigenvalues of the type-2 states lie in the
middle of those of the bulk states; the gap between the eigenvalues of these
large-IPR states and the bulk states goes to zero as the system size goes to 
infinity. We will see that the type-2 large-IPR states are actually made up
of a linear combination of some bulk states.

We now present some numbers to provide a more detailed understanding
of the large-IPR states which are actually not bound states at the edges.
For convenience, we will discuss in this paragraph what happens if the
phase of the hopping in Eq.~\eqref{ham4} is given by $\cos (\om t)$ rather 
than $\sin (\om t)$; then a Floquet eigenstate $\psi$ which exists at the 
left edge (starting at site $n=0$) will have $\psi (n) = 0$ for all odd values 
of $n$ (see the discussion about symmetry in item (iii) below). It then 
turns out that we can tune one of the driving parameters (say, $a$) to find a 
value for which there is a state which is almost perfectly localized at one 
end of the system. For instance, for a 100-site system with $g=1$, $\om = 1$, 
we find that for $a=8.439$, there is a state which is large at the left
end, with Floquet eigenvalue equal to 1 and IPR equal to $0.9982$. 
In this state, we find that the probabilities
at the different sites are given approximately by $|\psi (0)|^2 = 0.9991$, 
$|\psi (2)|^2 = 8.2 \times 10^{-4}$, $|\psi (4)|^2 = 1.8 \times 10^{-6}$, 
$|\psi (6)|^2 = |\psi (8)|^2 = |\psi (10)|^2 = \cdots = 1.2 \times 10^{-6}$, 
and $|\psi (n)|^2 = 0$ if $n$ is odd. The fact that the Floquet eigenvalue 
is equal to 1 implies that deep inside the bulk, this state must be a 
superposition of states with momenta $k = \pm \pi/2$, so that its energy is 
equal to $E_k = -2 g J_0 (a/\om) \cos k = 0$ (see Appendix B).
Next, we find that the above values of $|\psi (n)|^2$ 
remain unchanged even when $L$ is increased to, say, 1000. This can
be understood as follows. The numerical program automatically normalizes the 
wave functions for a finite-sized system; hence for a system with $L$ sites 
and therefore $L/2$ even-numbered sites (we will assume that $L \gg 1$), the 
probabilities at the even-numbered sites are given by
\bea && (|\psi (0)|^2, |\psi (2)|^2, |\psi (4)|^2, |\psi (6)|^2, \cdots) 
\non \\
&& \simeq~ \frac{1}{1 + 1.2 \times 10^{-6} (L/2)} \non \\
&& ~~\times ~(0.9991, 8.2 \times 10^{-4}, 1.8 \times 10^{-6}, 
1.2 \times 10^{-6}, \cdots). \non \\
&& \eea
We then see that the probabilities at the first few sites will not change much 
from the values they have for $L=100$ till $L$ starts becoming comparable to 
$1/(1.2 \times 10^{-6}) \sim 10^6$. Clearly, one needs to go to enormous 
system sizes to distinguish between a true edge state (type-1) and a state 
which is not a bound state but has an IPR close to 1 when $L$ is about 100. 

We thus conclude that periodic driving of a non-interacting tight-binding 
model with certain values of the driving amplitude $a$ can generate large-IPR
states. These are bound states at the edges for type-1 but not true edge 
states for type-2 (although they can be made almost indistinguishable 
from a true edge state by tuning the driving parameters as we have seen
above).

We will now discuss the symmetry properties of the Floquet operator $U_T$ 
which will, in turn, imply symmetries of the large-IPR states. The symmetries 
of $U_T$ follow from its definition as a time-ordered product (Appendix A). 
Using Eq.~\eqref{ham4}, we can write the Hamiltonian $H$ for one
particle as a $L \times L$ matrix in the basis of states $| n \ra$
(which denotes the state where the particle is at site $n$). The symmetries
of $U_T$ then follow from the symmetries of $H$ as follows.

\noi (i) The fact that $\sin (\om t) = - \sin (\om (T - t))$ implies
that $H^* (t) = H (T-t)$. This implies that $U_T^* = U_T^{-1}$. If $\psi$
is a Floquet eigenstate satisfying $U_T \psi = e^{i \ta} \psi$, this symmetry
implies that $\psi^*$ is also a Floquet eigenstate with the same eigenvalue.
We can then consider the superpositions $\psi + \psi^*$ and $i (\psi - \psi^*)$
to show that $\psi$ can be chosen to be real.

\noi (ii) If we combine the parity transformation $|n \ra \to | L - 1 -n \ra$
with $|n \ra \to (-1)^n |n \ra$ and complex conjugation, we find that 
$H (t) \to - H^* (t)$. This implies that $U_T$ is unitarily related
to $U_T^*$. This implies that if $\psi (n)$ denotes the $n$-th component
of an eigenstate of $U_T$ with eigenvalue $e^{i \ta}$, then a state $\psi'$ 
with $\psi' (n) = (-1)^n \psi^* (L-1-n)$ is an eigenstate of $U_T$ with 
eigenvalue $e^{- i\ta}$. This implies that if there is a Floquet eigenstate 
with a large weight near the left edge of the system with eigenvalue 
$e^{i \ta}$, there will be an eigenstate with a large weight near the right 
edge with eigenvalue $e^{-i \ta}$. It is clear that these two states
have the same IPR since $\sum_{n=0}^{L-1} |\psi (n)|^4$ is invariant under
$\psi (n) \to (-1)^n \psi (L-1-n)$.

\noi (iii) If we shift the time $t \to t + T/4$, the term in the phase of the 
hopping amplitude in Eq.~\eqref{ham4} changes from $\sin (\om t)$ to $\cos 
(\om t)$. If we combine this with the transformation $|n \ra \to (-1)^n |n>$,
we have that $H (t) \to - H(T-t)$. More specifically, the transformation
$|n \ra \to (-1)^n |n \ra$ is done by the unitary and diagonal matrix $W$ 
whose diagonal elements are given by $W_{nn} = (-1)^n$; since $W^2 = I$, the 
eigenvalues of $W$ are equal to $\pm 1$. Then 
\beq W H(t) W ~=~ - ~H(T-t), \label{wh} \eeq 
and this implies that $W U_T W = U_T^{-1}$. Hence, for every Floquet eigenstate
$\psi$ with eigenvalue $e^{i \ta}$, there will be an eigenstate $W \psi$ with 
eigenvalue $e^{-i \ta}$. We now recall that Floquet eigenvalues do not change 
under time shifts while eigenstates change by a unitary transformation; 
however, if there is a state with a large weight near one particular edge, 
its unitary transformation will give a state with a large weight 
at the same edge. We therefore conclude that near each edge, large-IPR states
will either come in pairs with Floquet eigenvalues equal to $e^{\pm i \ta}$
(if $e^{i \ta} \ne \pm 1$), or they can come singly if the eigenvalue is
equal to $\pm 1$. Further, the argument in the previous paragraph shows that
there will be corresponding states with large weight at the opposite edge 
with the same eigenvalues. Also, if there is a single large-IPR state $\psi$ 
near an edge with Floquet eigenvalue equal to $\pm 1$, $W \psi$ will have
the same eigenvalue and therefore must be identical to $\psi$ up to a sign.
Hence $\psi$ must be an eigenstate of $W$. This means that the components
of $\psi$, denoted as $\psi (n)$ must be zero if $n$ is odd (even) depending
on whether $W \psi$ is equal to $+ \psi$ or $- \psi$.

\subsection{Tight-binding model with a staggered potential}
\label{sec3b}

\begin{figure}[H]
\centering
\subfigure[]{\includegraphics[width=0.49\textwidth]{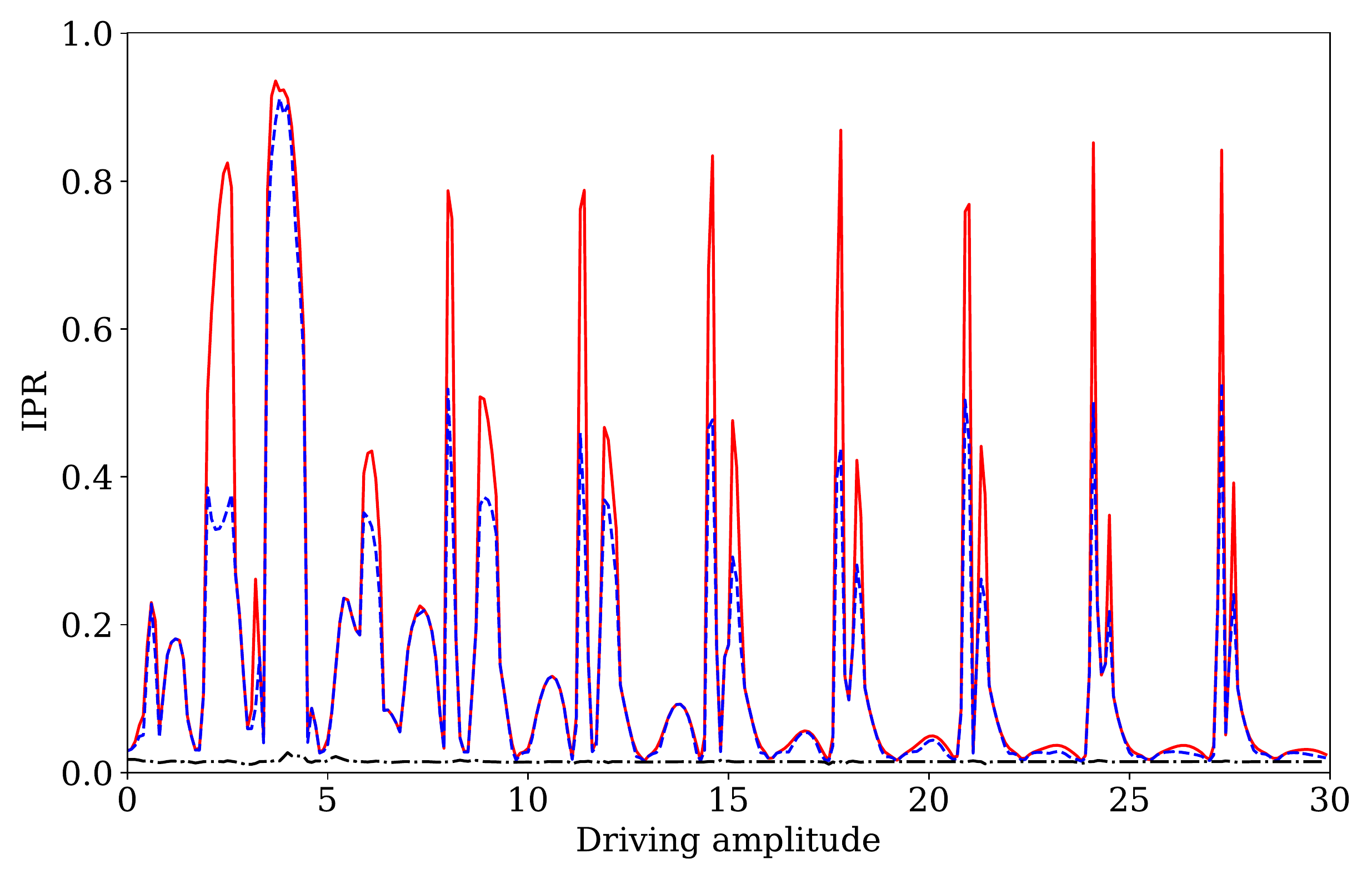}}
\subfigure[]{\includegraphics[width=0.49\textwidth]{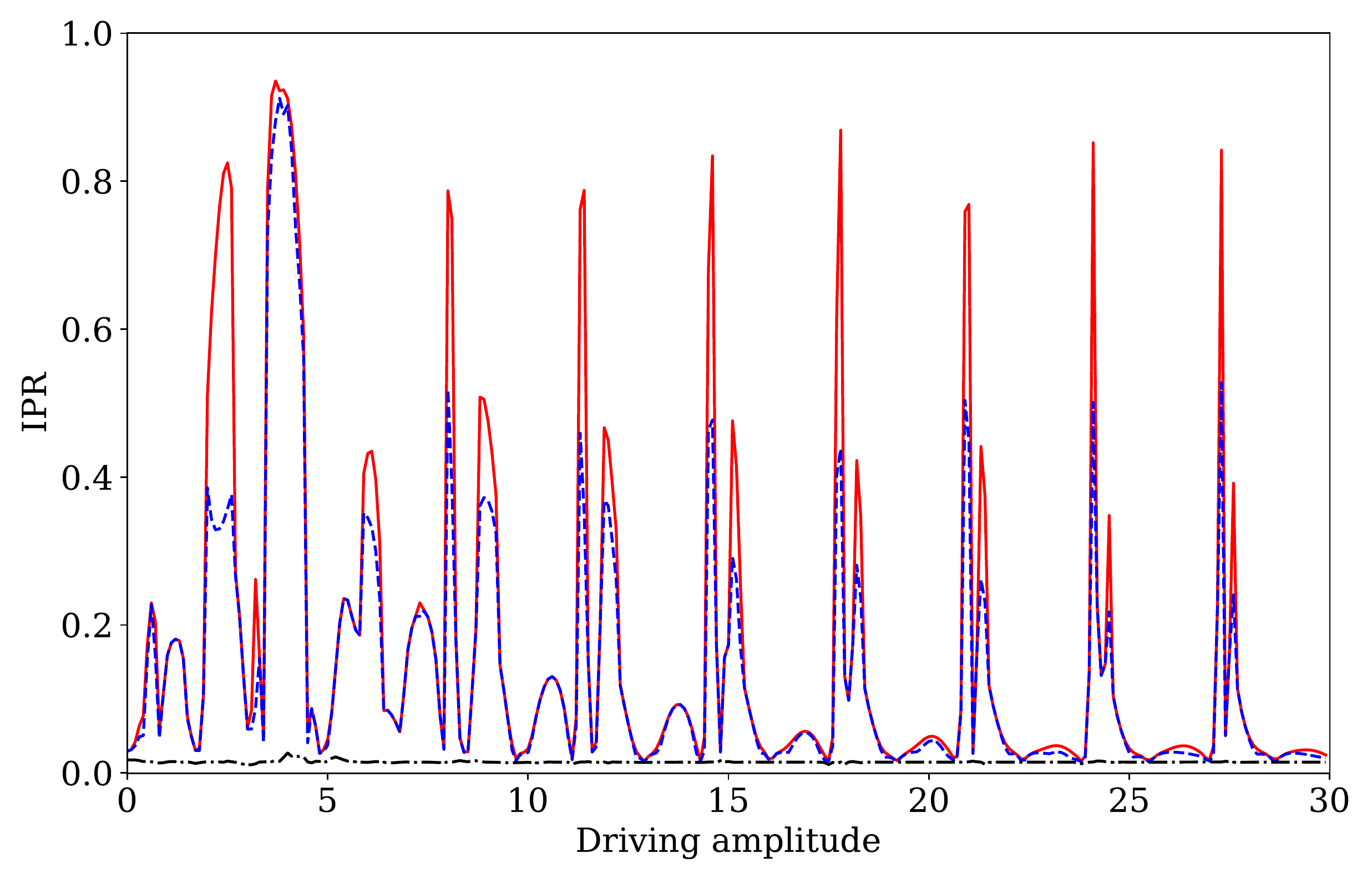}}
\caption{Largest two IPRs as a function of the driving amplitude $a$ for 
staggered potentials $v = 1$ and $-1$, for a 101-site system with $g=1$ and
$\om = 1$. The largest and second largest IPRs are shown as red solid and blue
dashed lines. The smallest IPR is shown by a black dash-dotted line near the
bottom. The two figures look identical due to a parity symmetry as discussed 
in the text.} \label{ss1fig06} \end{figure}

We will now study the effects of a staggered potential $v$. The Hamiltonian 
with driving is given by
\begin{eqnarray} H &=& - ~g~\sum_{n=0}^{L-2} ~(e^{\frac{ia}{\om}\sin(\om t)} 
c_{n}^{\dg}c_{n+1} ~+~ e^{-\frac{ia}{\om}\sin(\om t)}c_{n+1}^{\dg}c_{n})
\non \\
&& +~ v ~\sum_{n=0}^{L-1} ~(-1)^{n} c_{n}^{\dg}c_{n}, \label{ham5} 
\end{eqnarray}
where $v$ is the strength of the staggered potential. The numerical results 
that we obtain are as follows. We have considered a 101-site system with
$g=1$ and $\om=1$. We then find that $v$ and $-v$ give identical plots for the
IPRs. (We can show that the symmetry (ii) discussed above continues to hold if 
we also transform $v \to - v$, provided that $L$ is odd). We vary the driving 
amplitude $a$ from 0 to 30 in steps of $0.1$, and plot the largest two IPRs 
as a function of $v$. Figure~\ref{ss1fig06} shows the IPRs for $v=\pm 1$.
We find that large IPR values correspond to 
a pair of eigenstates which are localized at the opposite ends of the system 
and have the same Floquet eigenvalues. Since large values of IPRs imply the 
presence of edge states, we see that the regions where edge states exist 
are significantly larger compared to the case with $v=0$ (Fig.~\ref{ss1fig01}). 

We can understand why the edge states at the opposite ends have the same
Floquet eigenvalues as follows. When the number of lattice sites, $L$, is odd,
the Hamiltonian and therefore the Floquet operator $U_T$ are invariant under 
a combination of parity ($n \to L - 1 -n$) 
and $a \to - a$. We have seen earlier that changing $a \to - a$ does not change
the Floquet eigenvalues. The above symmetry therefore implies that if $\psi_L$ 
is a Floquet eigenstate localized near the left edge with a Floquet 
eigenvalue $e^{i \ta}$, there will be a Floquet eigenstate $\psi_R$ localized 
near the right edge with the same eigenvalue $e^{i \ta}$.

\begin{figure}[H]
\centering
\subfigure[]{\includegraphics[width=0.49\textwidth]{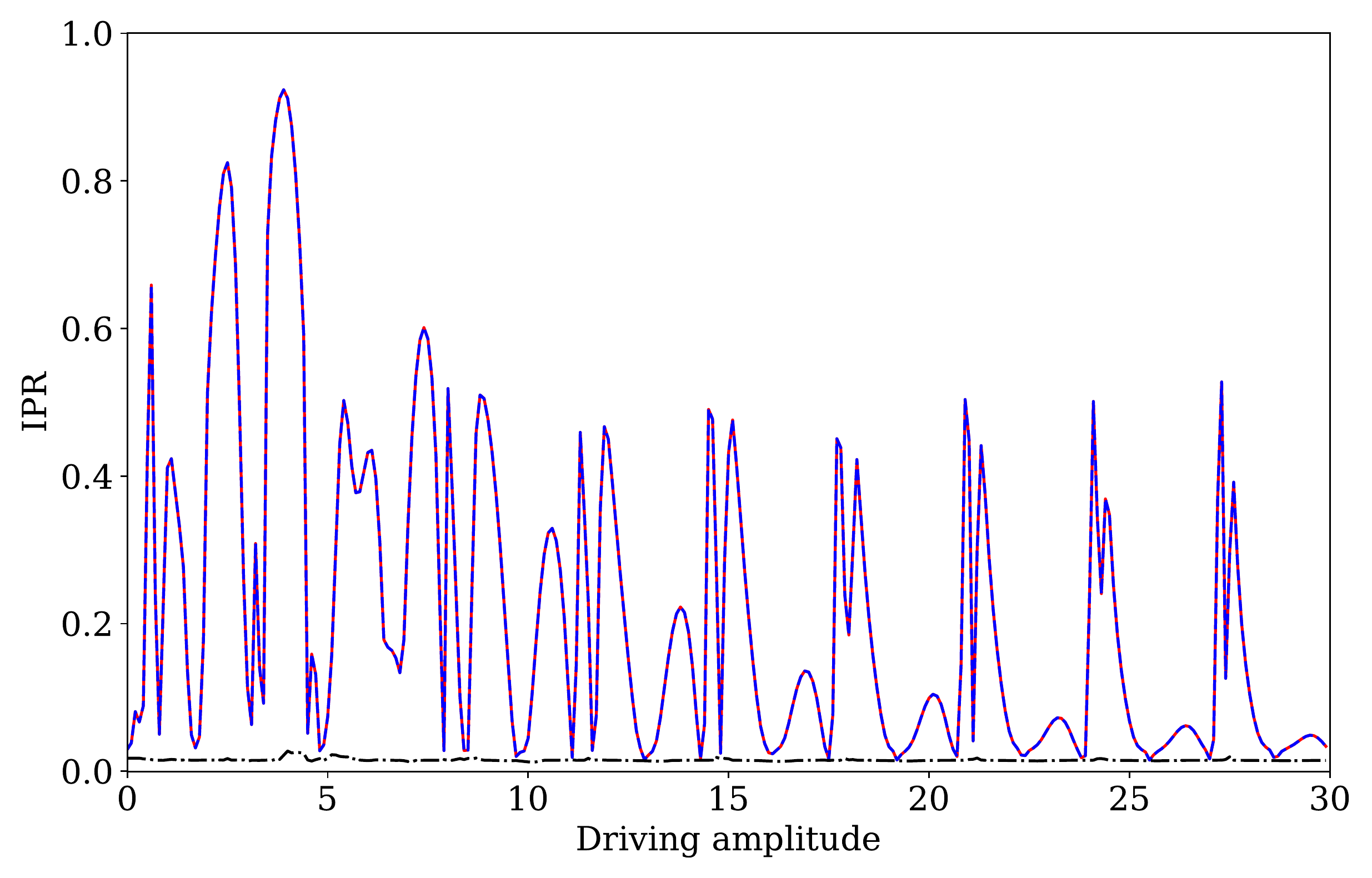}}
\subfigure[]{\includegraphics[width=0.49\textwidth]{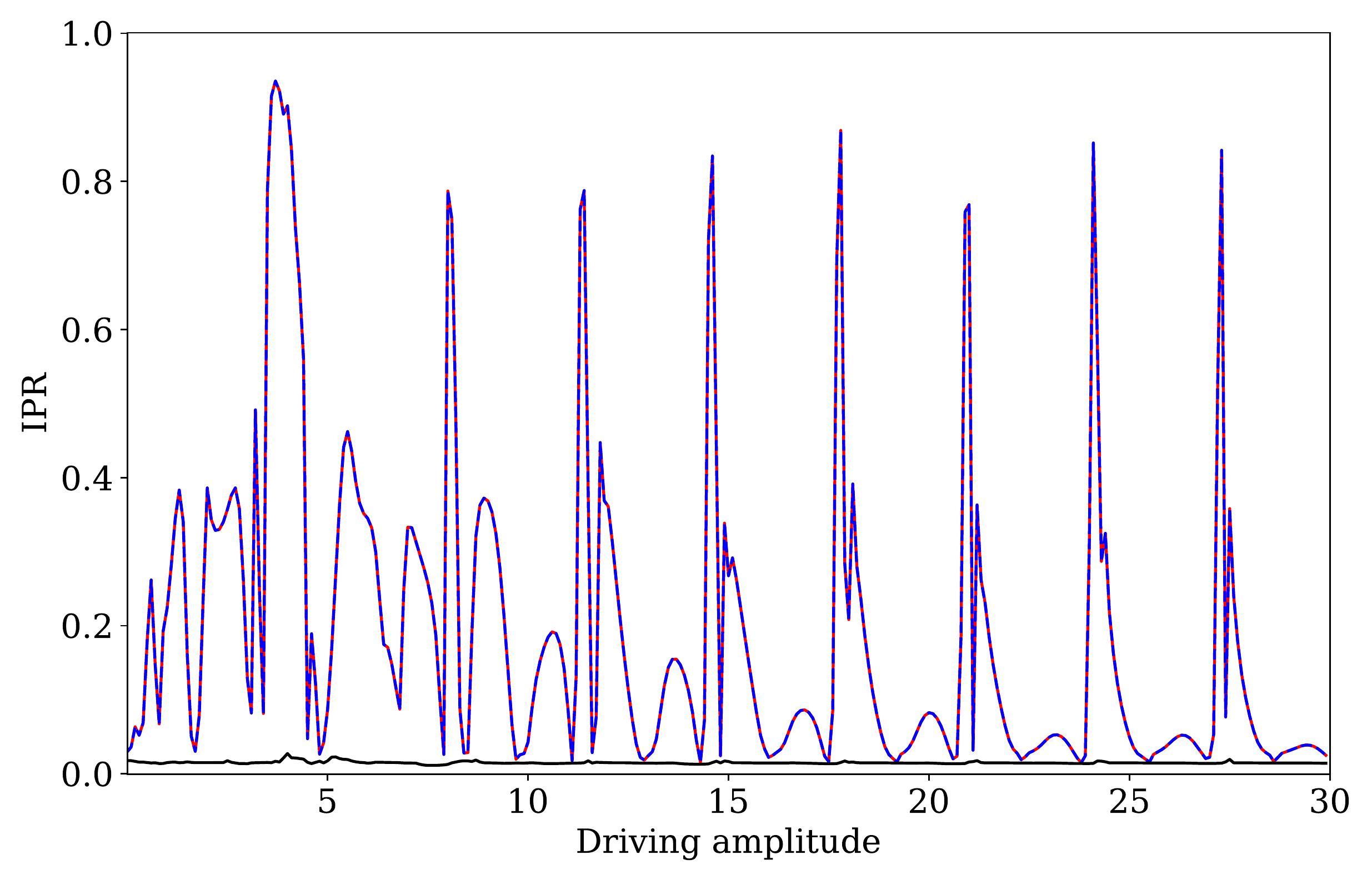}}
\caption{Largest two IPRs (which coincide with each other) as a function of 
the driving amplitude $a$ for staggered potentials $v = 1$ and $-1$, for a 
100-site system with $g=1$ and $\om = 1$. The smallest IPR is shown by a 
dash-dotted line near the bottom. The two figures are not identical as there 
is no parity symmetry when the number of sites is even.} 
\label{ss1fig07} \end{figure}

There is an interesting difference between the cases where the number
of sites is even and odd. In Fig.~\ref{ss1fig07}, we show the largest two 
eigenvalues for $v=\pm 1$ for a 100-site system in Fig.~\ref{ss1fig07}. 
We find once again that large values of the IPR correspond to a pair of 
eigenstates which are localized at opposite ends of the system; however their 
Floquet eigenvalues are complex conjugates of each other, unlike the case of 
an odd number of sites where the Floquet eigenvalues of the state at opposite 
ends are equal to each other.


\begin{figure}[H]
\centering
\includegraphics[width=0.49\textwidth]{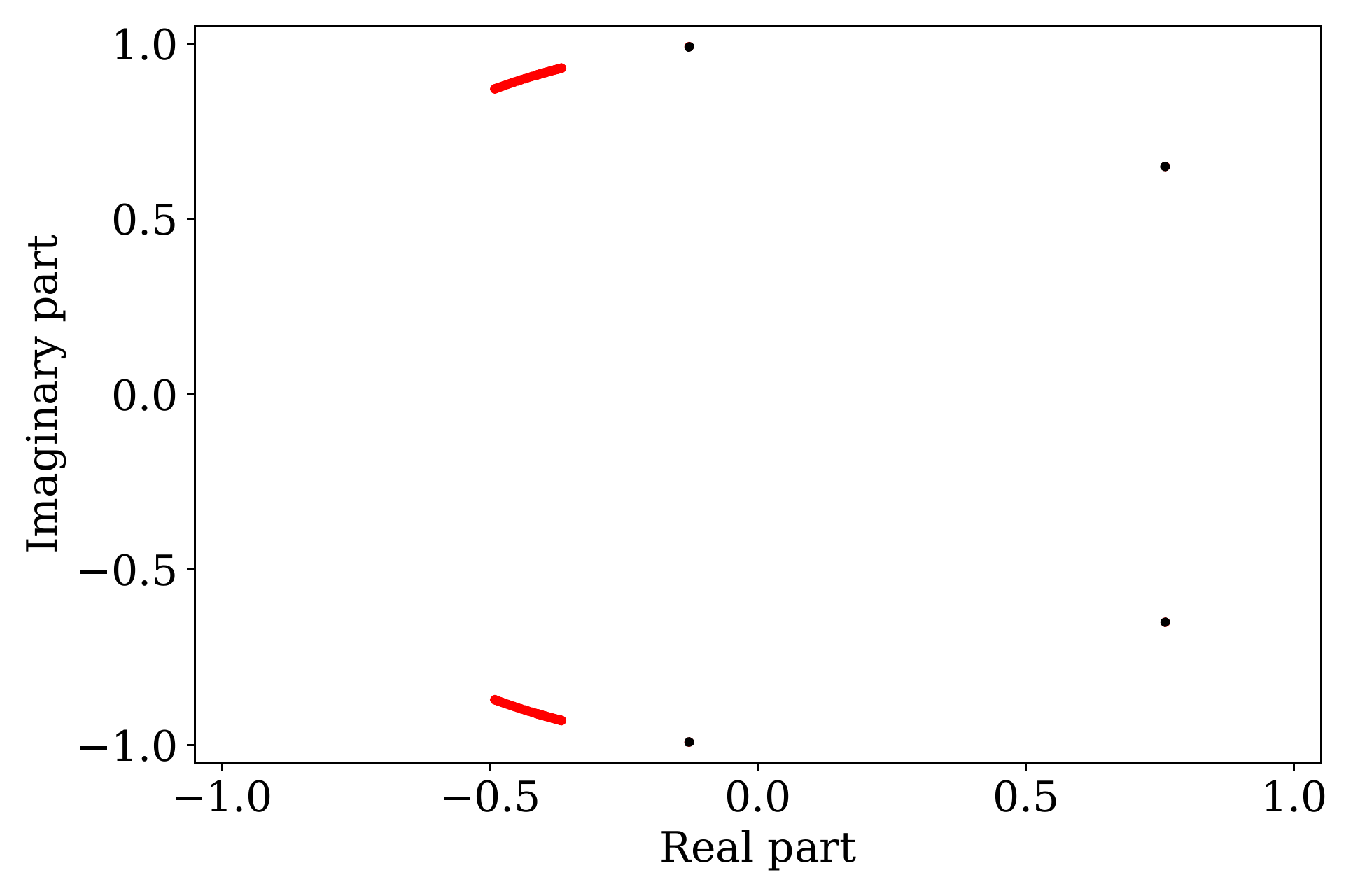}
\caption{Plot of the real and imaginary parts of the Floquet eigenvalues 
for a 100-site system with $g=1$, $v=1$, $\om =1$ and $a=4$. 
There are four large-IPR states (shown in black). These are well separated 
from the bulk states (shown by the red curves), and they correspond 
to bound states at the edges.} \label{ss1fig09} \end{figure}

In Fig.~\ref{ss1fig09} we show the Floquet eigenvalues for a 100-site 
system with $g=1$, $\om = 1$, $v=1$ and $a=4$. The four isolated points
correspond to edge states (two at each end). The value $a=4$ has been
chosen since it corresponds to a peak in the largest two IPRs as shown in
Fig.~\ref{ss1fig07} (a).

\begin{figure}[H]
\centering
\subfigure[]{\includegraphics[width=0.49\textwidth]{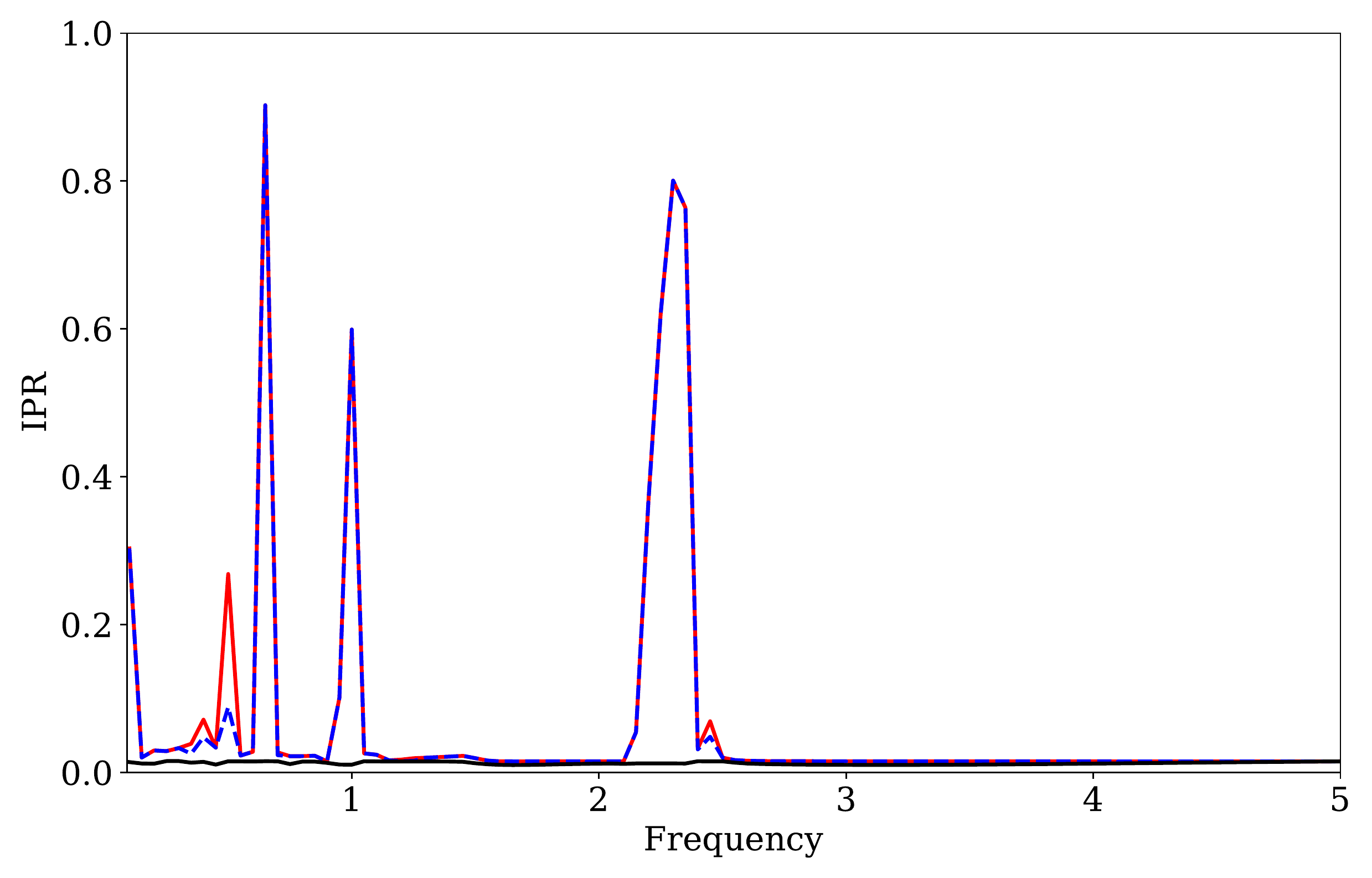}}
\subfigure[]{\includegraphics[width=0.49\textwidth]{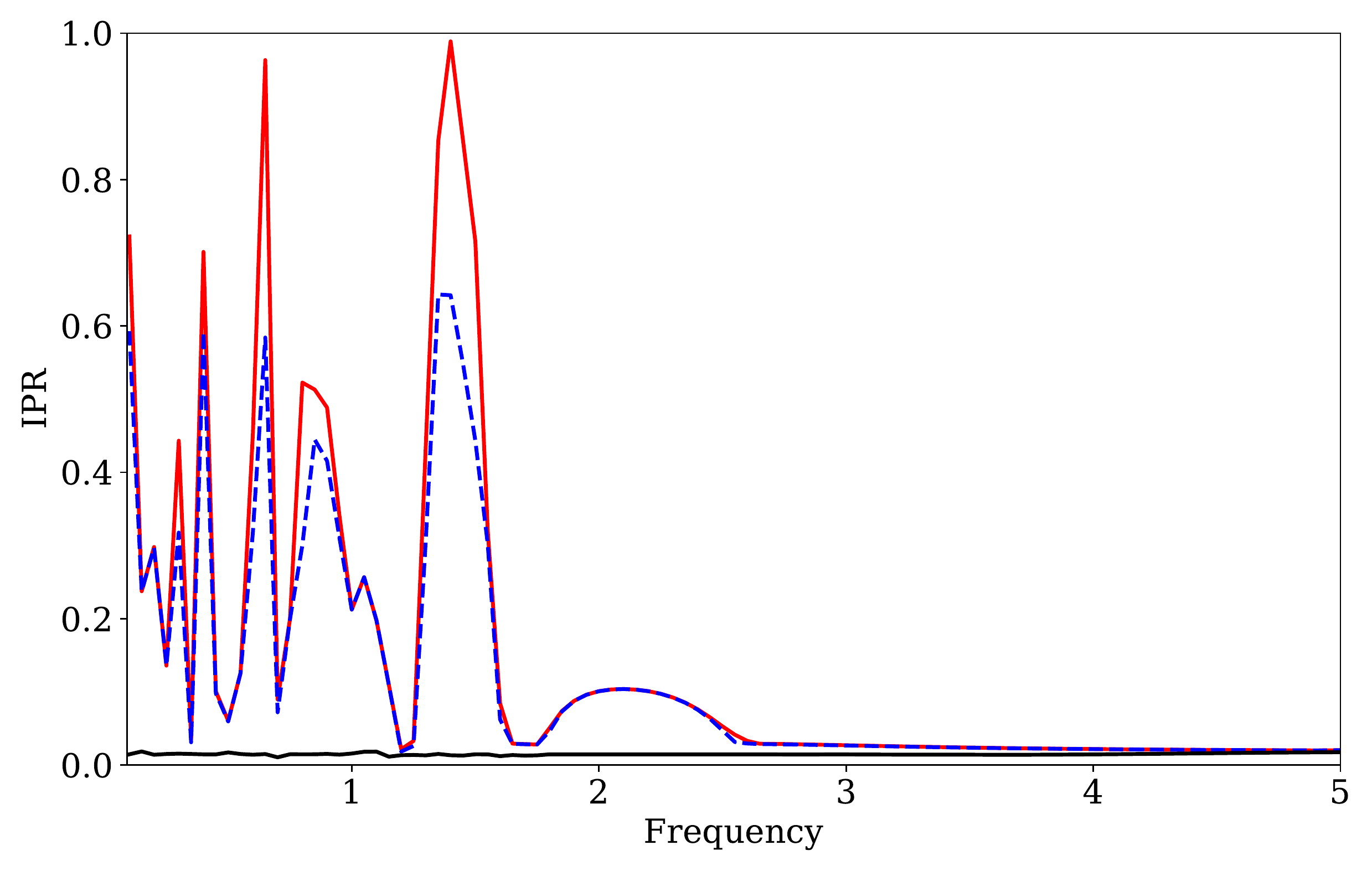}}
\caption{Largest two IPRs as a function of the driving frequency $\om$ for 
staggered potentials $v = 0$ and $1$, for a 101-site system with $g=1$ and 
$a = 5.6$. The smallest IPR is shown by a black solid line near the bottom.} 
\label{ss1fig08} \end{figure}

In Fig.~\ref{ss1fig08} we show the largest two IPRs as a function of $\om$ for
a 101-site system with $g=1$ and $a=5.6$; we have taken $v=0$ and 1 in 
Figs.~\ref{ss1fig08} (a) and (b) respectively. The figures show that edge
states appear in a larger range of values of $\om$ when $v$ is non-zero.
However, edge states do not appear in either case when $\om$ becomes large
enough.

\subsubsection{Floquet perturbation theory for $g \ll v$}
\label{sec3b1}

We have seen numerically that the introduction of a non-zero $v$ enhances the 
regions of $a$ where edge states exist. We would now like to understand
this analytically. Since we have taken the frequency $\om$ to be of the same
order as the hopping amplitude $g$, the Floquet-Magnus expansion in powers
of $1/\om$ would not be useful here~\cite{bukov,mikami}. We will therefore 
use a different approach to understand the appearance of edge states. Namely,
we will consider the limit $g \ll v$, and will use a Floquet perturbative 
expansion in $g$ to obtain a time-independent Hamiltonian. We will then
match the results obtained in this way to those found numerically.

We begin by briefly discussing the Floquet perturbation 
theory~\cite{soori,bhaskar,asmi}. We write the Hamiltonian in Eq.~\eqref{ham5}
as $H= H_0 + V$, where
\begin{eqnarray}
H_{0} &=& v ~\sum_{n=0}^{L-1} ~(-1)^{n} c_{n}^{\dg}c_{n}, \non \\
V &=& - ~g ~\sum_{n=0}^{L-2} ~[e^{\frac{ia}{\om} \sin(\om t)} c_{n}^{\dg}
c_{n+1} ~+~ e^{-\frac{ia}{\om}\sin(\om t)} c_{n+1}^{\dg}c_{n}]. \non \\
\end{eqnarray} 
The eigenstates of $H_{0}$ are the states localized at various sites 
$|n \rangle = | 0 \rangle, | 1 \rangle, | 2 \rangle, \cdots$, and the
eigenenergies are $E_{n} =+v$ or $-v$ depending on whether $n$ is even or odd. 
Now let $| \psi \rangle$ be a solution of the time-dependent Schr\"odinger 
equation. Then
\begin{equation} i \frac{d\psi}{dt} ~=~ (H_{0} ~+~ V(t)) ~\psi. 
\label{Scheq} \end{equation}
We now write $\psi$ in terms of the eigenstates of $H_{0}$ as 
\begin{equation}
| \psi(t) \rangle ~=~ \sum_{m} ~c_{m}(t) ~e^{-i E_{m} t} | n \rangle.
\end{equation}
Substituting this in Eq.~\eqref{Scheq}, we get an equation for the 
coefficients $c_{n}(t)$,
\begin{equation} \frac{dc_{m}}{dt} ~=~ -i \sum_{m' \ne m} \langle m| V(t) | 
m' \rangle ~e^{i(E_{m}-E_{m'})t} ~c_{m'}(t). \label{cnt1} \end{equation}
We now solve Eq.~\eqref{cnt1} up to terms of order $g^{2}$. Integrating the 
above equation, we obtain
\begin{eqnarray}
&& c_{m}(T) \non \\
&=& c_{m}(0) ~-~ i \sum_{m' \ne m} \int_{0}^{T} dt \langle m| V(t)
| m' \rangle e^{i(E_{m}-E_{m'})t} c_{m'}(0) \non \\
&& - \sum_{m'\ne m,m''} \int_{0}^{T} dt \langle m| V (t) | m' \rangle 
e^{i(E_{m}-E_{m'})t} \non \\
&& ~~~\times \int_{0}^{t} dt' \langle m' | V (t') | m'' \rangle 
e^{i(E_{m'}-E_{m''})t'} c_{m''}(0). \label{cnt2} \end{eqnarray}
We can re-write Eq.~\eqref{cnt2} as a matrix equation
\begin{equation}
c_m (T) e^{-iE_m T} ~=~ \sum_{m'} (I ~-~ i T H_{F}^{(1)} ~-~ 
\dfrac{T^{2}}{2} H_{F}^{(2)})_{mm'} c_{m'} (0), \label{cnt3} \end{equation}
where 
$I$ is the identity matrix. We can now write the matrix appearing
in Eq.~\eqref{cnt3} in the form
\begin{equation}
U_{eff} ~=~ e^{-i H_{eff} T} ~\equiv~ I ~-~ i T H_{F}^{(1)} ~-~ 
\dfrac{T^{2}}{2} H_{F}^{(2)} \label{cnt4} \end{equation}
up to order $T^2$.
Namely, we have a unitary matrix $U_{eff}$ which is related to an effective 
Hamiltonian $H_{eff}$ which is correct up to order $g^2$.
We find that $H_{eff}$ in terms $H_{F}^{(1)}$ and $H_{F}^{(2)}$ is
\begin{equation}
H_{eff} ~=~ H_{F}^{(1)} ~+~ \frac{iT}{2} ~[H_{F}^{(2)} ~-~ (H_{F}^{(1)})^{2}]. 
\end{equation}
We can now 
numerically compute the eigenvalues and eigenstates of $U_T= \mathcal{T} 
\exp (-i\int_{0}^{T} H(t)dt)$ and compare these with the same quantities 
for $U_{eff}$.



We note first that if $g=0$, the eigenstates of $U_T$ are just states
localized at different lattice sites; there is a large degeneracy as states
localized at even and odd number of sites have the same eigenvalues.
With the introduction of even a small $g$, we see a drastic change in the 
eigenstates. For $g/v \simeq 1/100$ we find two localized edge states (one 
at each end) while all the other states are extended over the whole system. 
We find this numerically for both the periodically driven system ($U_T$) and
the Floquet perturbative calculation ($U_{eff}$). For a 100-site system
with $g=1$, $\om =1$, $a=2.6$, and a large staggered potential $v =10.1$, 
we find that both $U_T$ and $U_{eff}$ 
have exactly two eigenstates localized at the edges (one at each edge), while 
all the other states are bulk states delocalized over the entire lattice. 
Further, the probabilities $|\psi (n)|^2$ for the edge states for $U_T$ and 
$U_{eff}$ look almost identical to each other (and are similar to the ones 
shown in Fig.~\ref{ss1fig02} (a)).

To summarize this section, we see that the introduction of a small 
hopping $g$ in the driven system with a staggered potential $v$ can change 
the properties of the eigenstates significantly. We observe this numerically 
for both the driven system and the effective Hamiltonian obtained from 
Floquet perturbation theory.

\subsubsection{Study of the maximum IPR versus $v$ for different values of $a$}
\label{sec3b2}

\begin{widetext}
\begin{figure}[H]
\centering
\subfigure[]{\includegraphics[width=0.47\textwidth,height=0.27\textwidth]{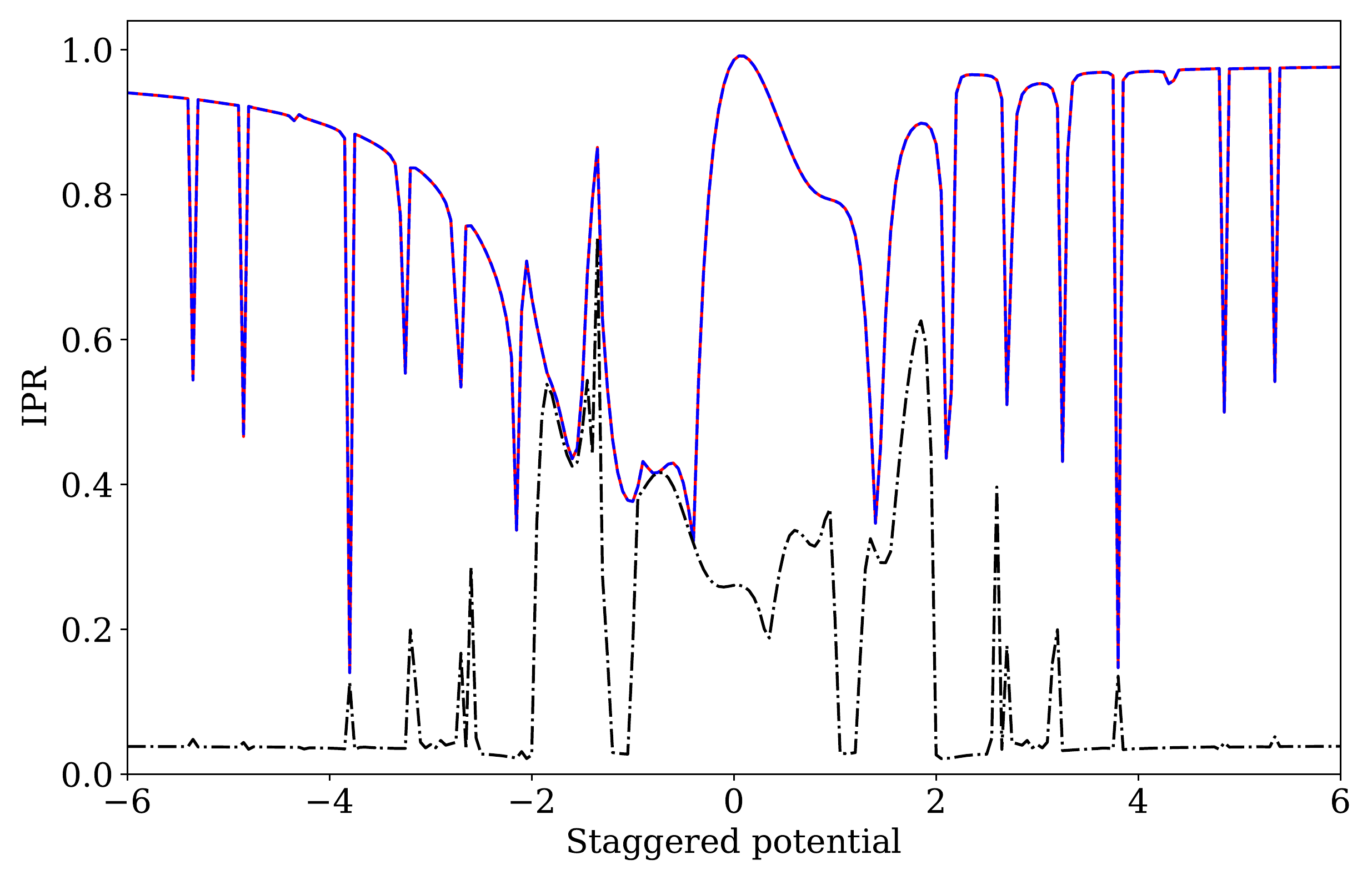}}
\quad \quad
\subfigure[]{\includegraphics[width=0.47\textwidth,height=0.27\textwidth]{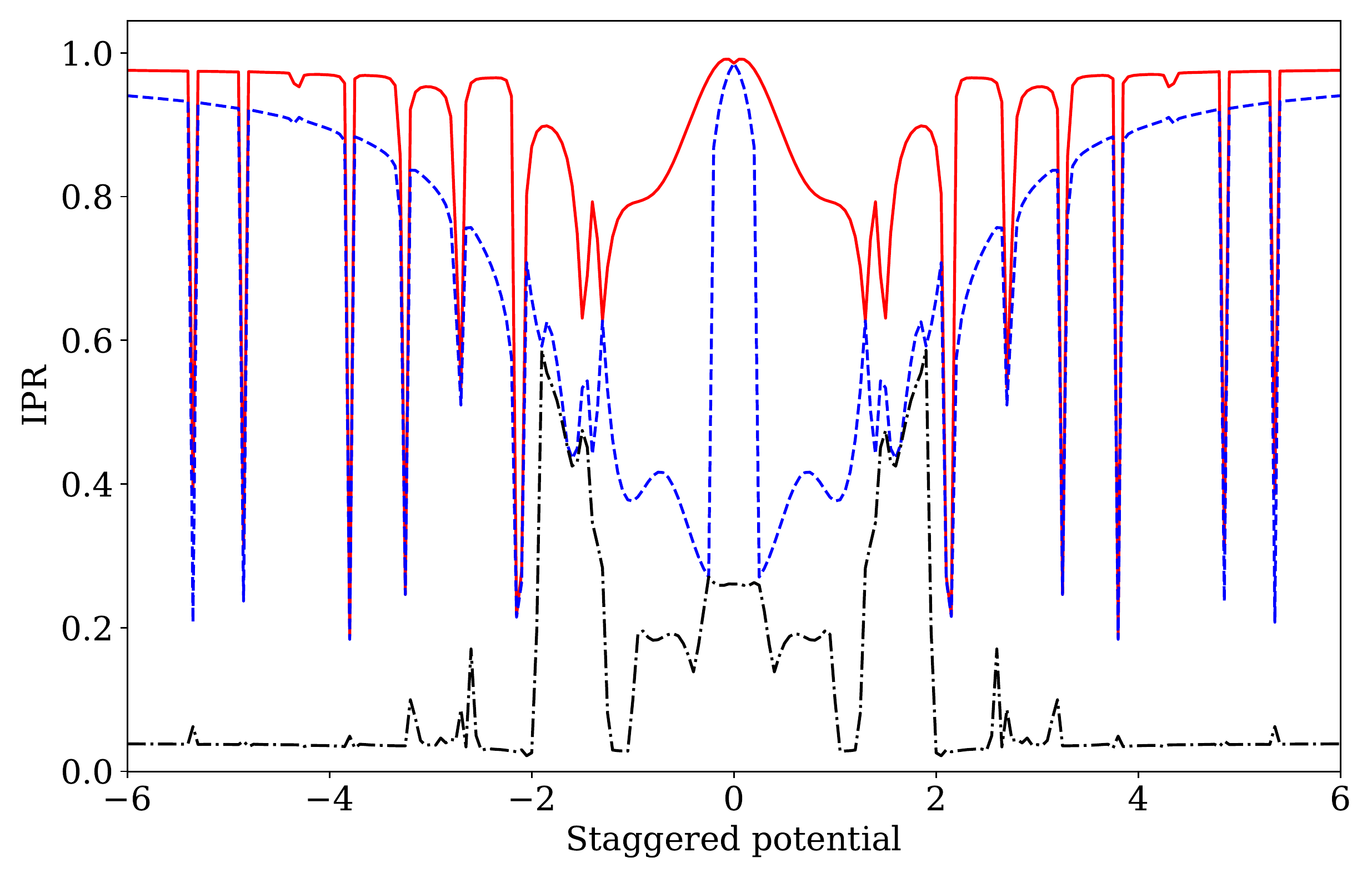}}
\subfigure[]{\includegraphics[width=0.47\textwidth,height=0.27\textwidth]{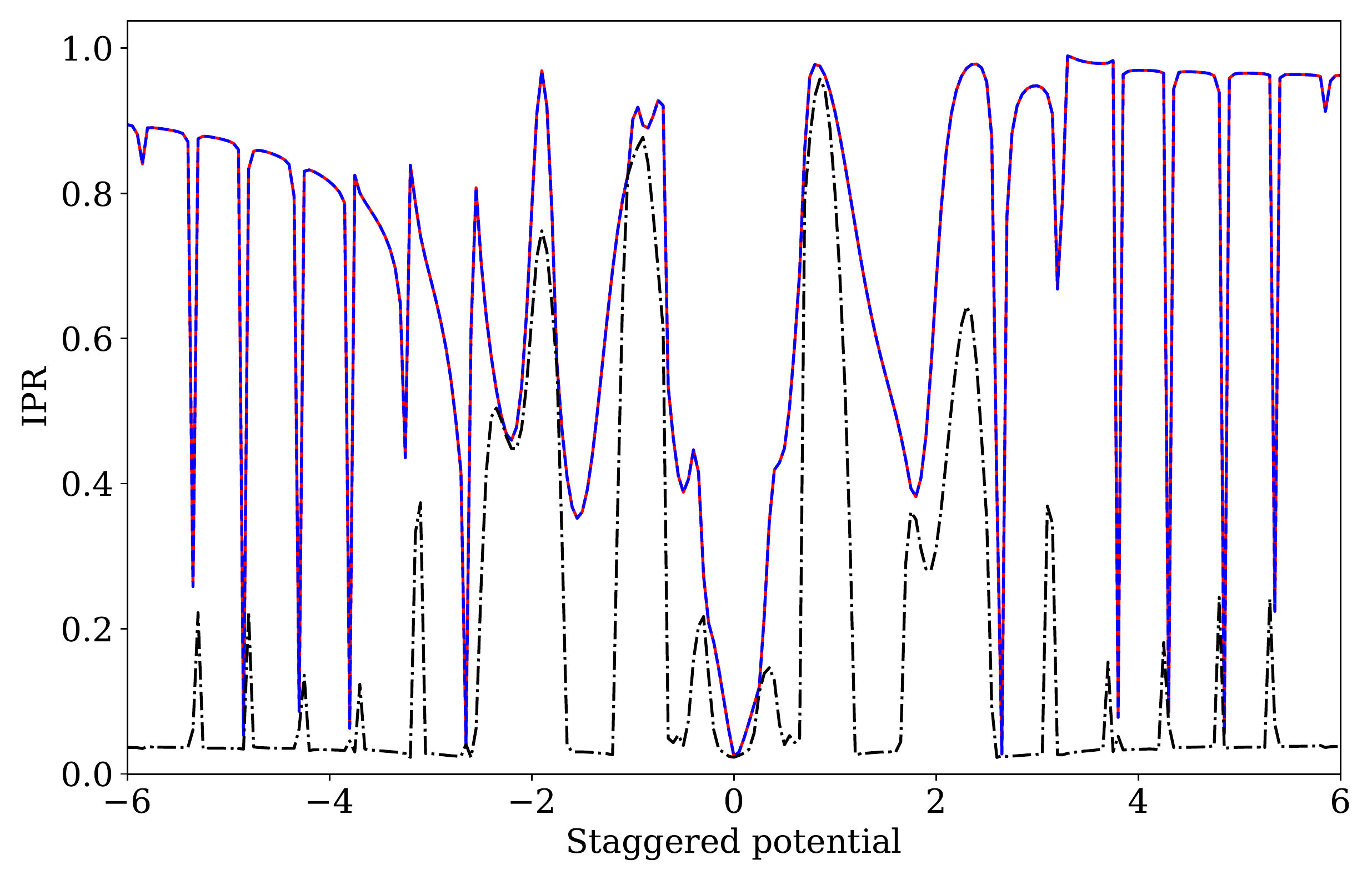}}
\quad \quad
\subfigure[]{\includegraphics[width=0.47\textwidth,height=0.27\textwidth]{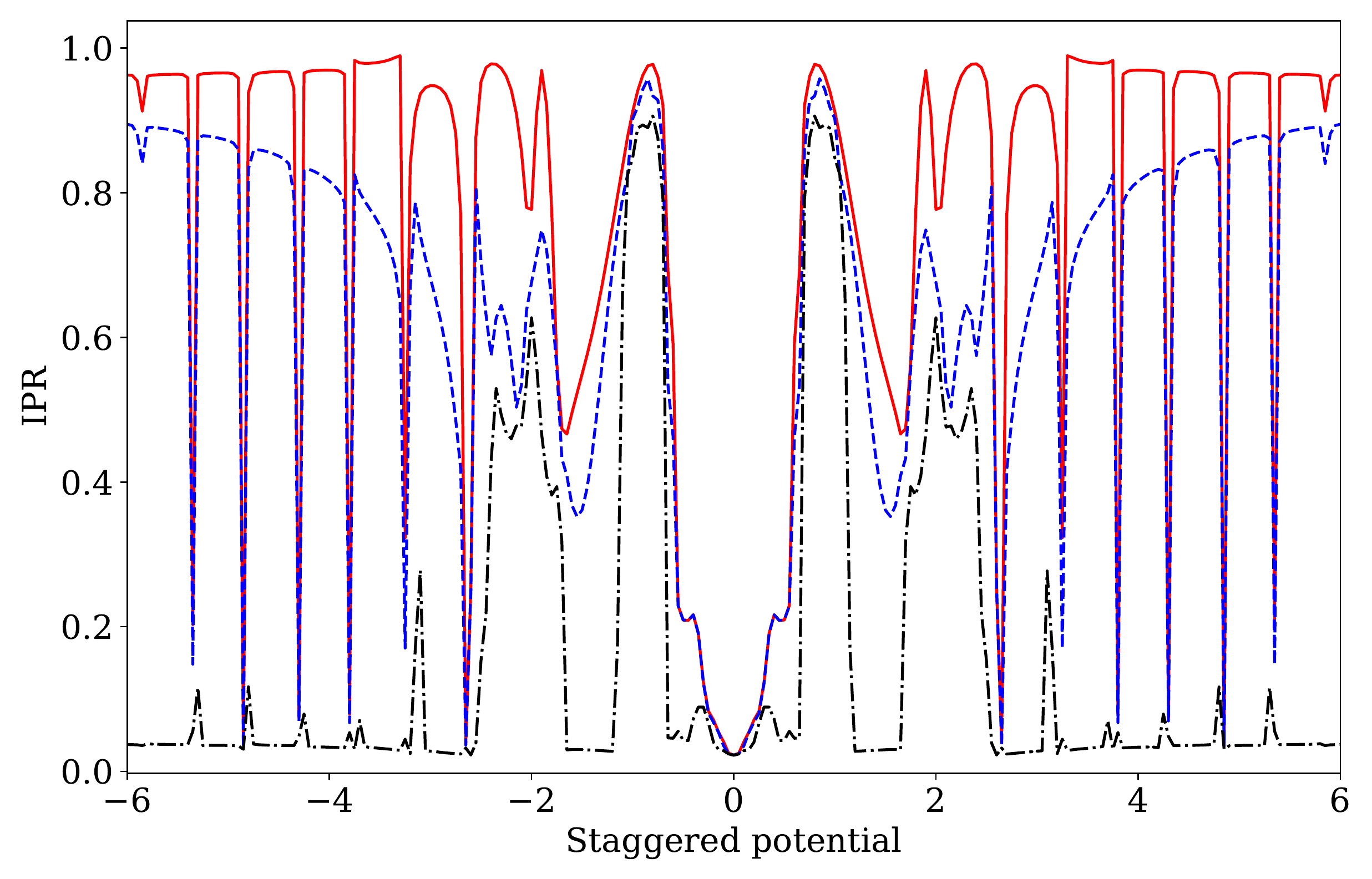}}
\caption{Largest three IPRs (shown as red solid, blue dashed and black dash 
dots, respectively) as a function of $v$ for different values of $a$ and 
number of sites, for $g=1$ and $\om = 1$. (a) $a=2.6$ and 100 sites, 
(b) $a=2.6$ and 101 sites, (c) $a=4$ and 100 sites, and (d) $a=4$ and 101 
sites. The results are seen to depend on whether the number of sites is 
even or odd.} \label{ss1fig10} \end{figure}
\end{widetext}

We will now study the variation of the largest three IPRs as a
function of $v$ for different values of $a$; this will give us information 
about the ranges of $v$ where localized edge states exist. In 
Fig.~\ref{ss1fig10} we show the results 
for 100-site and 101-site systems, with $g=1$ and $\om = 1$. 
It is interesting to look at the point $v=0$. We see that the maximum IPR in 
the vicinity of $v=0$ is small for values of $a$, like $a=4$, which do not 
correspond to large IPR regions in Fig.~\ref{ss1fig01}, while it is large for 
values 
of $a$, like $a=2.6$, which lie in the large IPR region in Fig.~\ref{ss1fig01}. 
Hence the results for $v=0$ are consistent between Figs.~\ref{ss1fig01} and 
\ref{ss1fig10}. For larger values of $v$, we see sharp drops in the IPR 
values; this is due to hybridization between the edge states at the two ends of 
system which reduces their probabilities and therefore their IPRs.

\onecolumngrid

\subsubsection{Time evolution of a state initialized at one edge}
\label{sec3b3}

The driven system has another symmetry if the driving of the
phase in Eq.~\eqref{ham5} is taken to be $\cos (\om t)$ instead of $\sin 
(\om t)$. We then see that $H(T-t) = H(t)$. If we also change $v \to - v$ and 
$c_n \to (-1)^n c_n$, using the operator $W$ defined around Eq.~\eqref{wh}, we 
have $H(T-t,-v) = - W H(t,v) W$. Following an argument similar to the one 
given there, we see that if $U_T (v)$ is the Floquet operator for a particular
value of $v$, the Floquet operator for $-v$ is given by $U_T (-v) = W (U_T 
(v))^{-1} W$. Hence for every Floquet eigenstate of $U_T (v)$ given by $\psi 
(v)$ with Floquet eigenvalue $e^{i \ta}$, there will be a Floquet eigenstate 
of $U_T (-v)$ given by $\psi (-v) = W \psi (v)$ with Floquet eigenvalue 
$e^{-i \ta}$. Thus the results for $+v$ and $-v$ will be similar, and it is 
therefore sufficient to study the case $v \ge 0$. In the rest of this section, 
we will take the phase in Eq.~\eqref{ham5} to be $\cos (\om t)$.

\begin{figure}[H]
\centering
\includegraphics[width=0.49\textwidth]{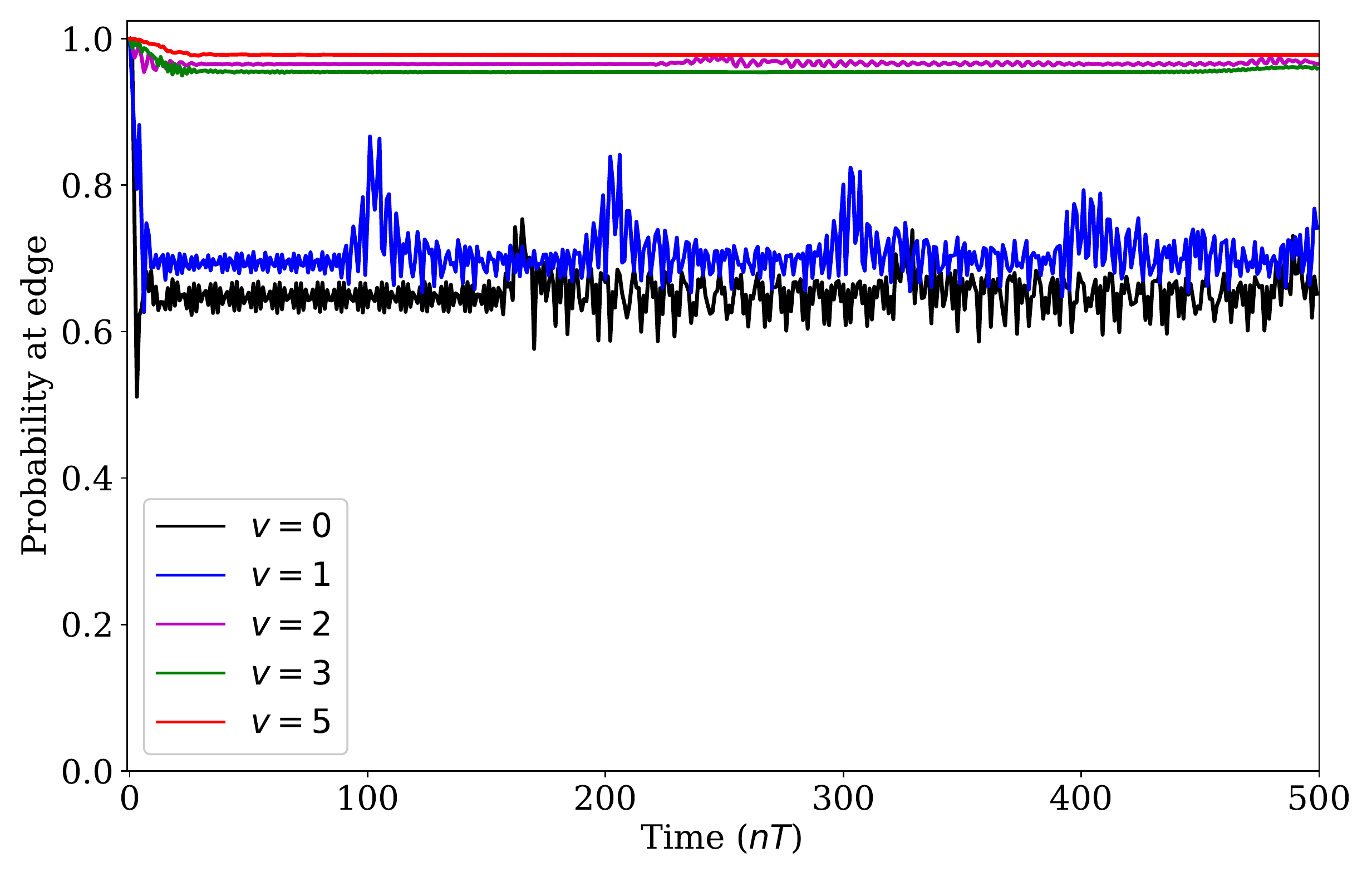}
\caption{Edge probability of a state initialized at the left edge for a 
100-site system with $g=1$, $\om =1$, and $a=2.6$.} \label{ss1fig11} 
\end{figure}

\begin{figure}[H]
\centering
\includegraphics[width=0.49\textwidth]{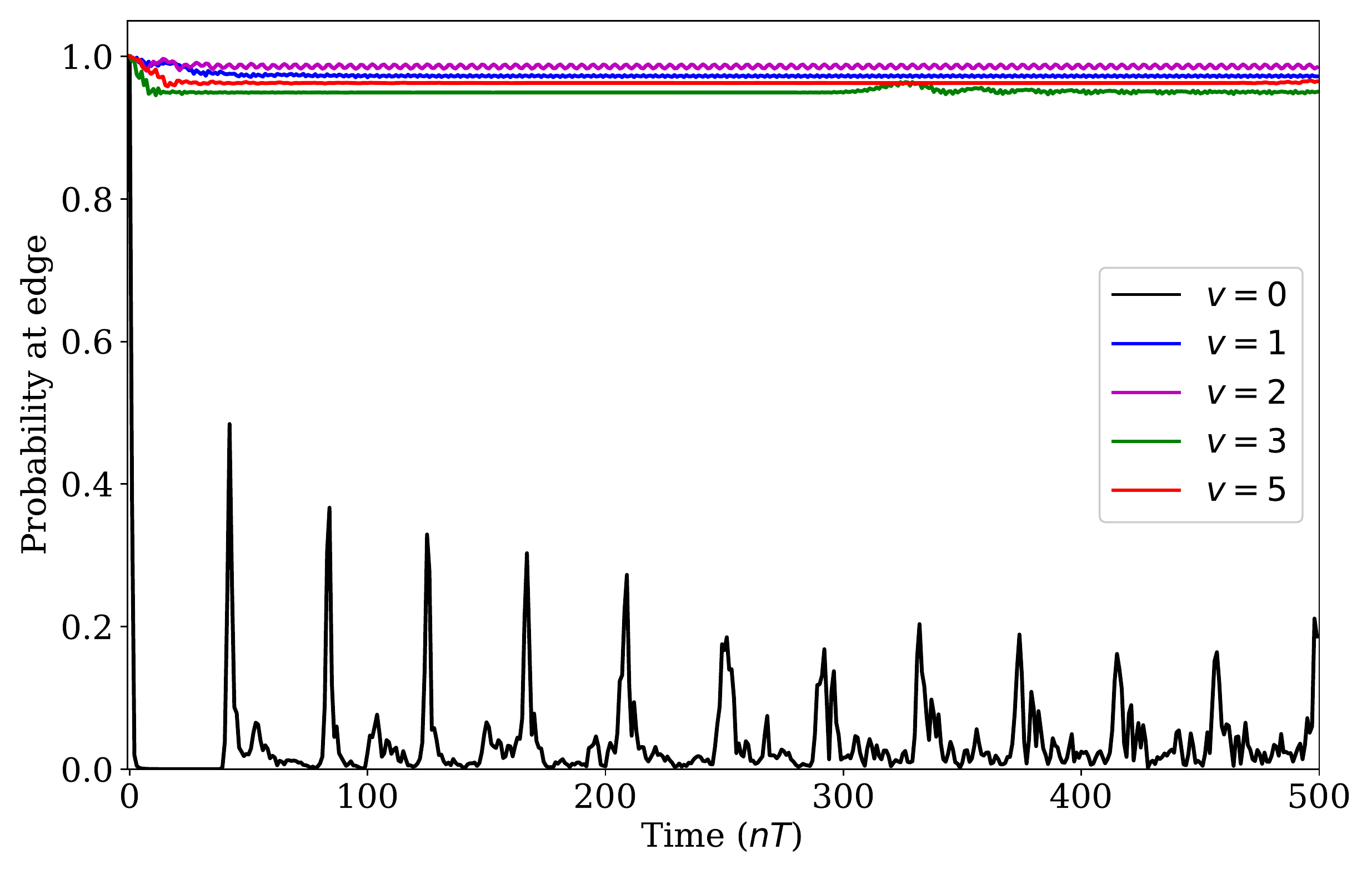}
\caption{Edge probability of a state initialized at the left edge for a 
100-site system with $g=1$, $\om =1$, and $a=4$.} \label{ss1fig12} \end{figure}

We will now study the dynamics of states initialized near one edge of
the system in the presence of periodic driving. 
We will start at time $t=0$ with a state which is localized at the left 
edge of the system and study how it evolves with time; the initial
state will not be taken to be an eigenstate of the Floquet operator. 
We will look at how the probability that the particle returns to that edge 
(we will consider the first three sites to be the edge, so the edge 
probability is equal to $|\psi (0)|^2 + |\psi (1)|^2 + |\psi (2)|^2$) varies 
with the stroboscopic time which will be taken to be integer multiples of the 
time period $T$. 
We have studied a 100-site system with $g=1$, $\om=1$, and $a=2.6$ (where 
we know that edge states exist for $v=0$) and $a=4$ (where there are no 
edge states for $v=0$). 
In Figs.~\ref{ss1fig11} and \ref{ss1fig12}, the sign of $v$ is 
the sign of the staggered potential at the first site from the left ($n=0$).

We conclude the following from Figs.~\ref{ss1fig11} and 
\ref{ss1fig12}.

\noi (i) For $a=2.6$ where we know that edge states exist at $v=0$, the 
probability of staying at the edge remains close to 1 for all times. When 
$|v|$ is increased to 1 and 2, the probability becomes smaller. However, 
when $|v|$ is increased further to $3$ and $5$, the probability of staying 
at the edge again comes close to 1 at all times. 

\noi (ii) For $a=4$ where there is no edge state at $v=0$, the probability of 
staying at the edge remains close to zero, except for sharp peaks at some
particular times. As $|v|$ is increased to 3 and 5, the probability rises 
and stays close to 1 at all times.

\noi (iii) Even for values of $v$ and $a$ where the probability of staying 
close to the edge is small at most times, we see sudden jumps in the 
probability at some regular time intervals.

These features can be explained as follows. If the system has an edge state 
at a particular value of $a$ and $v$, then an initial state which is close to
the edge will have a large overlap with that eigenstate and will therefore
have a large probability for the particle to stay near the edge. This matches 
with our earlier plots showing the regions of $a$ and $v$ where edge states
exist. When there are no edge eigenstates which are localized at the edges,
the probability of staying near the edges is low. However there are peaks
in the probability of coming close to the edge; this occurs because the
particle initially starts at one edge, moves into the bulk, and then 
repeatedly gets reflected back and forth between the two edges. The 
probability becomes large whenever it returns to the original edge. This
occurs at time intervals given by the recurrence time $2L/v_{max}$, where 
$L$ is the size of the system and $v_{max}$ is the maximum velocity of the 
particle when it is subjected to periodic driving. For instance, when the 
staggered potential $v=0$, the maximum velocity if $v_{max} = |2 g J_0 (a/
\om)|$. This is because the effective Hamiltonian has a nearest-neighbor 
hopping amplitude equal to $-g J_0 (a/\om)$ (see Appendix B). Hence the energy 
dispersion in the bulk is given by $E_k = -2 g J_0 (a/\om) \cos k$; the group 
velocity is then given by $v_k = 2 g J_0 (a/\om) \sin k$, which has a maximum 
value of $|2 g J_0 (a/\om)|$.

\section{Bose-Hubbard model with periodic driving}
\label{sec4}

We will now consider a model with interacting particles. Specifically we will 
consider the Bose-Hubbard model and investigate if periodic driving can give 
rise to two-particle bound states which are localized at one edge of the 
system. On a lattice of size $L$, the Bose-Hubbard model subjected to periodic
driving by an electric field has a Hamiltonian of the form
\begin{eqnarray}
H &=& - ~g ~\sum_{n=0}^{L-2} ~(e^{\frac{ia}{\om}\sin(\om t)} b_{n}^{\dg}
b_{n+1} ~+~ e^{-\frac{ia}{\om}\sin(\om t)} b_{n+1}^{\dg}b_{n}) \non \\
&& + ~\frac{u}{2} ~\sum_{n=0}^{L-1} \rho_{n}(\rho_{n}-1), \label{ham6} 
\end{eqnarray}
where $\rho_n = b_n^\dg b_n$ is the particle number operator at site $n$. To 
study a system with two bosons numerically, we construct the Hamiltonian in 
the basis $|n_{1}, n_{2} \rangle$, where $n_1$ and $n_2$ denote the positions 
of the two bosons, and we can assume that $n_{1} \le n_{2}$ since the 
particles are indistinguishable. For a system with $L$ sites, the Hamiltonian 
in this basis will be a $L(L+1)/2$-dimensional matrix.
After constructing the Hamiltonian we calculate the Floquet operator $U_T$
as explained in Appendix A. We then look at the Floquet eigenstates and
their IPRs to identify bound states in which both the particles are
localized near one edge of the system.
In our calculations, we will define the edge as consisting of the three states 
$|0,0 \rangle, ~|0,1 \rangle$ and $|1,1 \rangle$, i.e., the probability at 
the edge will mean the sum of the probabilities of these states. 
In all our numerical calculations we will take $g=1$ and $\om=1$, and vary
the driving amplitude $a$ and the interaction strength $u$.

Before presenting the numerical results, we note a useful symmetry which
relates the Floquet operators for positive and negative values of $u$. This can
be seen most clearly if we take the phase of the hopping to be $\cos (\om t)$ 
instead of $\sin (\om t)$ in Eq.~\eqref{ham6}; then the symmetry $\cos (\om 
(T-t)) = \cos (\om t)$ implies that the Hamiltonian $H(t) \to - H (T-t)$
if we change $u \to - u$ and $b_n \to (-1)^n b_n$. The latter corresponds to 
a transformation by a unitary and diagonal matrix $W_2$ which gives 
$W_2 | n_1, n_2 \ra = (-1)^{n_1 + n_2} | n_1, n_2 \ra$; note that 
$W_2^2 = I$. This implies that the Floquet operators for interaction 
strengths $\pm u$ are related as $U_T (-u) = W_2 (U_T (u))^{-1} W_2$. Hence, 
if there is a two-particle bound state $\psi (u)$ at one edge of the system 
with a Floquet eigenvalue $e^{i \ta}$ for interaction $u$, there will be a 
two-particle bound state $\psi (-u)$ at the same edge with a Floquet eigenvalue
$e^{-i \ta}$ for interaction $-u$; the wave functions for the two eigenstates 
will be related as $\psi (-u) = W_2 \psi (u)$. This implies that the 
probabilities of the different basis states $|n_1,n_2 \ra$ will be identical 
in the two eigenstates. This symmetry between positive and negative values of 
$u$ implies that it is sufficient to study the case $u \ge 0$.

\subsection{Edge probability versus $u$ for different values of $a$}
\label{sec4a}

\begin{figure}[H]
\subfigure[]{\includegraphics[width=0.45\textwidth]{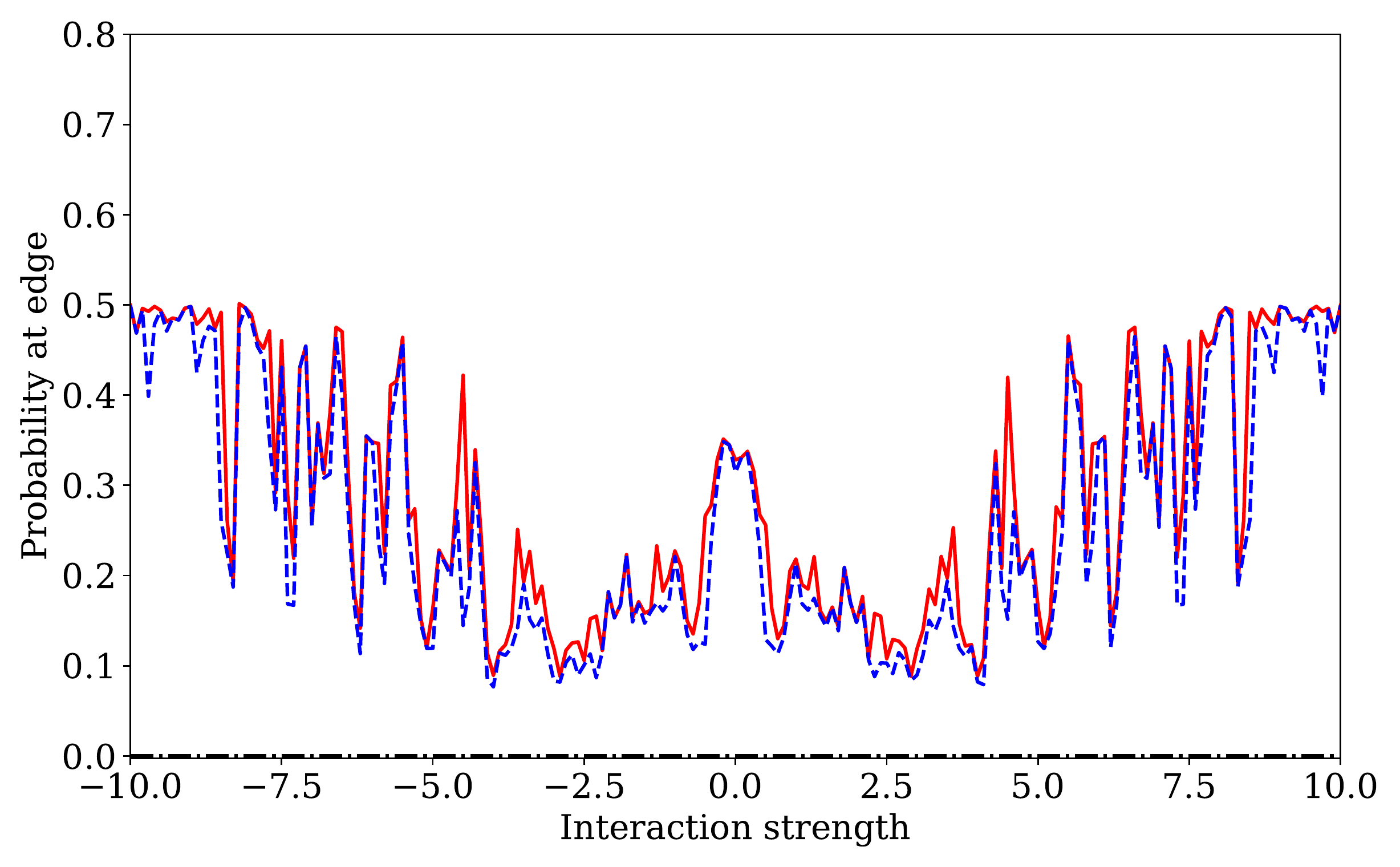}}
\subfigure[]{\includegraphics[width=0.45\textwidth]{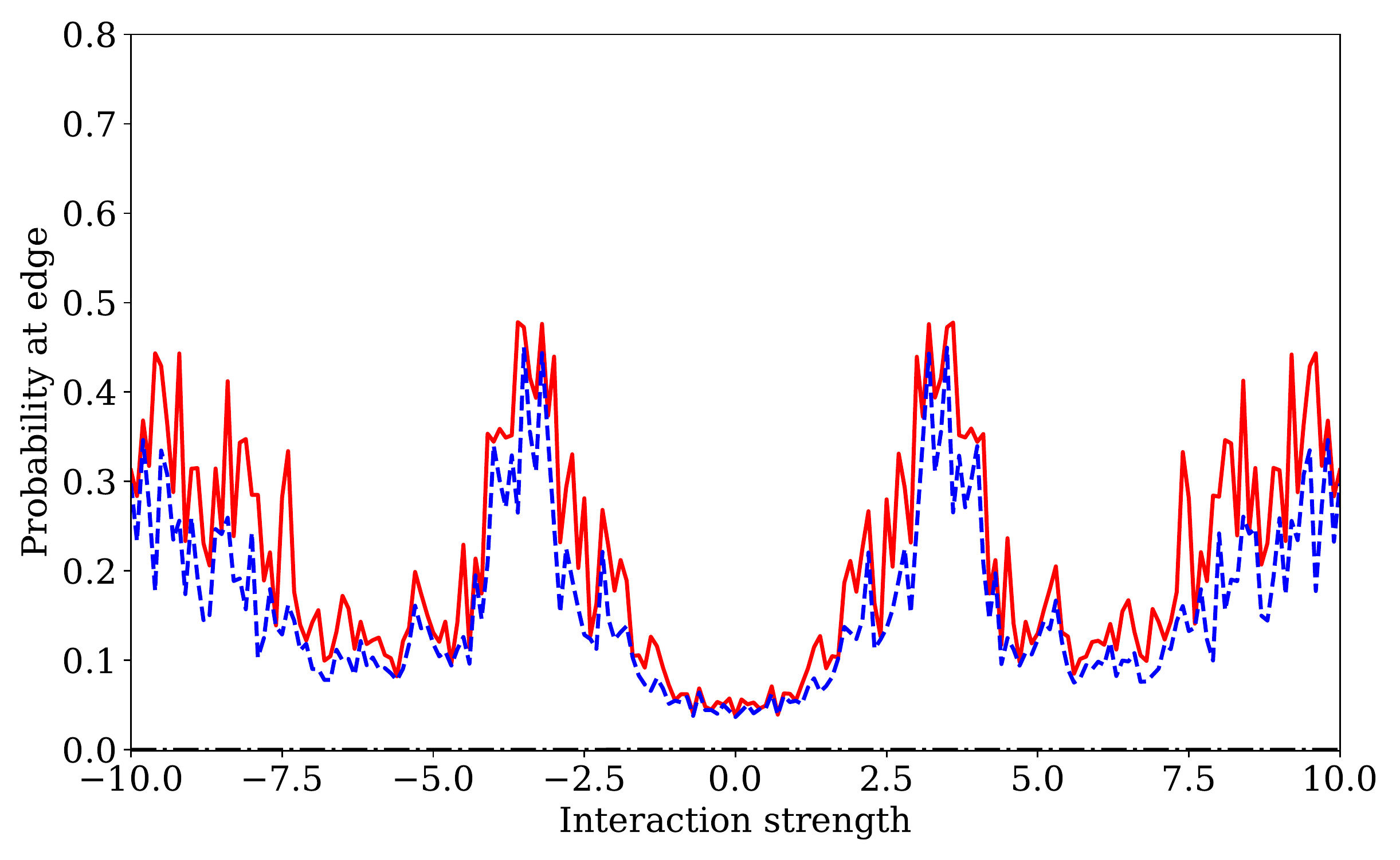}}
\caption{Largest two edge probabilities at edge as a function of $u$. For 
$a=2.6$, $u=0$ gives a large probability at the edge, while for $a=4$, $u=0$ 
gives an edge probability close to 0. The overall trend is that as $u$ 
increases the edge probability increases. Note that a large edge probability 
does not necessarily imply a bound state.} \label{ss1fig13} \end{figure}

We recall from Fig.~\ref{ss1fig01} that driving the non-interacting model with 
$a=2.6$ gives rise to edge states, while $a=4$ does not lead to edge states.
To study the possibility of interactions producing two-particle bound states 
which are localized near one of the edges, we will specifically
choose these two values of $a$ and study the probability at the edge as a 
function of the interaction strength $u$; we define the edge probability
to be the sum of the probabilities of the two particles being at the
locations $|0,0 \ra$, $|0,1 \ra$ and $|1, 1 \ra$. The numerical results 
obtained for a 30-site system are shown in Fig.~\ref{ss1fig13}. 
We note the following features in the figure.

\noi (i) For $a=2.6$, the edge probability is fairly large at $u=0$. 
This is because the non-interacting system has edge states as we saw in 
Fig.~\ref{ss1fig01}, and therefore there are states in which both the particles
occupy those states. However, as $u$ is turned on, either to positive or 
negative values, the edge probability first drops and then starts rising 
again with increasing $u$. Thus a two-particle bound state at the edge 
becomes more likely as $u$ increases.

\noi (ii) For $a=4$, the edge probability is small at $u=0$; this is because
the non-interacting system does not have edge states as shown in
Fig.~\ref{ss1fig01}. As $u$ is turned on, the probability at the edge first 
rises, then drops and then starts rising again as $u$ increases.

\subsection{Floquet perturbation theory for $g \ll u$}
\label{sec4b}

\begin{widetext}
\begin{figure}[H]
\subfigure[]
{\includegraphics[width=0.47\textwidth,height=0.3\textwidth]{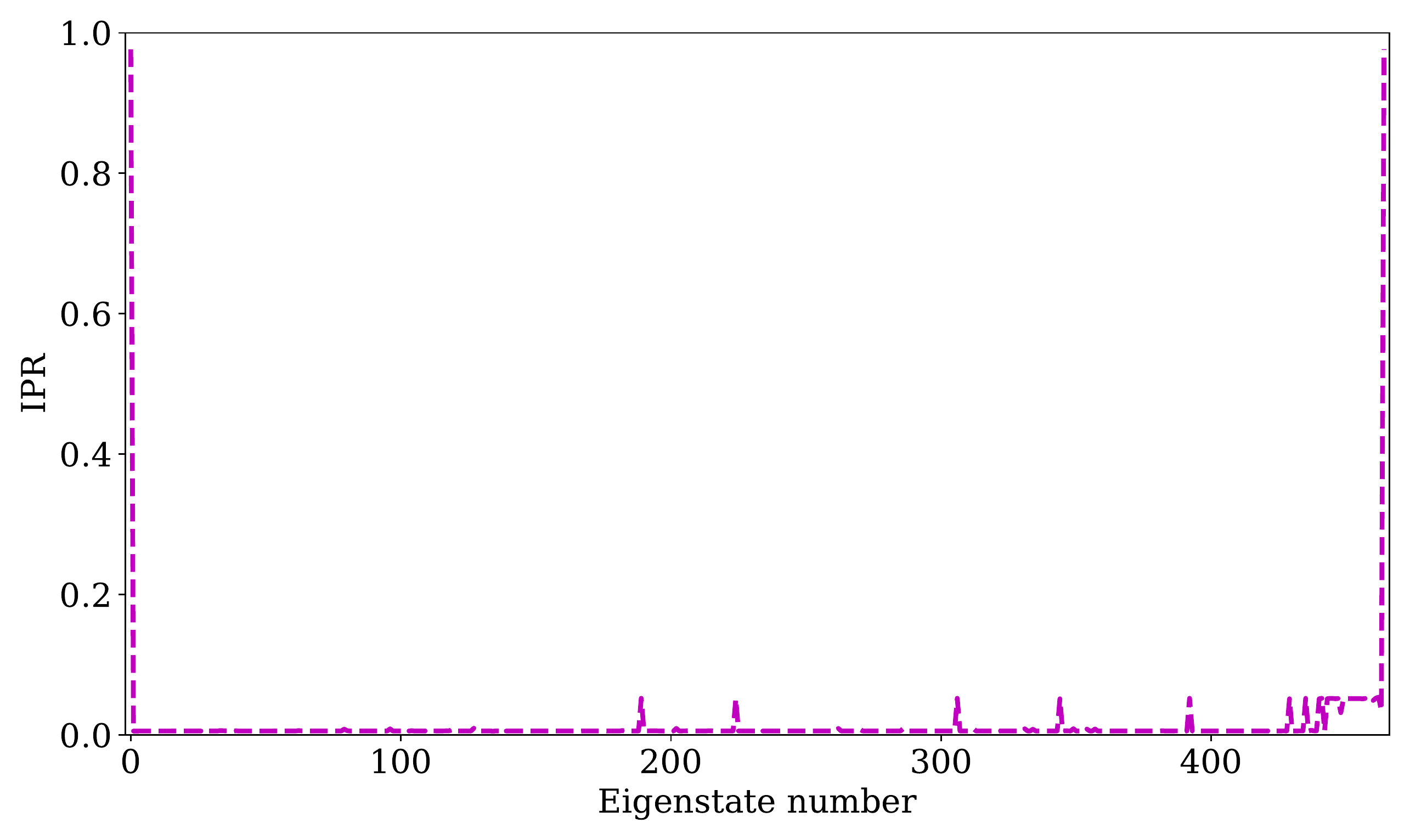}} \quad
\quad
\subfigure[]{\includegraphics[width=0.47\textwidth,height=0.3\textwidth]{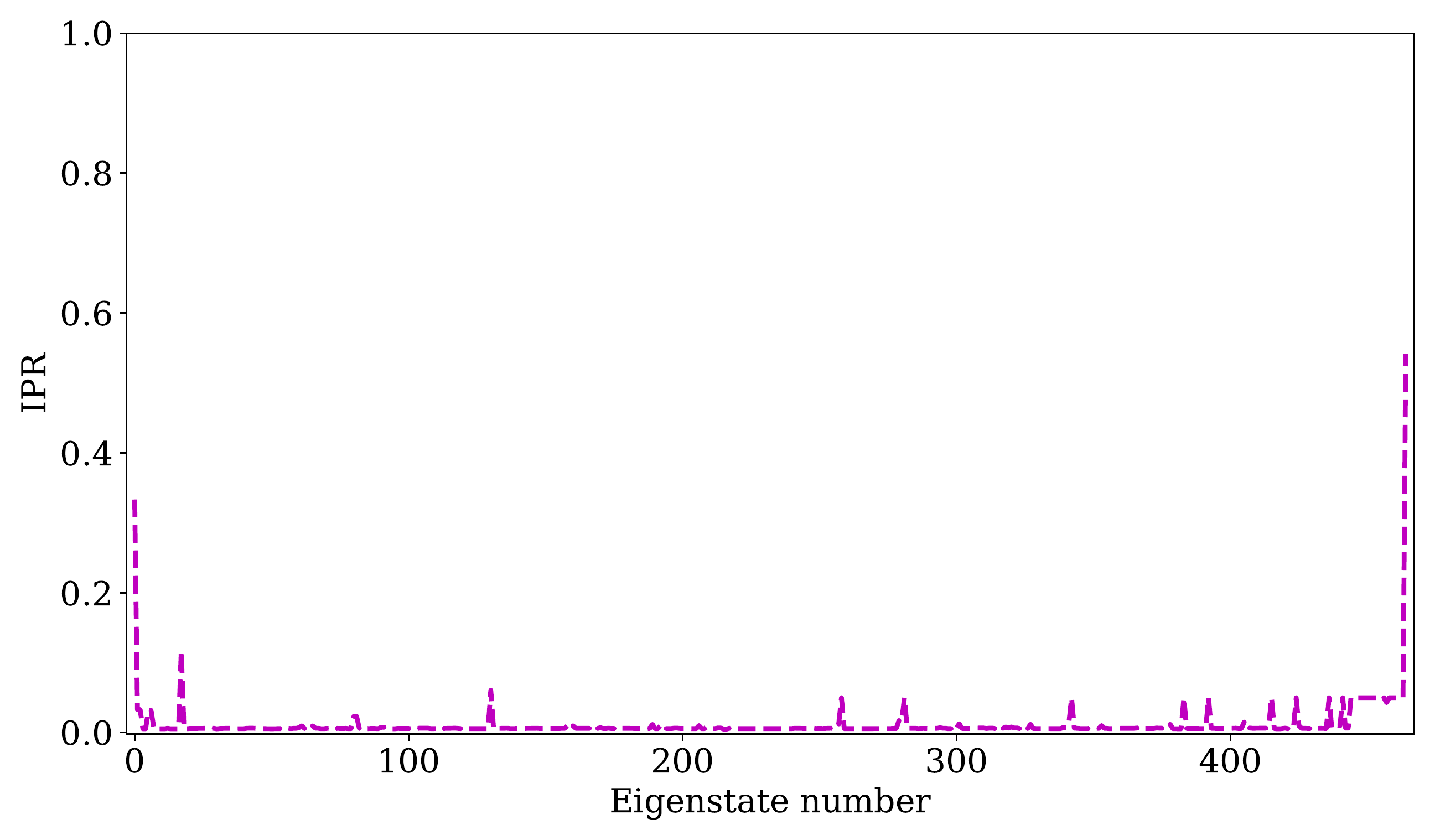}}
%
%
\caption{Comparison between IPR plots for the driven system and 
the perturbatively obtained Hamiltonian respectively for a 30-site system with 
$g=0.1$, $u=20.2$, $\om =1$, and $a=2.6$. The IPR plots 
show two states with large IPRs implying a localized bound state at each 
of the edges.}
\label{ss1fig14} \end{figure} 
\end{widetext}

\begin{widetext}
\begin{figure}[H]
%
\subfigure[]{\includegraphics[width=0.47\textwidth]{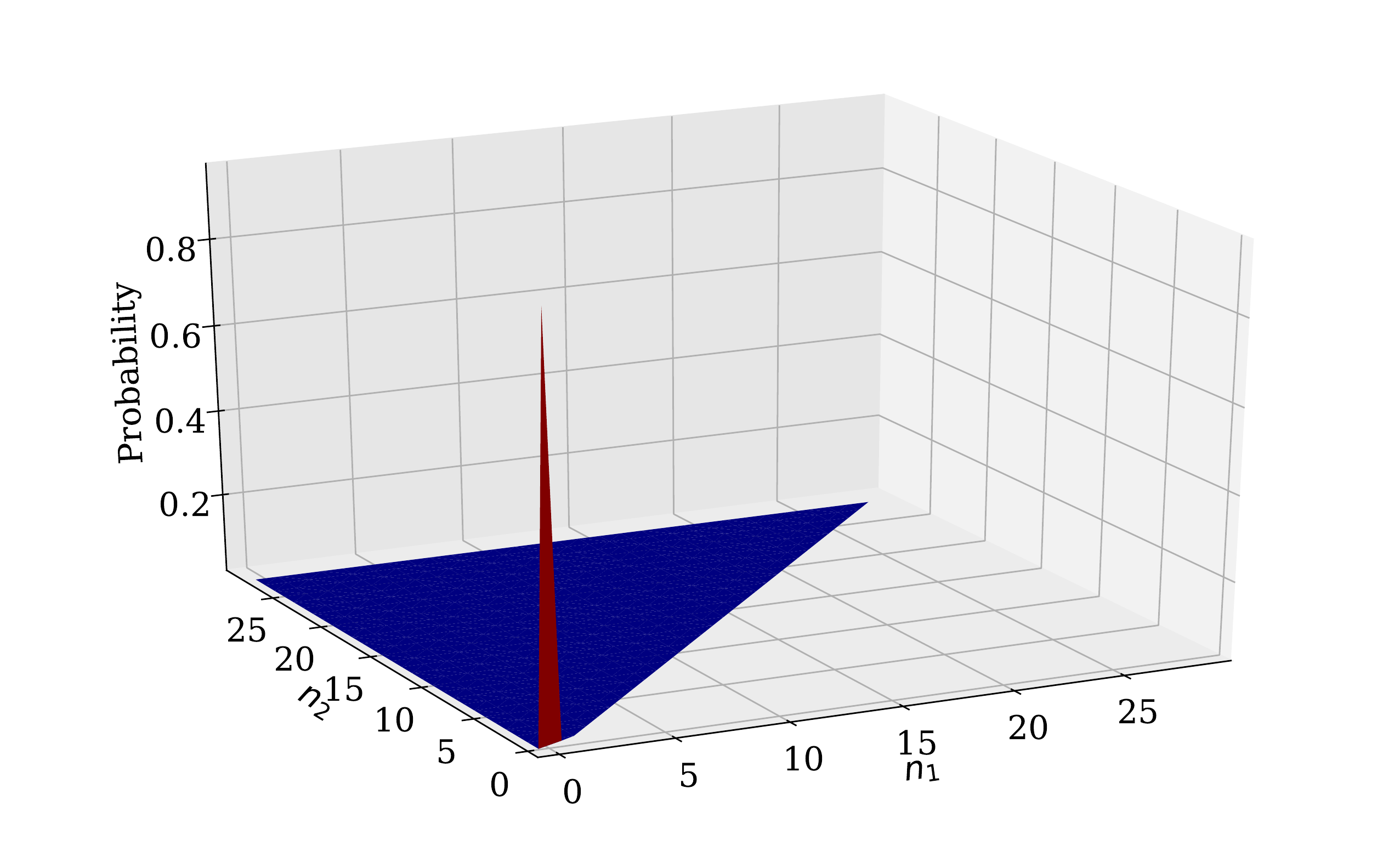}} 
\quad \quad
\subfigure[]{\includegraphics[width=0.47\textwidth]{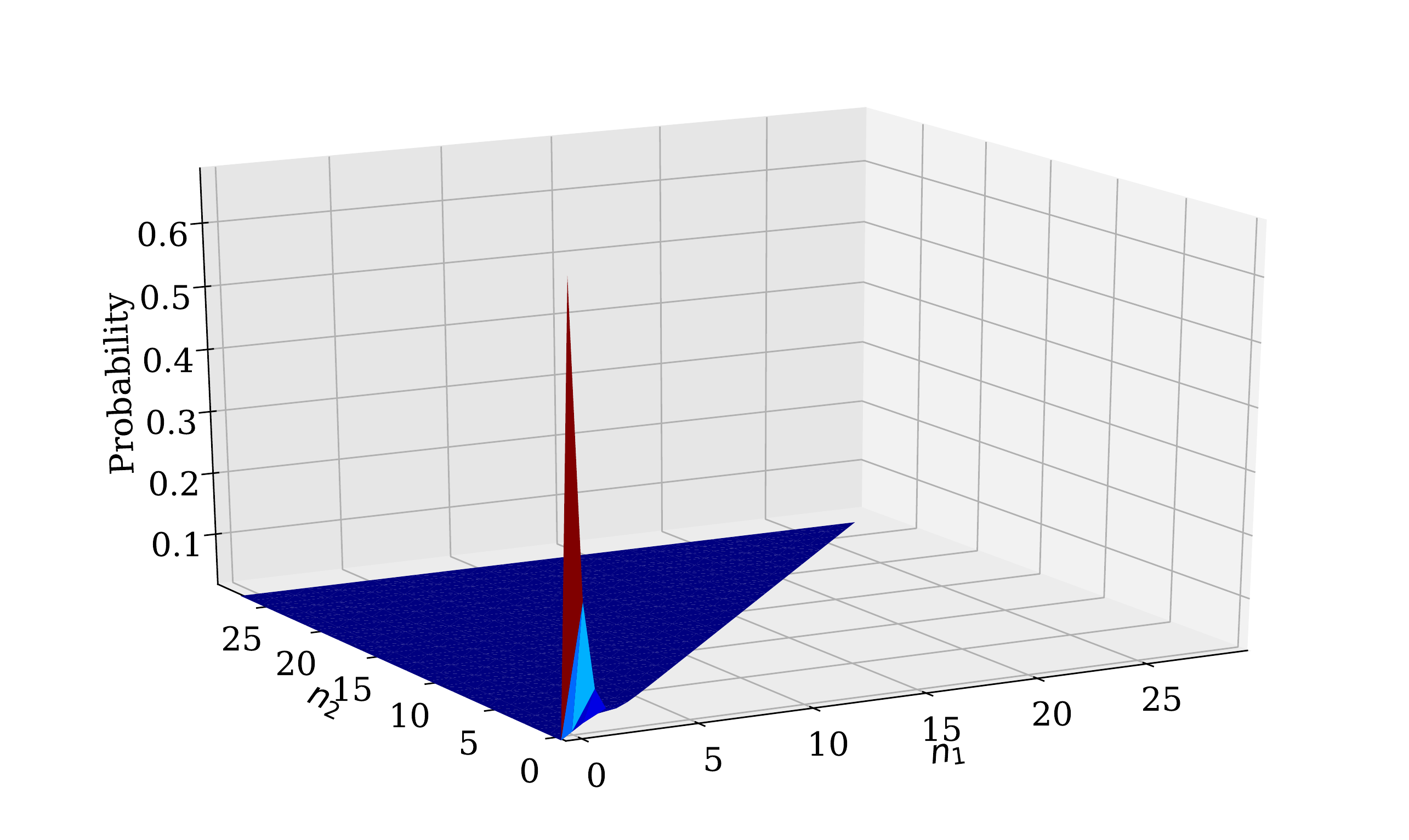}}
%
\subfigure[]{\includegraphics[width=0.47\textwidth]{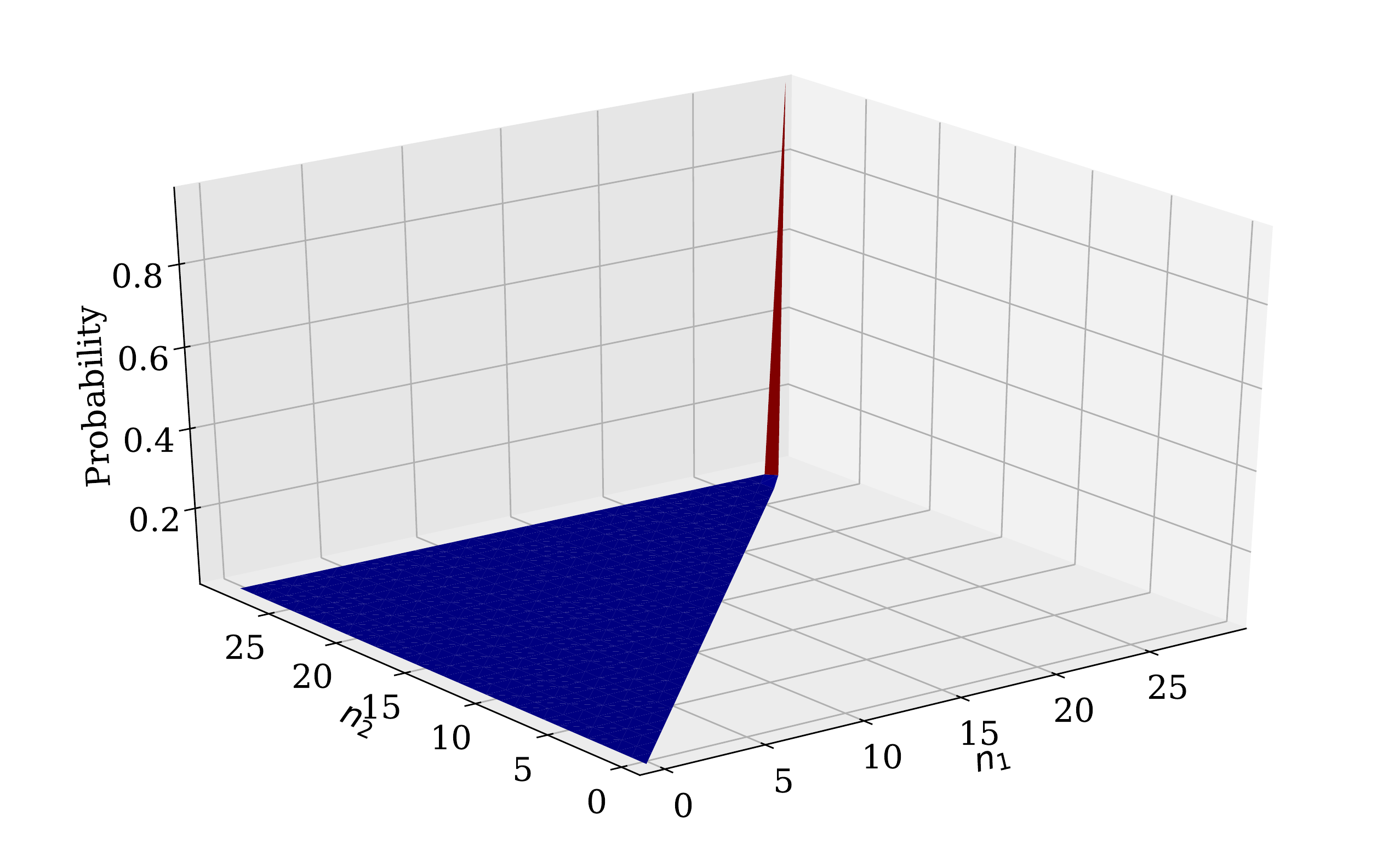}}
\quad \quad
\subfigure[]{\includegraphics[width=0.47\textwidth]{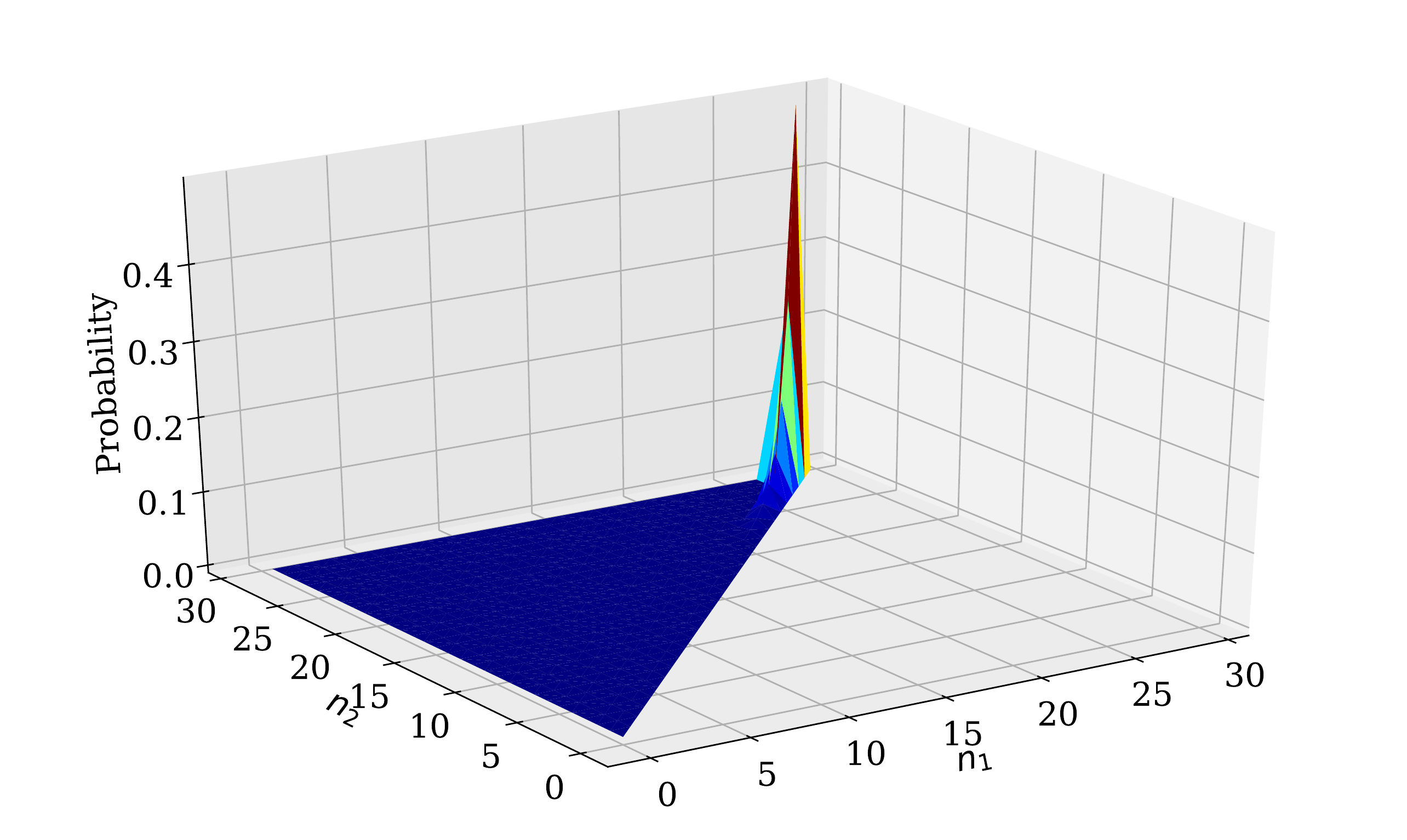}}
\caption{Comparison between numerical results for the driven system and 
the perturbatively obtained Hamiltonian respectively for a 30-site system with 
$g=0.1$, $u=20.2$, $\om =1$, and $a=2.6$. 
The two-particle probability $\vert \psi(n_{1}, n_{2}) \vert^{2}$ vs 
$(n_{1}, n_{2})$, where $n_{1} \le n_{2}$, obtained numerically and 
perturbatively are shown for the two states in plots (a), (c) and (b), (d)
respectively.} \label{ss1fig15} \end{figure} 
\end{widetext}

We now present a Floquet perturbation theory in the limit that $g \ll u$ along
the same lines as in Sec.~\ref{sec3b1}. We write the Hamiltonian as a sum 
\begin{eqnarray}
H &=& H_{0} + V, \non \\
H_0 &=& \frac{u}{2} ~\sum_{n=0}^{L-1} ~\rho_n (\rho_{n} -1), \non \\
V &=& - ~g ~\sum_{n=0}^{L-2} ~(e^{\frac{ia}{\om}\sin(\om t)} b_{n}^{\dg}
b_{n+1} ~+~ e^{-\frac{ia}{\om} \sin(\om t)}b_{n+1}^{\dg}b_{n}), \non \\
\end{eqnarray}
and treat $V (t)$ as a perturbation. We first consider the limit $g = 0$, so
the Hamiltonian is $H_{0} = (u/2) \sum_{n=0}^{L-1} \rho_n (\rho_{n} -1)$.
Then the eigenstates of the Hamiltonian are simply the basis states $|n_1,
n_2 \ra$, and driving has no effect on this system.
We now find that introducing even a small amount of hopping ($g \ll u$)
has a drastic effect.
Namely, we find that all bulk states are delocalized and there are only two 
two-particle bound states which are localized at one of the ends of the system.
We can use Floquet perturbation theory to first order in $g$ to understand the
emergence of two-particle edge states for small hopping. 
Before showing the results we first discuss Floquet perturbation theory for 
this system.
For convenience, we will denote the eigenstates of $H_{0}$, $| n_{1}, n_{2} 
\ra$ 
by a single symbol $| m \rangle$, where $m$ takes $L(L+1)/2$ possible
values. The eigenenergies of $H_{0}$ are $E = 0$ or $u$ depending on whether 
the two bosons are on the same site or on different sites. 
Since there are degeneracies in the eigenenergies of $H_{0}$, we have to use 
degenerate Floquet perturbation theory. We look for solutions of 
the time-dependent Schr\"odinger equation $| \psi \rangle$ for the 
time-periodic Hamiltonian $H$. We can write 
\begin{equation} | \psi(t) \rangle= \sum_{m} c_{m}(t) e^{-i E_{m} t} 
| m \rangle. \end{equation} 
Then the Schr\"odinger equation leads to the following equations for the 
amplitudes $c_m (t)$,
\begin{equation} \frac{dc_{m}}{dt} ~=~ -i \sum_{m' \ne m} \langle m| V (t) | 
m' \rangle e^{i(E_{m}-E_{m'})t} ~c_{m'}(t). \end{equation}
Up to first order in $g$ and $T$, the effective time-evolution 
operator relating $c_m (T) e^{-i E_m T}$ to $c_m (0)$ is then given by 
\begin{equation} U_{eff} ~=~ I ~-~ i T H_{F}^{(1)}, \end{equation}
where
\begin{eqnarray} && (H_{F}^{(1)})_{mm'} \non \\
&=& (H_0)_{mm'} ~+ ~\dfrac{1}{T} ~\int_{0}^{T} dt ~e^{i(E_{m}- E_{m'})t} ~
\langle m | V (t) | m' \rangle. \non \\
\end{eqnarray}
We will now show a comparison of the numerical results obtained for the 
driven system and those obtained from the effective Hamiltonian $H_F^{(1)}$
obtained from Floquet perturbation theory. We consider a $30$-site system
with $g=0.1$, $u=20.2$ and $\om = 1$, so that $g \ll u$.
We have shown the results for $a=2.6$ in Figs.~\ref{ss1fig14} and 
\ref{ss1fig15}. In Fig.~\ref{ss1fig14} we see two states with large IPRs 
which are two-particle bound states localized at one of the two ends of the 
system. The probabilities $|\psi (n_1,n_2)|^2$ of these states are 
shown in Figs.~\ref{ss1fig15} (a-d). We find a good match between the results 
obtained numerically (Figs.~\ref{ss1fig15} (a) and \ref{ss1fig15} (c)) 
and perturbatively (Figs.~\ref{ss1fig15} (b) and \ref{ss1fig15} (d)).

\onecolumngrid

\onecolumngrid

\subsection{Time evolution of a two-particle state initialized at one edge}
\label{sec4c}

\begin{figure}[H]
\includegraphics[width=0.49\textwidth]{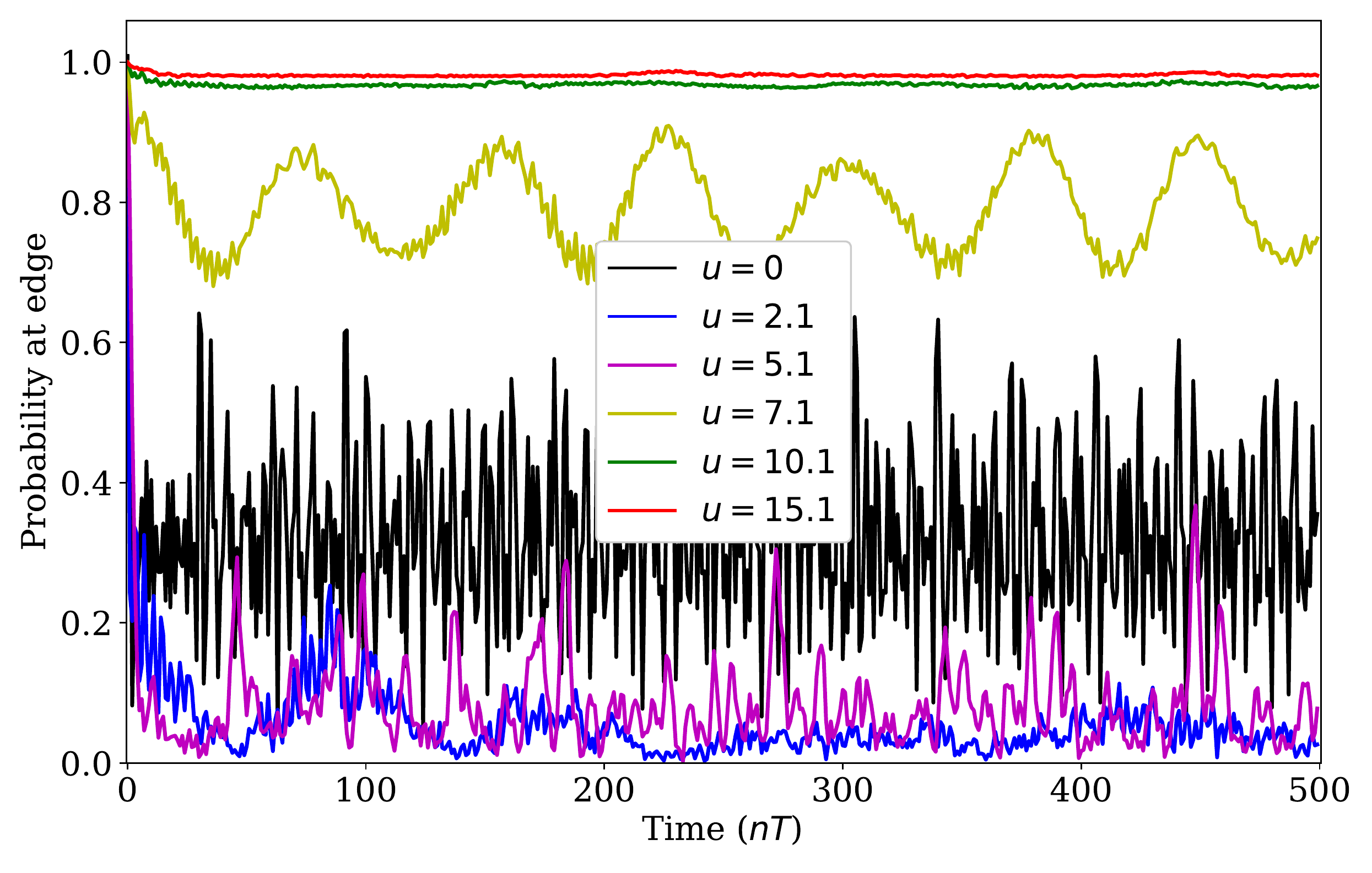}
\caption{Return probability of a state initialized with both particles at 
left edge. We have taken a 20-site system with $a=2.6$ and $g=1$.} 
\label{ss1fig16} \end{figure}

\begin{figure}[H]
\includegraphics[width=0.49\textwidth]{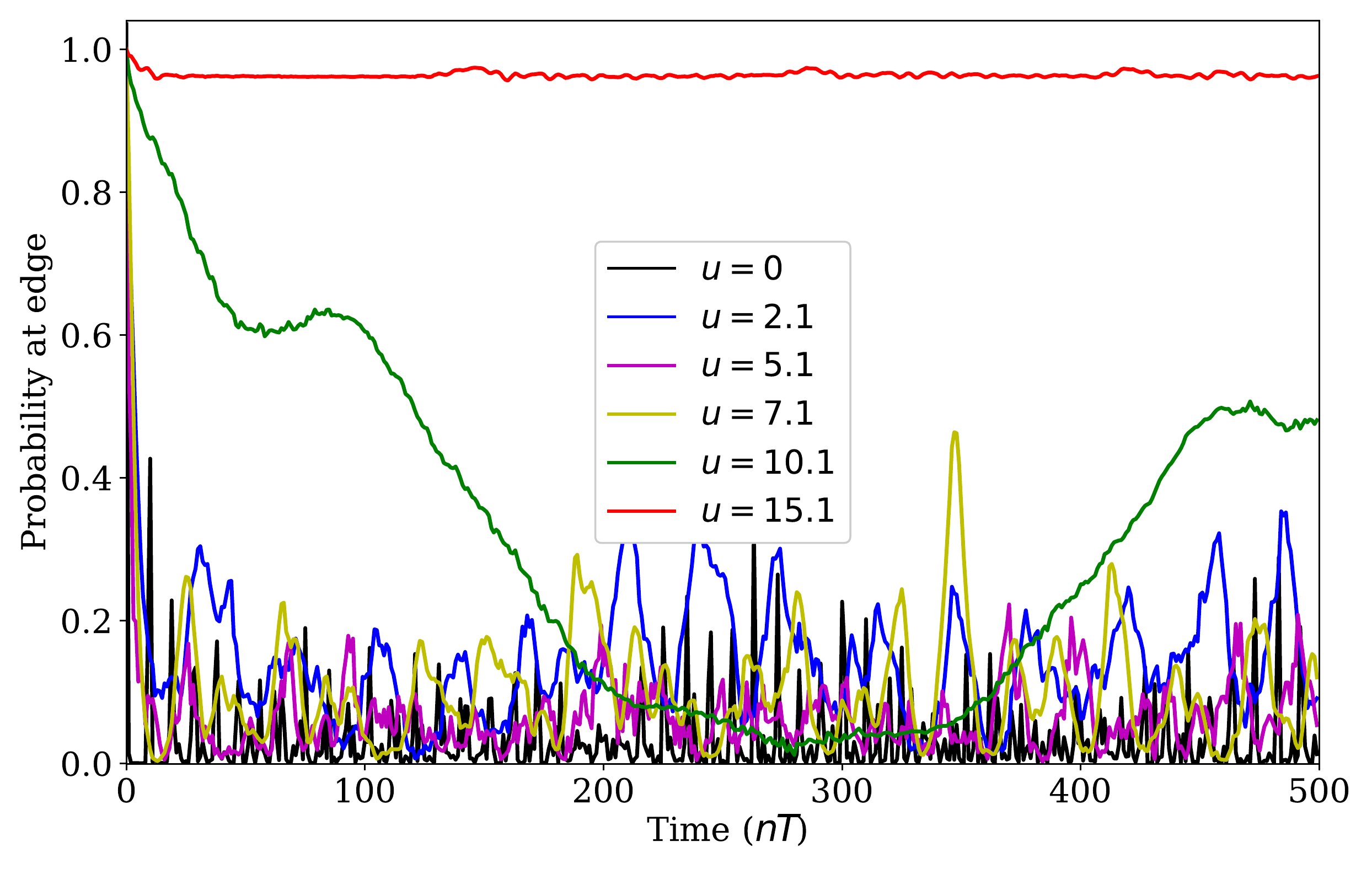}
\caption{Return probability of a state initialized with both particles at 
edge. We have taken a 20-site system with $a=4$ and $g=1$.} 
\label{ss1fig17} \end{figure}

We will now study the time evolution of the probability of two particles 
remaining close to the edge where they are initialized. We will take the
initial state to be one where both particles are at the site labeled 0.
We define the probability of the two particles to remain near the edge as the 
sum $|\psi (0,0)|^2 + |\psi (0,1)|^2 + |\psi (1,1)|^2$ and track this as a 
function of time. We study this for different values of the driving amplitude 
$a$ and the interaction strength $u > 0$. We again choose two values of $a$ 
given by $2.6$ and $4$, one lying within the region where single-particle 
edge states exist, and one lying in the region where there are no edge 
states. We consider a 20-site system with $g=1$ and $\om=1$. The numerical
results are shown in Figs.~\ref{ss1fig16} and \ref{ss1fig17} 
for $a=2.6$ and 4 respectively. We note the following.

\noi (i) For $a=2.6$ where the non-interacting driven model hosts edge states, 
we see that for $u=0$ the particle always stays near the edge, hence 
the probability remains non-zero for all times. For $a=4$ which does not give
edge states when driven, the probability for $u=0$ drops to zero quickly.
Thus the two bosons quickly spread out into the bulk of the system. 

\noi (ii) For a moderately strong interaction strength $u=\pm 2, ~\pm 4$
(of the same order as the hopping $g$), the probability to remain at the 
edge remains small for all times. Hence a moderate amount of either 
attraction or repulsion makes the bosons delocalize into the bulk. 

\noi (iii) For large $u= \pm 10, ~\pm 20$ the two particles remain localized 
near the edge for all times. For large and attractive $u$ this is easy to 
understand as the two particles being at the zeroth site form a deep 
attractive well leading to a bound state. This is also true for large and 
repulsive $u$ since the particles form a deep repulsive well which 
leads to a bound state on a lattice. (It is interesting to note, for instance,
that a large attractive or repulsive potential at one site on a lattice
can host a bound state localized near that site, whereas in a continuum
model, a $\de$-function potential can host a bound state only if it is
attractive). Hence a small hopping ($g \ll |U|$) cannot delocalize the 
two bosons for either sign of $u$.

\noi (iv) An interesting feature in some of the plots is the oscillation of 
the probability with a large time period for certain parameter values. 
For instance, for $a=2.6$ this happens for $u = \pm 6$ as we see in
Fig.~\ref{ss1fig16}. This can be understood as follows.
If there are two Floquet eigenstates which are localized at the edge, have
slightly different Floquet eigenvalues, and have a large overlap with the
initial state where both particles are at the zeroth sites, then the 
time-evolved state will oscillate back forth with a time period with is 
inversely proportional to the difference of the quasienergies of the two 
bound state. 
Indeed we find that for the parameter values given above there are two
such eigenstates at the edge with closely spaced Floquet eigenvalues.



\section{Detection of edge states using transport measurements}
\label{sec5}

In this section we will discuss how it may be possible to detect the edge 
states studied in the earlier sections by looking at transport across
the driven system. In particular, if one attaches metallic leads to the two 
ends of the system, we find that signatures of the edge states may appear as
peaks in the differential conductance when one applies a voltage bias 
across the system which is equal to the quasienergy of one of the edge states.
To find the differential conductance across a periodically driven system
we have to calculate use Floquet scattering theory~\cite{moskalets,agarwal}.


Floquet scattering theory works most easily if we consider a model without
interactions. We will consider a system of non-interacting electrons described 
by the driven tight-binding system discussed in Sec.~\ref{sec3}, which is
attached to two leads which are not driven. We will ignore the spin of the 
electron in this section; to include the effect of spin we will only need to 
multiply our final results for the conductance by a factor of 2. 
The Hamiltonian of our system will have the following parts.
The periodically driven part in the middle (called the wire $W$) will have 
$L$ sites going from $n=0$ to $L-1$; the Hamiltonian in this region will be 
\beq H_{W} ~=~ - ~g ~\sum_{n=0}^{L-2} ~(e^{i\frac{a}{\om}\sin(\om t)} 
c_{n}^{\dg}c_{n+1} ~+~ e^{-i\frac{a}{\om}\sin(\om t)}c_{n+1}^{\dg}c_{n}).
\label{hw} \eeq
(We will ignore the staggered potential $v$ in this section).
The leads on the left and right sides of the wire, $L$ and $R$, consist of 
sites from $n=-\infty$ to $-1$ and from $n=L$ to $\infty$ respectively, with 
the Hamiltonians
\bea H_L &=& - ~g_{l} ~\sum_{n=-\infty}^{-2} ~(c_{n}^{\dg}c_{n+1}+ 
c_{n+1}^{\dg}c_{n}), \non \\
H_R &=& - ~ g_{l} ~\sum_{n=L}^{\infty}(c_{n}^{\dg}c_{n+1}+ c_{n+1}^{\dg}
c_{n}). \label{hlr} \eea
(Note that the energy bands in the wire and leads will lie in the ranges
$[-2g,2g]$ and $[-2g_l,2g_l]$, respectively, and we are allowing these to 
differ from each other). Finally, there will be couplings between the left 
and right leads and the wire given by the Hamiltonian
\beq H_C ~=~ -~ g_c ~(c_{-1}^\dg c_0 ~+~ c_0^\dg c_{-1} ~+~ c_{L-1}^\dg
c_L ~+~ c_L^\dg c_{L-1}). \label{hc} \eeq

We will now consider an electron which is incident from the left lead 
with an energy $E_0 = - 2 g_l \cos (k_0)$ and wave function $\psi (n) = 
e^{i (k_0 n - E_0 t)}$; the energy must lie in the range $[-2g_l,2 g_l]$.
When the electron enters the wire region which is being driven with a 
frequency $\om$, it may lose or gain energy in multiples of $\om$.
Hence it may get reflected back to the left lead or transmitted to the
right lead with an energy $E_{p} = E_{0}-p\om$, where $p$ is an integer. 
This must be related to the momentum $k_p$ by the relation
$E_{p}= -2g_{l}\cos(k_{p})$. For this to describe a propagating wave, we must 
have $k_p$ real which means that $E_p$ must lie in the range 
$[-2g_{l}, 2g_{l}]$. If $E_p$ lies outside this range, the corresponding 
wave functions $e^{ik_p n}$ should decay exponentially as we go away from
the wire into the leads; hence $k_p$ will be complex and we have to 
choose the imaginary part of $k_p$ appropriately. We find that $k_p$ has 
to be chosen as follows.

\noi (i) For $-2 g_l \le E_{p} \le 2g_{l}$, we have $k_{p} = 
\cos^{-1}[-E_p/(2g_{l})]$, and we choose $0 \le k_p \le \pi$.
The group velocity for this case is given by $v_p = dE_p/dk_p = 2 g_l 
\sin (k_p)$ which satisfies $v_p \ge 0$. This will appear in the expressions
for the currents in the leads.

\noi (ii) For $E_{p} < -2g_{l}$, we have $k_{p} = 
i\cosh^{-1}(-\frac{E_p}{2g_{l}})$.
This corresponds to a decaying wave function which does not contribute to the
current in the leads.

\noi (iii) For $E_{p} > 2g_{l}$, we have $k_{p} = 
i\cosh^{-1}(\frac{E_p}{2g_{l}})+\pi$.
This also corresponds to a decaying wave function and does not contribute to 
the current.

Next, we find the reflection and transmission amplitudes, $r_p$ back to the
left lead and $t_p$ to the right lead respectively, for different values of 
$E_p$. To do this, we write down the wave functions in the wire and the leads 
and use their continuity at the junctions between the different regions. 
The wave functions are as follows.

\noi (i) Region I (left lead): 
\bea \psi (n) &=& e^{i(k_0 n - E_0t)} \non \\
&& +~ \sum_{p=-\infty}^{\infty} r_{p} ~e^{i(-k_{p} n - E_{p}t)} 
~~~{\rm for}~~~ n \le -1. \eea

\noi (ii) Region II (wire):
\beq \psi (n) ~=~ \sum_{p=-\infty}^{\infty} c_{n,p} ~e^{-iE_{p}t} 
~~~{\rm for}~~~ 0 \le n \le L-1. \eeq

\noi (iii) Region III (right lead):
\beq \psi (n) ~=~ \sum_{p=-\infty}^{\infty} t_{p} ~e^{i(k_{p} n- E_{p}t)}
~~~{\rm for}~~~ n \ge L. \eeq

We now solve the Schr\"odinger equation $i d \psi /dt = H \psi$, where $\psi$ 
denotes all the $\psi (n)$'s combined into a column, and the Hamiltonian $H$ 
in this equation can be obtained from the second-quantized Hamiltonians in 
Eqs.~(\ref{hw}-\ref{hc}) in the usual way.
Solving these equations, which naturally involves matching the wave
functions at the junctions between the different regions, and equating
the coefficients of $e^{-iE_p t}$ on the two sides of every equation for all
values of $p$, we obtain the following $L+2$ equations for each value of $p$,
\begin{eqnarray}
&& g_c ~c_{0,p} ~-~ g_l ~r_{p} ~=~ g_l ~\de_{p,0}, \non \\
&& g_c ~r_{p} ~e^{ik_{p}} ~+~ E_p ~c_{0,p} \non \\
&& +~ g ~\sum_{m=-\infty}^{\infty} J_{m}(a/\om) ~c_{1,p+m} ~=~ - g_c ~
e^{-i k_0} ~\de_{p,0}, \non \\
&& g ~\sum_{m} (-1)^{m} J_{m}(a/\om) ~c_{n-1,p+m} ~+~ E_{p} ~c_{n,p} \non \\
&& +~ g ~\sum_{m} J_{m}(a/\om) ~c_{n+1,p+m} ~=~ 0 ~~~{\rm for}~~~ 1 \le n
\le L-2, \non \\
&& g ~\sum_{m} (-1)^{m} J_{m}(a/\om) ~c_{L-2,p+m} ~+~ E_{p} ~c_{L-1,p} \non \\
&& +~ g_{c} ~t_{p} ~e^{ik_{p}L} ~=~ 0, \non \\
&& g_c ~c_{L-1,p} ~-~ g_l ~t_{p} ~e^{i k_p (L-1)}~=~ 0. 
\label{fleqs} \end{eqnarray}
We can write Eqs.~\eqref{fleqs} as a matrix equation where the 
left-hand side consists of a matrix acting on a column of $r_{p}, ~t_{p}$ and 
$c_{n,p}$'s, and the right-hand side is given by a column formed out of the 
right-hand sides of the same equations. 
Inverting this matrix equation we can obtain the $r_{p}, t_{p}$ and 
$c_{n,p}$'s in principle. However, this is a infinite-dimensional matrix 
equation and we must therefore truncate it to find the solutions numerically. 
If we keep only $2n_{H}+1$ values of $p$, going from $p=-n_H$ to $+n_H$, 
we will obtain a $(L+2)(2n_{H}+1)$-dimensional matrix from which we can
find $r_{p}, t_{p}$ and $c_{n,p}$. To see if the truncation error is small
enough, we can verify how well the current conservation relation
\begin{equation}
v_{0}= \sum_{p} ~v_p ~(| r_{p}|^{2} + | t_{p} | ^{2}) \label{cons} 
\end{equation}
is satisfied, where $v_p$ is the group velocity for energy $E_p$, and the sum 
over $p$ in Eq.~\eqref{cons} only runs over values for which $E_p$ lies in 
the range $[-2g_p,2g_p]$. 

In the above analysis, we have assumed that the electron is incident from the 
left lead with an energy $E_0$. We can similarly consider what happens if an 
electron is incident from the right lead with the same energy $E_0$. We denote 
the corresponding reflection and transmission amplitudes by $r_p'$ back to 
the right lead and $t_p'$ to the left lead respectively. We now note that the 
Hamiltonians in Eqs.~(\ref{hw}-\ref{hc}) have a parity symmetry if we shift 
$t \to t + \pi/\om$ (this interchanges the factors of $e^{\pm i (a/\om) 
\sin (\om t})$ appearing in Eq.~\eqref{hw}). Hence we will have $|t_p|^2
= |t_p'|^2$ and $|r_p|^2 = |r_p'|^2$. This implies that the outgoing 
current in the right lead will be given by~\cite{moskalets}
\beq I_R ~=~ \frac{e}{h} \int_{-2g_l}^{2g_l} ~dE_0 ~\sum_{p=-\infty}^\infty
\frac{v_p}{v_0} ~|t_p|^2 ~[f(_L (E_0) ~-~ f_R (E_0)], \label{ir} \eeq
where $f_{L/R} (E_0) = [e^{(E_0 - \mu_{L/R})/(k_B T)} + 1]^{-1}$ denotes the 
Fermi-Dirac functions for the left (right) leads respectively, the chemical 
potentials are $\mu_{L/R} = - e V_{L/R}$ where $V_{L/R}$ denote the voltages 
applied to the leads, $-e$ denotes the electron charge, $h = 2 \pi \hbar$, 
and we have ignored the electron spin in writing Eq.~\eqref{ir}.
At zero temperature ($T=0$), we get $I_R = 0$ if $V_L = V_R$. If we set
$V_R = -E_0/e$ and take the limit $V_L \to V_R$, the differential 
conductance $G = I_R/(V_L - V_R)$ is given by
\beq G ~=~ \frac{e^2}{h} ~\sum_p ~\frac{v_p}{v_0} ~|t_p|^2, \eeq
where $|t_p|^2$ is evaluated at the energy $E_0$. We can now plot $G$ as a 
function of $E_0$ to see if any peaks appear due to the 
edge states produced by the driving in the wire region.



\begin{figure}[H]
\centering
\subfigure[]{\includegraphics[width=0.49\textwidth]{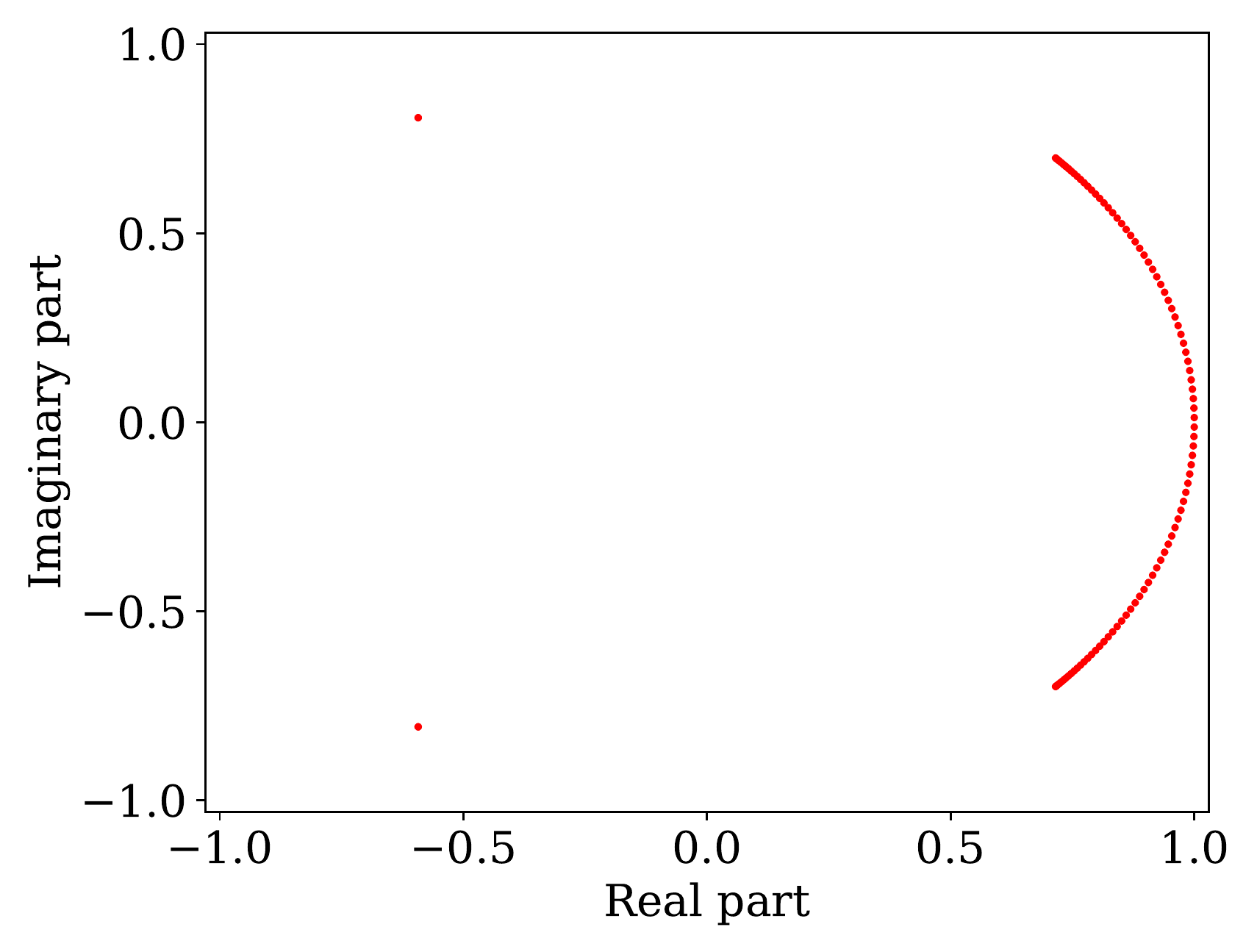}}
\subfigure[]{\includegraphics[width=0.49\textwidth]{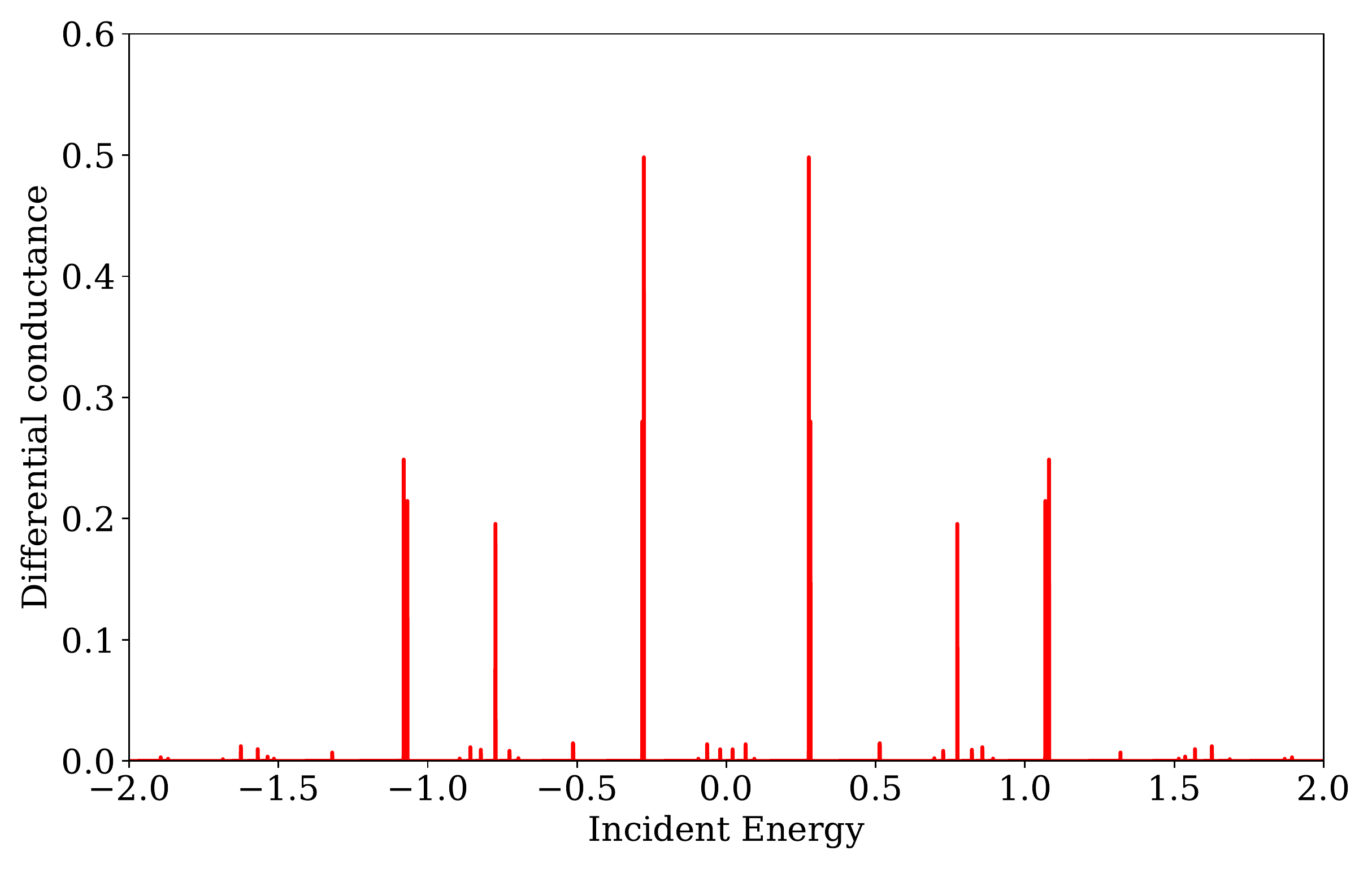}}
\caption{(a) Floquet eigenvalues of edge states (isolated red dots): 
$-0.5925 \pm 
0.8056 i$ corresponding to quasienergy = $\pm 0.2756$. (b) We see peaks in the 
differential conductance $G$ (in units of $e^2/h$) when the incident electron 
has the same energy. The other peaks in $G$ correspond to side bands with 
energies equal to $\pm 0.2756 \pm p \om$.} \label{ss1fig18} \end{figure}

\begin{figure}[H]
\subfigure[]{\includegraphics[width=0.49\textwidth]{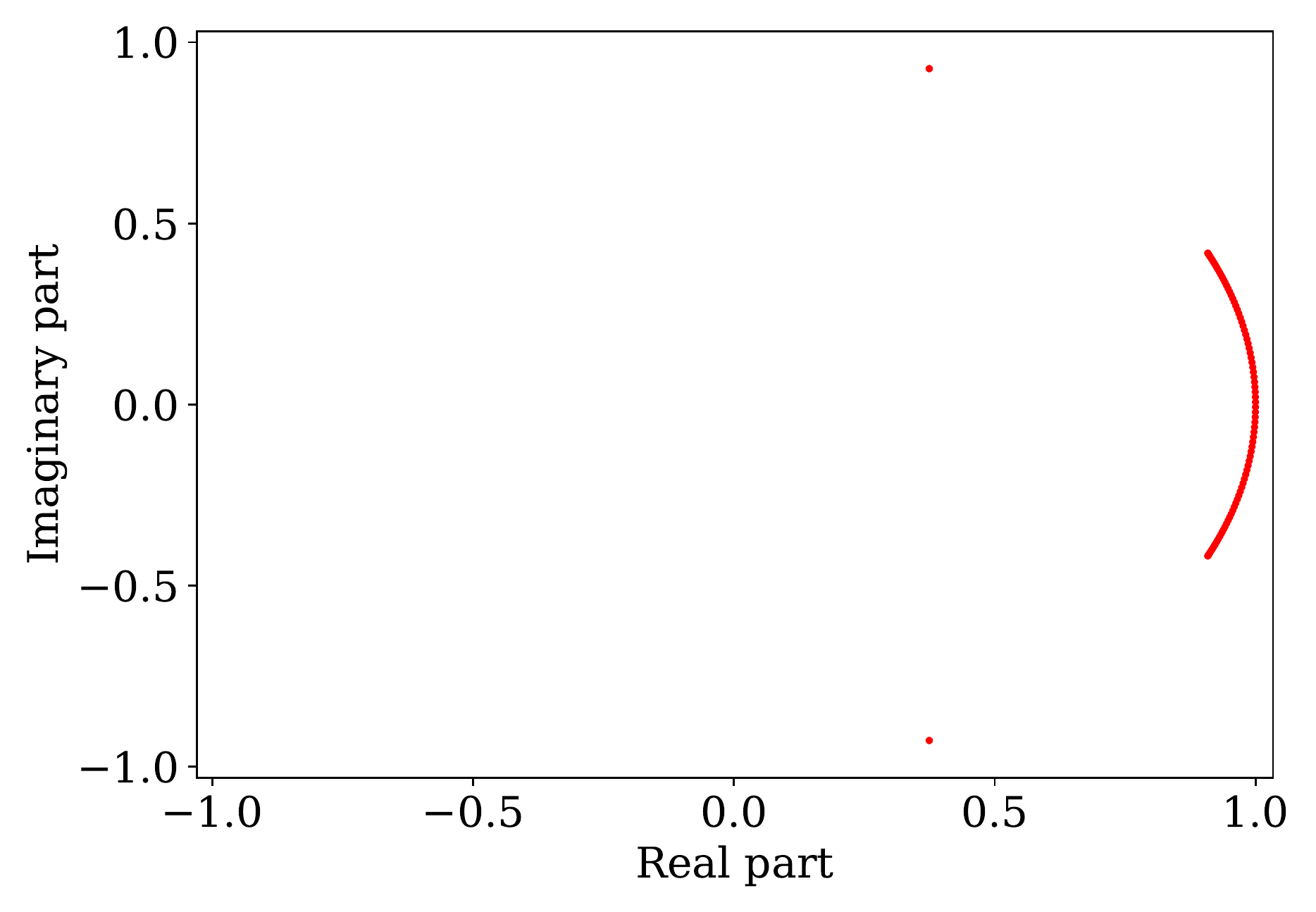}}
\subfigure[]{\includegraphics[width=0.49\textwidth,height=0.35\textwidth]{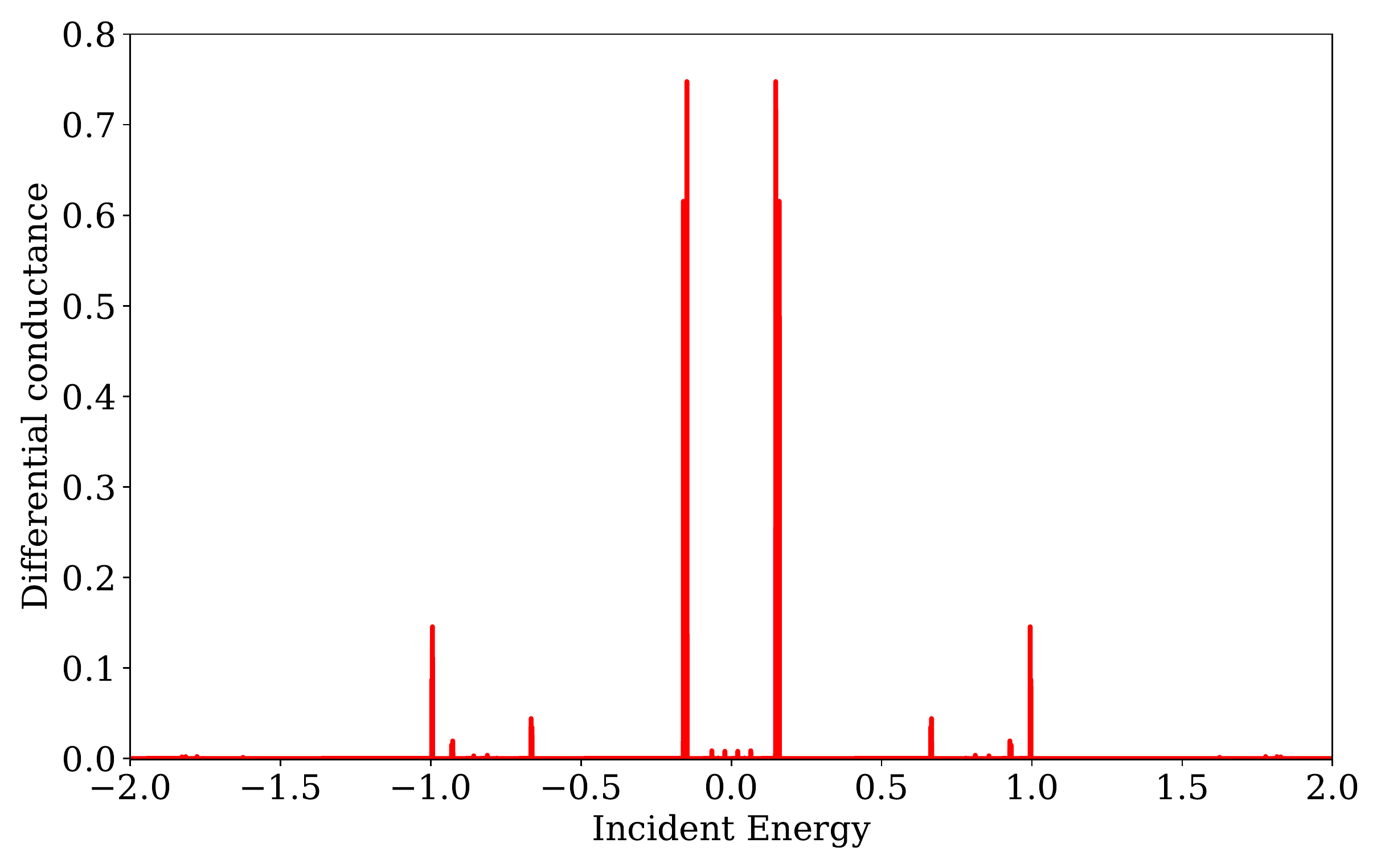}}
\caption{(a) Floquet eigenvalues of edge states (isolated red dots): 
$0.3755 \pm 
0.9268 i$ corresponding to quasienergy = $\pm 0.1482$. (b) We see peaks in the 
differential conductance $G$ (in units of $e^2/h$) when the incident electron 
has the same energy. The other peaks in $G$ correspond to side bands with 
energies equal to $\pm 0.1482 \pm p \om$} \label{ss1fig19} \end{figure}

We will now present our numerical results for the edge states and their
effects in a plot of $G$ versus of $E_0$ in Figs.~\ref{ss1fig18} and 
\ref{ss1fig19}. We take the coupling between the wire and the leads to 
be small,
i.e., $g_c \ll g_l$, so that the edge states are not significantly disturbed
by this coupling. We choose $g = g_l =1$, $g_c = 0.01$ and $\om = \pi /4$,
and consider two values of $a/\om = 2.5$ and $5.6$ where we know that edge 
states exist. We indeed see that there are peaks when $E_0$ is equal to
the quasienergy of any of the edge states or $E_0$ differs from the 
quasienergies by integer multiples of $\om$. In general we also find
contributions to $G$ from the bulk states in the wire, but in the limit 
$g_{c} \ll g$, these vanish, and we only see contributions from the edge 
states. It is important to note here that although the band width in the 
leads, $4g_l$, is equal to the bare band width in the wire, $4g$, the driving 
reduces the effective band width in the wire to $4 g |J_0 (a/\om)|$, as
explained in Appendix B (for $a/\om = 2.5$ and $5.6$, $|J_0 (a/\om)| = 0.048$ 
and $0.027$ respectively). As a result, we have edge states whose Floquet 
eigenvalues lie well outside the effective band width of the wire but inside 
the band width of the leads. This makes it possible to detect these edge states
by sending in an electron with the appropriate energy from the leads, while
easily distinguishing their contributions from those of the bulk 
states in the wire.

Figure~\ref{ss1fig18} (a) shows that the Floquet eigenvalues for a 
driven 100-site system (with no leads) for $g=1$, $\om =\pi/4$, and 
$a/\om = 2.5$. The edge states are clearly distinguishable from the bulk 
states; their Floquet eigenvalues are $-0.5925 \pm 0.8056 i$, and the
the corresponding quasienergies are $\pm 0.2756$. To calculate the
differential conductance, we have chosen a smaller system with $L=10$, so that
the states at the two edges can hybridize with each other and thereby lead to 
transmission across the wire. We choose $g=g_l=1$ and $g_{c}=0.01$; the small 
values of $g_c$ (which is equivalent to having a large barrier between the wire 
and the leads) ensures that the bulk states in the wire contribute very little
to the conductance. 
Figure~\ref{ss1fig18} (b) shows a plot of $G$ versus $E_0$. We see that 
there are exactly at the quasienergies $\pm 0.2756$ corresponding to the edge
states and also at side bands whose energies differ from the edge states
by integer multiples of $\om$.

In Figs.~\ref{ss1fig19} (a) and (b), we show the same results for
$a/\om = 5.6$; all the other parameters have the same values as in 
Figs.~\ref{ss1fig18}. The edge states now have Floquet eigenvalues 
$0.3755 \pm 0.9268 i$, corresponding to quasienergies $\pm 0.2756$. 
Once again we see peaks in $G$ at these quasienergies and other energies
differing from them by integer multiples of $\om$.

\begin{figure}[H]
\centering
\includegraphics[width=0.49\textwidth]{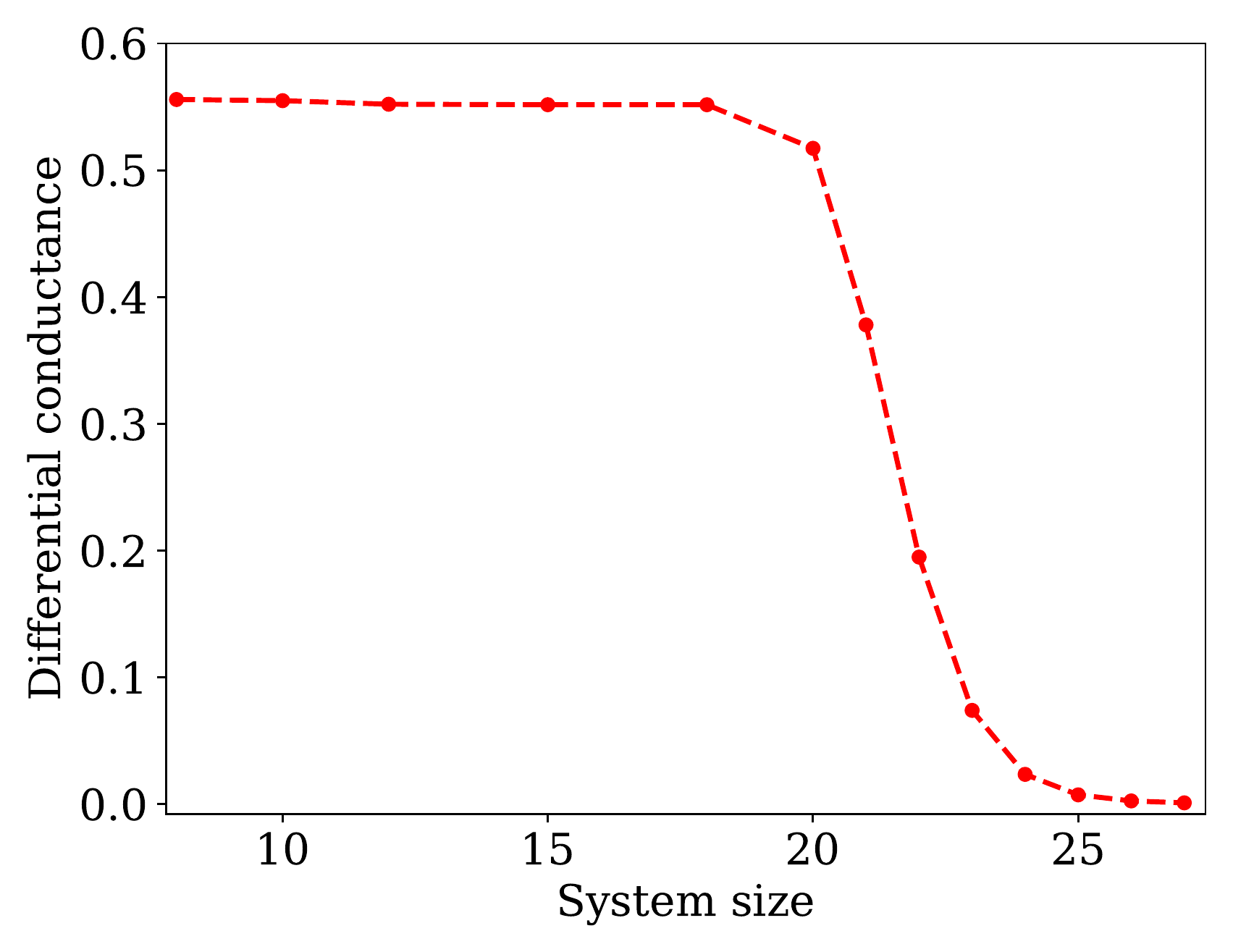}
\caption{Maximum differential conductance $G$ (in units of $e^2/h$) versus 
system size $L$.
We have chosen $\om = \pi/4$, $a/\om = 2.5$ (so that $J_0 (a/\om)$ is close
to zero), $g=g_l=1.0$, and $g_c =0.01$} \label{ss1fig20} \end{figure}

We emphasize that the detection of edge states through a measurement of the
conductance requires that the states at the two edges must hybridize with each 
other by a significant amount; if they do not hybridize, the electron will 
not be transmit from one end to the other.
The hybridization between the two edges is crucially dependent on the
system size $L$. We find numerically that the hybridization becomes very small
beyond about $L=20$ for the parameter values that we have chosen. This is 
because the wave function decreases exponentially with some decay length
as we go away from the edge; hence the overlap between the edge states at the 
two ends will become very small if $L$ becomes larger than the decay length.
Figure~\ref{ss1fig20} shows a plot of the maximum value of $G$ 
(i.e., maximized as a function of the incident energy) versus the system size 
$L$, for the same parameter values as in Figs.~\ref{ss1fig18}. We see 
that there is a sharp drop beyond about $L=20$. 

\section{Discussion}
\label{sec6}

We begin by summarizing our results. We find that a tight-binding model in 
one dimension can host edge states when the phase of the hopping amplitude 
is periodically driven in time by applying an oscillating electric field. 
The presence of a staggered potential or an on-site Bose-Hubbard 
interaction generally enhances the regions where such states appear.
The edge states only appear when the driving frequency is of the order of 
the hopping. For frequencies much larger than the hopping, we find that there 
are no edge states; the reason for this is explained at the end of Appendix 
C 1. Hence we cannot use the Floquet-Magnus expansion~\cite{bukov,mikami}, 
which is valid at high frequencies, to study the edge states. 
We have used a Floquet perturbation theory to show that when the staggered
potential or the interaction strength is much larger than the hopping, 
periodic driving can generate states localized at the edges.
The results obtained by this method agree well with those found numerically.
Finally, we have shown that a measurement of the differential conductance
across a periodically driven wire with non-interacting electrons can detect the 
edge states; the conductance has peaks when the voltage bias coincides with
the quasienergy of one of the edge states.

We now recall our most interesting findings.
In the case of a non-interacting model, we have studied the ranges of the 
various parameters for which one or more Floquet eigenstates exist near each 
edge of a long but finite system. In some cases, we find that these states 
are truly localized at the edges; their wave functions decay rapidly to zero 
as one moves away from the edges and are therefore normalizable. In other 
cases, we find that the wave functions are much larger at the edges than in 
the bulk; however, they do not go to zero as we go deep into the bulk, and the
wave functions are therefore not normalizable. (By tuning the driving
parameters, however, the wave functions of these states can be made to go 
to almost zero in the bulk and therefore look very similar to 
true edge states). We find that the two kinds of
states are respectively associated with Floquet eigenvalues which lie outside 
or within the continuum of eigenvalues of the bulk states. We have then 
studied the time evolution of a state which is not a Floquet eigenstate and is 
initially localized at the edge of the system. Depending on the system
parameters, we find that the state can, with a finite probability, remain 
localized near the edge for all times or can move away completely into the 
bulk. The former happens if the system has Floquet eigenstates which are
localized at the edges. A similar phenomenon occurs in the periodically
driven Bose-Hubbard model with two particles when we study the time
evolution of a state which initially has both particles at the edge. Once
again we find that the particles can remain at the edge with a finite
probability or move away into the bulk, depending on the system parameters,
and the former happens if there are Floquet eigenstates which are localized at
the edges. At the end, we have studied how the edge states can be detected 
using transport measurements. We consider a tight-binding model of 
non-interacting fermions in which semi-infinite leads 
are weakly coupled to a finite length wire in the middle. 
The hopping phase is periodically driven in the wire; this effectively
reduces the value of the hopping in the wire and this can make the band width
in the wire much smaller than in the leads.
We find that when the isolated wire (i.e., without any leads) 
has edge states whose Floquet eigenvalues lie outside the range of eigenvalues 
of the bulk states of the wire, the differential conductance across the system 
with leads has peaks when the chemical potential of the leads is equal to the 
quasienergies of the edge states. Hence the conductance can provide clear 
evidence of the presence of edge states.

We would like to point out two directions for detailed investigations in the 
future. First, we do not know if the edge states have a 
topological significance. There does not seem to be a topological invariant 
which can tell us how many such states should appear at each edge for a given 
set of system parameters. Second, it would be interesting to examine the 
effects of disorder on the various edge states. We have found that the edge 
states are robust to some amount of disorder if their Floquet eigenvalues are 
separated by a gap from the eigenvalues of the bulk states. This gap-induced 
protection has been found in other driven systems as well~\cite{deb,saha}.

It may be possible to test the results presented in this paper in 
systems of cold atoms trapped in an optical lattice which is periodically 
shaken~\cite{eckardt,lignier}. In such systems, the driving parameters can 
be experimentally varied over a wide range which would allow one to change 
the ratio $a/\om$ across several zeros of the Bessel function as in 
Fig.~\ref{ss1fig01}. Regarding the Bose-Hubbard model, the on-site 
two-body interaction $U$ can be modulated by magnetic Feshbach 
resonance~\cite{chin}. There are experiments on static systems where the 
ratio $U/g$ could be varied over a large range~\cite{trotzky}. Periodic 
driving of such a system would allow one to test our results for this model.
Finally, transport measurements for detecting the edge states can be carried 
out in quantum wire systems where a portion of the wire is subjected to 
electromagnetic radiation which is associated with an oscillating electric 
field.

\vspace{.8cm}
\centerline{\bf Acknowledgments}
\vspace{.5cm}

S.S. thanks MHRD, India for financial support through a PMRF.
D.S. thanks DST, India for Project No. SR/S2/JCB-44/2010 for financial support.

\appendix

\section{Basics of Floquet theory}
\label{basics}

In this Appendix we will briefly present the basics of Floquet 
theory~\cite{bukov,mikami}.
Consider a time-periodic Hamiltonian with period $T$, so $H(t+T) = H(t)$. 
According to Floquet theory, the solutions $\psi_n (t)$ of the time-dependent
Schr\"odinger equation $i d\psi_n /dt = H \psi_n$ can be taken to satisfy 
the condition 
\beq \psi_n (T) ~=~ e^{-i\ep_n T} \psi_n (0), \label{flo1} \eeq
where $e^{-i\ep_n T}$ is the $n$-th Floquet eigenvalue and $\psi_n (0)$ 
is the corresponding Floquet eigenstate. The quantity $\ep_n$ is called the 
quasienergy. Since changing $\ep \to \ep + j\om$, where $\om = 2\pi/T$ and 
$j$ can be any integer, does not change the value of $e^{-i\ep T}$, we
can take $\ep_n$ to lie in the range $[-\om/2,\om/2]$. Next, we can write 
$\psi_n (t)$ in the form
\begin{equation}
\psi_n (t) ~=~ e^{-i\ep_n t} ~\sum_{m=-\infty}^{\infty} ~e^{-im\om t} 
\phi_{n,m}. \end{equation}
Similarly, the time-periodic Hamiltonian can be written as 
\beq H ~=~ \sum_{p=-\infty}^{\infty} ~e^{-ip\om t} H_p, \label{hp1} \eeq
Substituting the above expressions in the Schr\"odinger equation, we
obtain an infinite set of equations~\cite{shirley}
\begin{equation}
\sum_{p=-\infty}^{\infty} ~H_p ~\phi_{n,m-p} ~=~ (\ep_n ~+~ m\om) ~\phi_{n,m}, 
\end{equation}
where $m$ can take any integer value. This matrix eigenvalue equation can be
written as
\begin{eqnarray}
&& \begin{pmatrix} 
\ddots & & & & \\
& H_0+\om & H_{-1} & H_{-2} & \\
& H_{1} & H_0 & H_{-1} & \\
& H_2 & H_{1} & H_0-\om & \\
& & & & \ddots \\
\end{pmatrix} \begin{pmatrix}
\vdots \\
\phi_{n,-1}\\
\phi_{n,0}\\
\phi_{n,1}\\
\vdots \end{pmatrix} \non \\
&& = ~\ep_n \begin{pmatrix}
\vdots \\
\phi_{n,-1}\\
\phi_{n,0}\\
\phi_{n,1}\\
\vdots \end{pmatrix}. \end{eqnarray}
We can truncate this infinite dimensional matrix to a suitably large size
solve the equation numerically to obtain the quasienergies $\ep_n$ and 
Floquet states $\phi_{n,m}$.

There is another approach to solving a Floquet problem. Instead 
of doing a Fourier expansion of $H$, we define a Floquet time-evolution 
operator $U_T= \tau \exp (-i\int_{0}^{T} H(t)dt)$, where $\tau$ denotes
time-ordering. To compute $U_T$ numerically, we divide the interval 
$0$ to $T$ into $N$ steps of size $\Delta t$ each, with $N\Delta t= T$, and 
define $t_{j}=(j-1/2)\Delta t$, where $j = 1,2,\cdots,N$. Then we define
\begin{equation}
U_{T} ~=~ e^{-i\Delta t H(t_{N})} ~\cdots ~e^{-i\Delta t H(t_{2})} ~
e^{-i\Delta t H(t_{1})}, \end{equation}
where we eventually have to take the limit $\Delta t \to 0$ and $N \to \infty$
keeping $N\Delta t= T$ fixed. Since $U_T$ is a unitary operator, its
eigenvalues must be of the form $e^{i\theta_n}$, where the $\theta_n$'s are 
real. Since $\psi_n (T) = U_T \psi_n (0)$, we see from Eq.~\eqref{flo1} that 
$e^{i \theta_n}$ is equal to the Floquet eigenvalue $e^{-i\ep_n T}$. Thus the 
Floquet eigenvalues and eigenstates can be found by diagonalizing $U_T$.

The Floquet eigenvalues have the property that they do not change if the time 
is shifted by an arbitrary amount $t_0$, i.e., if we define the generalized 
time-evolution operator $U (t_2,t_1)= \tau \exp (-i\int_{t_1}^{t_2} H(t)dt)$,
then $U_T = U(T,0)$ and $U(T+t_0,t_0)$ have the same eigenvalues. This is 
because the periodicity of the Hamiltonian, $H(t+T)=H(t)$, implies that 
$U(T,0)$ and $U(T+t_0,t_0)$ are related to each other by a unitary 
transformation. Namely, $U(T+t_0,t_0)= U(T+t_0,T) U(T,0) U(0,t_0) = 
[U(0,t_0)]^{-1} U(T,0) U(0,t_0)$ since $U(T+t_0,T) = U(t_0,0)= 
[U(0,t_0)]^{-1}$. Note that the eigenstates of $U(T,0)$ and $U(T+t_0,t_0)$
differ by a unitary transformation given by $U(t_0,0)$.

\section{Floquet-Magnus expansion}

In this Appendix, we will use the Floquet-Magnus expansion in the high-frequency
limit $\om \to \infty$ to find the effective Hamiltonian $H_{eff}$ for some of 
our models. Given the form of the time-periodic Hamiltonian in Eq.~\eqref{hp1},
the effective Hamiltonian is given, up to order $1/\om$, by~\cite{bukov,mikami}
\beq H_{eff} ~=~ H_0 ~+~ \sum_{p \ne 0} ~\frac{1}{2p \om} ~[H_{-p},H_p] ~+~
\sum_{p \ne 0} ~\frac{1}{p \om} ~[H_p,H_0]. \label{heff1} \eeq

We now evaluate the expression in Eq.~\eqref{heff1} for the Hamiltonian
given in Eq.~\eqref{ham4} for a semi-infinite chain in which the site label 
goes from $n=0$ to $\infty$; we do this to study the structure of 
the effective Hamiltonian near the left end of the system assuming that the 
right end is infinitely far away. We use the identity~\cite{abram}
\beq e^{\frac{ia}{\om} \sin (\om t)} ~=~ \sum_{p=-\infty}^\infty~ J_p \left(
\frac{a}{\om} \right) ~e^{i p \om t}. \label{bessel} \eeq
Using the Bessel function identities $J_{-p} (z) = J_p (-z) = (-1)^p J_p (z)$ 
for all integers $p$, we find from Eqs.~\eqref{ham4}, \eqref{hp1} and 
\eqref{bessel} that
\beq H_p ~=~ - ~g ~(-1)^p J_p \left( \frac{a}{\om} \right) ~\sum_{n=0}^\infty
~(c_n^\dg c_{n+1} ~~+~ (-1)^p c_{n+1}^\dg c_n). \label{hp2} \eeq
It then follows that 
\beq H_0 ~=~ - ~g ~J_0 \left( \frac{a}{\om} \right) ~\sum_{n=0}^\infty
~(c_n^\dg c_{n+1} ~~+~ c_{n+1}^\dg c_n), \eeq
\bea [H_p, H_0] &=& - ~2 g^2 ~J_0 \left( \frac{a}{\om} \right) ~J_p \left( 
\frac{a}{\om} \right) ~c_0^\dg c_0 ~~{\rm if}~~ p ~~{\rm is~ odd}, \non \\
&=& 0 ~~{\rm if}~~ p ~~{\rm is~ even}, \eea
and $[H_p, H_{-p}] = 0$. Equation~\eqref{heff1} therefore gives 
\bea H_{eff} &=& - ~g ~J_0 \left( \frac{a}{\om} \right) ~\sum_{n=0}^\infty
~(c_n^\dg c_{n+1} ~~+~ c_{n+1}^\dg c_n) \non \\
&& - ~\frac{4g^2}{\om} ~J_0 \left( \frac{a}{\om} \right) \left( \sum_{p=1,3,
5,\cdots} ~\frac{J_p \left( \frac{a}{\om} \right)}{p} \right) 
c_0^\dg c_0, \non \\
&& \label{heff2} \eea
up to order $1/\om$. This is a tight-binding model in which the 
nearest-neighbor hopping amplitude is $-g J_0 (a/\om)$ (instead of the
original value of $-g$) and there is a potential at the leftmost site $n=0$.
We note here that for a chain which is infinitely long in both directions, the
effective Hamiltonian is simply given by 
\beq H_{eff} ~=~ - ~g ~J_0 \left( \frac{a}{\om} \right) ~\sum_{n=-
\infty}^\infty ~(c_n^\dg c_{n+1} ~~+~ c_{n+1}^\dg c_n) \label{heff3} \eeq
to all orders in $1/\om$. (The energy-momentum dispersion is therefore
$E_k = - 2 g J_0 (a/\om) \cos k$). The potential at $n=0$ in Eq.~\eqref{heff2}
appears only because the chain ends at that site. 

We emphasize that the expression given in Eq.~\eqref{heff2} is
only valid in the high-frequency
limit, and is therefore not directly applicable to the numerical results
reported in the earlier sections where $\om$ is of the same order as the
other parameters of the system such as $g$ and $v$. However, Eq.~\eqref{heff2}
demonstrates an interesting qualitative effect that arises when a system
ends at one site, namely, the driving gives rise to a potential at that 
site. It would therefore be interesting to study the effect of such 
a potential in a time-independent system to gain some understanding of
the conditions under which such a potential can host an edge state.

\section{Edge states for a static system with an edge potential}

Motivated by the results in Appendix B, we will now study whether a
semi-infinite chain with a time-independent Hamiltonian with a 
potential at the leftmost site can host an edge state localized near
that site. We will then look at the effect that the addition of a 
staggered potential can have. These are interesting problems in themselves,
quite apart from the fact that they can give us some understanding of
why edge states can appear in a periodically driven system.

We will now study two time-independent models on a semi-infinite system
in which the site labels go from $n=0$ to $\infty$. In each case, we will
study if there are eigenstates of the Hamiltonian which are localized near 
$n=0$.

\noi (i) A non-interacting tight-binding model with a potential at the leftmost
site and a staggered on-site potential. The Hamiltonian is
\begin{eqnarray}
H &=& - ~g ~\sum_{n=0}^{\infty} ~(c_{n}^{\dg}c_{n+1} ~+~ c_{n+1}^{\dg}c_{n})
~+~ v ~\sum_{n=0}^{\infty} ~(-1)^{n} ~c_{n}^{\dg}c_{n} \non \\
&& + ~A ~c_{0}^{\dg}c_{0}, \label{c1} \end{eqnarray}
and we will look for a single-particle eigenstate localized near $n=0$.

\noi (ii) A tight-binding Bose-Hubbard model with a potential at the leftmost 
site and an on-site interaction. The Hamiltonian is
\begin{eqnarray}
H &=& - ~g ~\sum_{n=0}^{\infty} ~(b_{n}^{\dg}b_{n+1} ~+~ b_{n+1}^{\dg}b_{n}) ~
+~ \frac{u}{2} ~\sum_{n=0}^{\infty} \rho_{n} (\rho_{n}-1) \non \\
&& + ~A ~b_{0}^{\dg}b_{0}, \label{c2} \end{eqnarray}
with $\rho_n = b_n^\dg b_n$, and we will look for a two-particle bound state 
localized near $n=0$.

In the models described by Eqs.~\eqref{c1} and \eqref{c2} respectively, we 
will numerically find the regions of parameter space $(v,A)$ and $(u,A)$ 
where single-particle and two-particle bound states exist near $n=0$. For 
the non-interacting model, we will provide an analytical derivation
of the results using the Lippmann-Schwinger method. 

\subsection{Non-interacting model with a staggered potential}

\begin{figure}[H]
\centering
\hspace*{-.6cm}\subfigure[]{\includegraphics[width=0.52\textwidth]{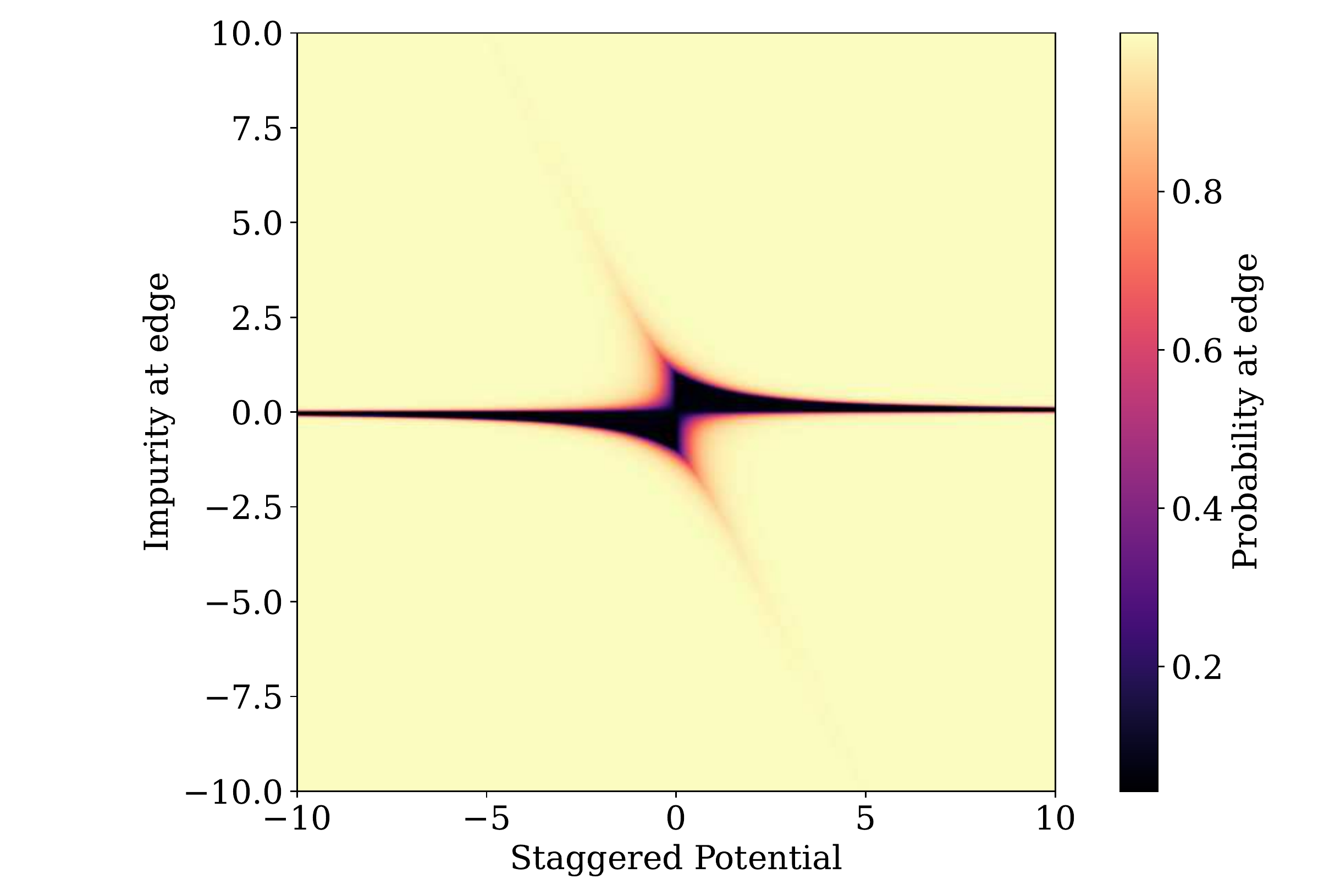}}
\label{nonintstate1}
\hspace*{-.6cm}\subfigure[]{\includegraphics[width=0.52\textwidth]{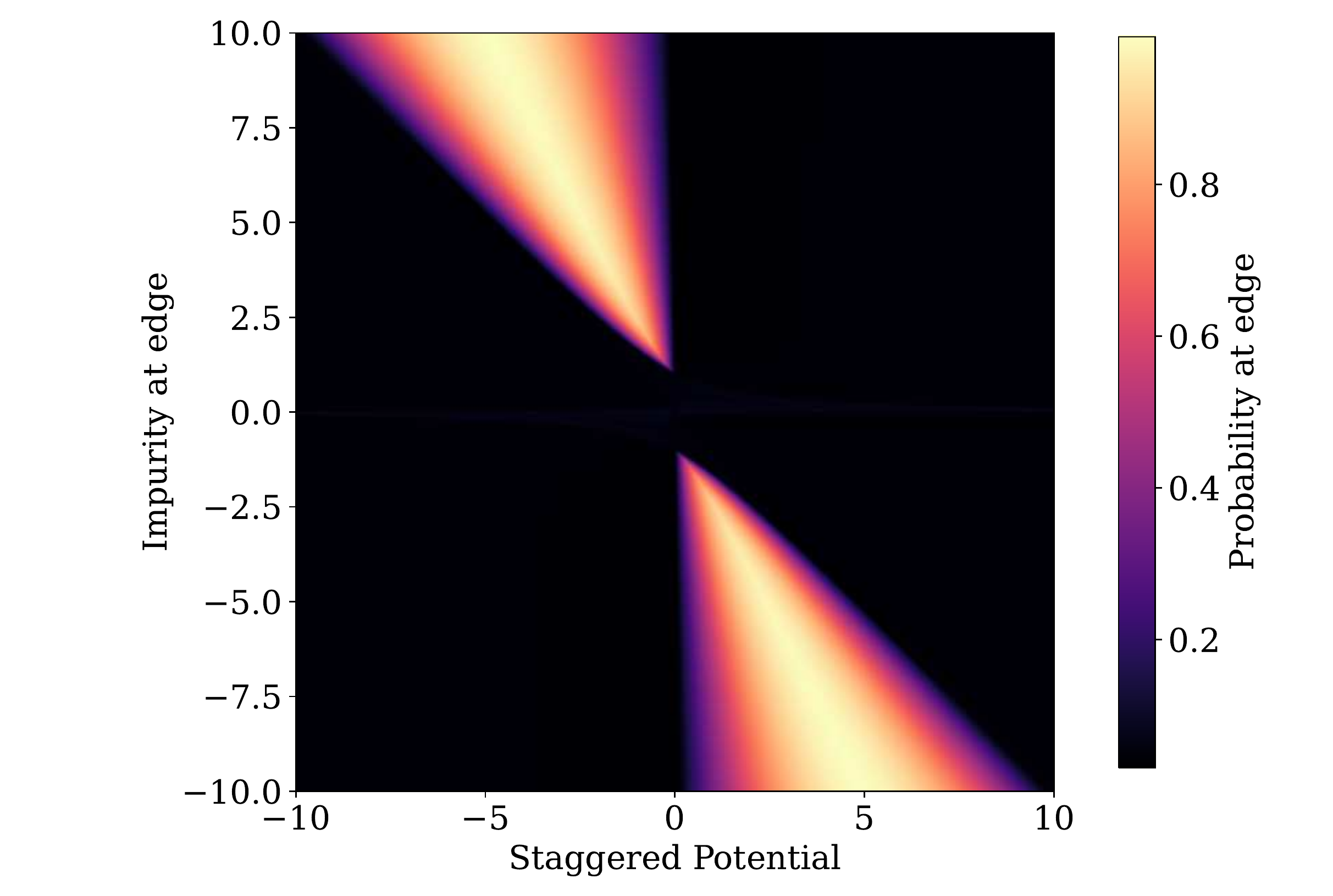}}
\label{nonintstate2}
\caption{Probabilities $|\psi (0)|^2$ of the states with the largest two IPR 
values at the 
left edge of a 100-site system versus $(v,A)$. The probability is larger in 
the lighter colored regions. Plot (a) shows that one edge state exists for a 
large range of parameters values, while plot (b) shows that a second edge state
exists in some smaller regions of parameters. We have set $g=1$.}
\label{ss1fig21} \end{figure}

To numerically find edge states as a function of the parameters $(v,A)$, we 
consider a 100-site system and look at the two states with the largest values 
of the IPR. The probabilities of the two states at the leftmost site, 
$|\psi(0)|^2$, are plotted versus $(v,A)$ in Figs.~\ref{ss1fig21} (a) and 
(b). A large value of the probability corresponds to an 
edge state localized near $n=0$. We find that this model can have zero,
one, or two edge states as we vary $(v,A)$. One edge state exists for a large
range of values of $(v,A)$ as we see in Fig.~\ref{ss1fig21} (a) while a
second edge state exists for a smaller range of $(v,A)$ as shown in 
Fig.~\ref{ss1fig21} (b). The two figures have the symmetry that they 
look the same under the inversion $(v,A) \to (-v,-A)$. This is because
the Hamiltonian in Eq.~\eqref{c1} changes sign if we flip the signs
of $v$ and $A$ and transform $c_n \to (-1)^n c_n$. Thus if there
is a bound state with energy $E$ and wave function $\psi (n)$ for parameters
$(v,A)$, there will be a bound state with energy $-E$ and wave function
$(-1)^n \psi (n)$ for parameters $(-v,-A)$; the probability $|\psi(0)|^2$
remains the same under this transformation.



We will now show that the regions of bound states (light regions) 
in Figs.~\ref{ss1fig21} (a) 
and (b) can be analytically understood using the Lippmann-Schwinger method. 
The analysis proceeds as follows. We consider the Hamiltonian $H$ in 
Eq.~\eqref{c1} for a semi-infinite system. This can be written as a sum 
$H= H_0 + V$, where
\bea H_0 &=& - ~g ~\sum_{n=0}^{\infty} ~(c_{n}^{\dg}c_{n+1} ~+~ c_{n+1}^{\dg}
c_{n}) ~+~ v ~\sum_{n=0}^{\infty} ~(-1)^{n} ~c_{n}^{\dg}c_{n}, \non \\
V &=& A ~c_{0}^{\dg}c_{0}. \label{ls1} \eea
We now write the equation $H | \psi \ra = (H_0 + V) | \psi \ra$ in the form
\beq |\psi \ra = \frac{1}{E ~I ~-~ H_0} V | \psi \ra. \label{ls2} \eeq
Working in the basis of states $|n \ra$, where 
$n=0,1,2,\cdots$, we can use the resolution of identity,
$I = \sum_{n=0}^\infty ~| n \ra \la n |$, to write 
\bea V | \psi \ra &=& \sum_{n=0}^\infty ~| n \ra \la n | \psi \ra \non \\
&=& A | 0 \ra \la 0 | \psi \ra. \label{ls3} \eea
Combining Eqs.~(\ref{ls2}-\ref{ls3}), we obtain 
\beq \la 0 | \psi \ra ~=~ A~ \la 0 | \frac{1}{E ~I ~-~ H_0} | 0 \ra \la 0 | 
\psi \ra. \label{ls4} \eeq
Assuming that we are looking for a state for which $\psi (0) = \la 0 | \psi 
\ra$ is non-zero, Eq.~\eqref{ls4} implies that
\beq \frac{1}{A} ~=~ \la 0 | \frac{1}{E ~I ~-~ H_0} | 0 \ra. \label{ls5} \eeq

We will now evaluate the right-hand side of Eq.~\eqref{ls5} by using the
resolution of identity written in terms of the basis of eigenstates of
$H_0$. Since $H_0$ has a staggered potential, its unit cell has two sites
which we will call $(a,b)$. The sites which were earlier labeled as 
$n=0,1,2,\cdots$ will now denote $(a,n/2)$ if $n$ is even and 
$(b,(n-1)/2)$ if $n$ is odd. We can then write
\bea H_0 &=& \sum_{m=0}^\infty ~[(a_m^\dg b_m ~+~ b_m^\dg a_m ~+~ b_m^\dg 
a_{m+1} ~+~ a_{m+1}^\dg b_m) \non \\
&& ~~~~~~~~~+~ v ~(a_m^\dg a_m ~-~ b_m^\dg b_m)]. \label{ls6} \eea
We can find the eigenvalues and eigenstates of this Hamiltonian as follows.
If the sum over $m$ went from $-\infty$ to $\infty$, we could use translation 
invariance and find that the energy eigenvalues are given by
\beq E_{k,\pm} ~=~ \pm ~\sqrt{4 g^2 \cos^2 k ~+~ v^2}, \label{ls7} \eeq
where $\pm$ denote the positive and negative energy bands respectively,
the eigenstates have the plane wave form
\bea \psi (a,m,k,\pm) &=& \al_{k,\pm} ~e^{ikm}, \non \\
\psi (b,m,k,\pm) &=& \beta_{k,\pm} ~e^{ikm}, \label{ls8} \eea
where 
\bea \al_{k,\pm} &=& -~ \frac{2 g \cos k}{\sqrt{(E_{k,\pm} ~-~ v)^2 ~+~ 4 g^2
\cos^2 k}}, \non \\
\beta_{k,\pm} &=& \frac{E_{k,\pm} ~-~ v}{\sqrt{(E_{k,\pm} ~-~ v)^2 ~+~ 4 g^2
\cos^2 k}}, \label{ls9} \eea
and the momentum $k$ lies in the range $[-\pi /2, \pi /2]$ since the unit
cell spacing is equal to 2 in terms of the original lattice spacing. Now, 
since our system ends on the left at the site $(a,0)$, we can find its 
eigenstates by appropriately superposing the momentum eigenstates 
corresponding to $\pm k$ in such a way that the wave function vanishes at 
the `phantom' sites given by $(a,m=-1)$ and $(b,m=-1)$. Noting that the
expressions in Eqs.~\eqref{ls9} are even functions of $k$, we then find that 
the eigenstates of the semi-infinite chain are given by
\bea \psi (a,m,k,\pm) &=& \sqrt{2} ~\al_{k,\pm} ~\sin (k(m+1)), \non \\
\psi (b,m,k,\pm) &=& \sqrt{2} ~\beta_{k,\pm} ~\sin(k(m+1)), \label{ls10} \eea
where the $\sqrt{2}$ has been put in to ensure orthonormality, and $k$ now 
lies in the range $[0,\pi /2]$. The wave function for the state $|k,\pm \ra$ 
at the leftmost site, $(a,m=0)$, is therefore given by $\sqrt{2} \al_{k,\pm} 
\sin k$.
 

We can now use the above eigenstates $| k,\pm \ra$ to write the resolution
of identity:
\begin{equation}
\int_{0}^{\pi/2} \frac{dk}{\pi/2} ~(| k,+ \rangle \langle k,+ | ~+~ 
| k,- \rangle \langle k,- |) ~=~ I. \label{ls11} \end{equation}
Using this in Eq.~\eqref{ls5}, we obtain 
\bea \frac{1}{A} &=& \int_{0}^{\pi/2} \frac{dk}{\pi/2} ~[\la 0 | \frac{1}{
E ~I ~-~ H_0} | k, +\ra \la k,+ | 0 \ra \non \\
&& ~~~~~~~~~~~~~~~~~~~+~ \la 0 | \frac{1}{E ~I ~-~ H_0} | k, -\ra \la k,- | 
0 \ra] \non \\
&=& \int_{0}^{\pi/2} \frac{dk}{\pi/2} ~2 \sin^2 k ~\left[ \frac{|\al_{k,+}|^2}{
E ~-~ E_{k,+}} ~+~ \frac{|\al_{k,-}|^2}{E ~-~ E_{k,-}} \right]. \non \\
&& \label{ls12} \eea
(The above equation is only valid for $E$ lying outside the energy bands
$E_{k,\pm}$, otherwise the denominators can vanish and we would have to 
evaluate the integral more carefully). Substituting the expression for 
$\al_{k,\pm}$ in Eq.~\eqref{ls9} in Eq.~\eqref{ls12}, we obtain 
\bea \frac{1}{A} &=& \int_{0}^{\pi/2} \frac{dk}{\pi} ~16 g^{2} \sin^2 k 
\cos^2 k \non \\
&& \times ~[ \frac{1}{(E-E_{k,+}) ~[(E_{k,+}-v)^{2} ~+~ 4g^{2}\cos^{2} k]} 
\non \\ 
&& ~~~+ ~\frac{1}{(E-E_{k,-}) ~[(E_{k,-}-v)^{2} ~+~ 4g^{2}\cos^{2} k]}]. \non \\
&& \label{ls13} \eea

Now, if an edge state exists, its energy $E$ must lie either above the
upper band ($E > \sqrt{4g^2 +v^2}$), or below the lower band ($E < -
\sqrt{4g^2 +v^2}$), or in the gap between the two bands ($-v < E < v$).
In the first two cases, the integral in Eq.~\eqref{ls13} gives the result
\begin{equation}
A ~=~ \dfrac{E^{2} ~-~ v^{2} ~+~ \sqrt{(E^{2} ~-~ v^{2})(E^{2} ~-~ 4g^{2} ~-~
v^{2})}}{2(E ~+~ v)}. \label{ls14} \end{equation}
In the third case, Eq.~\eqref{ls13} gives
\begin{equation}
A ~=~ - ~\dfrac{v^{2} ~-~ E^{2} ~+~ \sqrt{(v^{2} ~-~ E^{2})(4g^{2} ~+~
v^{2} - E^2)}}{2(v ~+~ E)}. \label{ls15} \end{equation}
Equations~\eqref{ls14} and \eqref{ls15} implicitly give the energy $E$ of 
an edge state in terms of $g$, $v$, and $A$. We find that these give certain 
conditions on the allowed values of $A$ for a given value of $(g,v)$.


\noi (i) $E < - \sqrt{4g^2 +v^2}$ implies that we must have $A < A_1$, where
\beq A_1 ~=~ \frac{2g^2}{v ~-~ \sqrt{v^2 ~+~ 4 g^2}}. \label{a1} \eeq

\noi (ii) $E > \sqrt{4g^2 +v^2}$ implies that $A > A_2$, where
\beq A_2 ~=~ \frac{2g^2}{v ~+~ \sqrt{v^2 ~+~ 4 g^2}}. \label{a2} \eeq

\noi (iii) $-v < E < v$ implies that if $v >0$, we must have $A < 0$,
while if $v <0$, we must have $A > 0$. Namely, we must have 
\beq Av ~<~ 0. \label{a3} \eeq

\begin{figure}[H]
\centering
\includegraphics[width=0.48\textwidth]{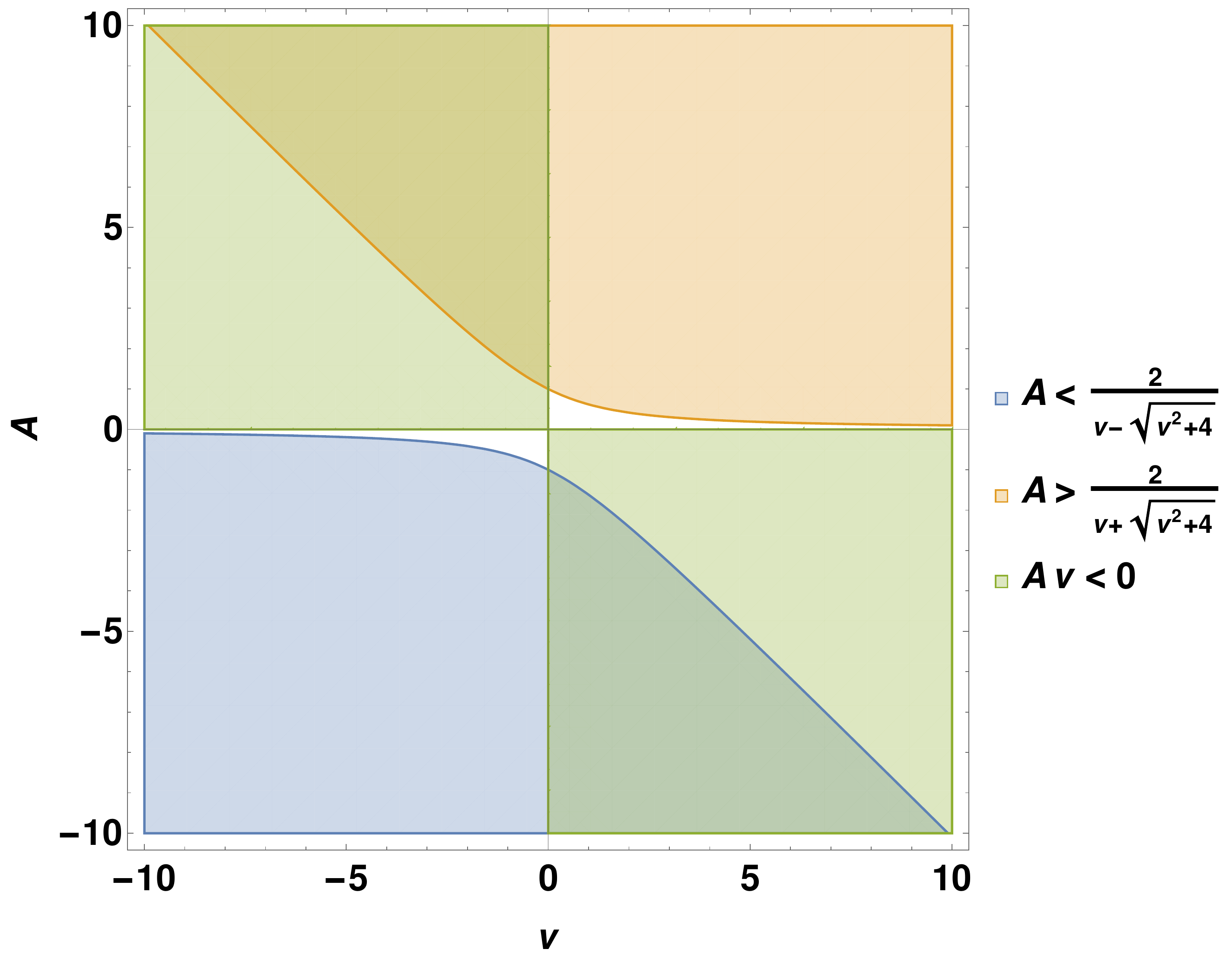}
\caption{Three colors showing regions of $A$ as a function of $v$ where edge 
states exist. The regions where two colors overlap
have two edge states. We have taken $g=1$.} \label{ss1fig22} \end{figure}

The regions described by Eqs.~(\ref{a1}-\ref{a3}) are shown in 
Fig.~\ref{ss1fig22}. We see that these agree well with the regions of bound
states (light regions) shown in Figs.~\ref{ss1fig21} (a) and (b). In 
particular, they correctly tell us that there are two bound states in the 
regions shown in Fig.~\ref{ss1fig21} (b).


As a special case of the above results, it is interesting to consider what 
happens if we set $v=0$; this will also be useful for the next section. Namely,
we only have a potential $A$ at the site $n=0$. Equations~(\ref{a1}-\ref{a2}) 
then imply that there will be a bound state if either 
$A < - g$ or $A > g$. The bound state energy and wave function can be derived 
easily. For $A < -g$, we find that the wave function and energy are given by
\bea \psi (n) &=& e^{-\ka n}, ~~~{\rm where}~~~ n ~\ge~ 0, \non \\
E &=& - 2 g \cosh \ka, ~~~{\rm and}~~~ e^{\ka} ~=~ - ~\frac{A}{g},
\label{psin1} \eea
where $\ka > 0$. For $A > g$, we have
\bea \psi (n) &=& (-1)^n ~e^{-\ka n}, \non \\
E &=& 2g \cosh \ka, ~~~{\rm and}~~~ e^{\ka} ~=~ \frac{A}{g}. \label{psin2} \eea

For $v=0$, we have seen above that the condition $|A| > |g|$ is required in
order to have an edge state. We can now understand why there are no edge 
states in the periodically driven system if $\om$ is very large. 
Equation~\eqref{heff2} shows that the effective hopping is $-g J_0 (a/\om)$, 
and the effective edge potential is given by $-(4 g^2/\om) J_0 (a/\om) 
\sum_{p=1,3,5,\cdots} J_p (a/\om)/p$. Clearly, when $\om$ becomes sufficiently 
large, the effective edge potential will become smaller in magnitude than the 
effective hopping, and there will not be any edge states.

\subsection{Bose-Hubbard model}

\begin{figure}[H]
\centering
\hspace*{-.6cm}\subfigure[]{\includegraphics[width=0.52\textwidth]{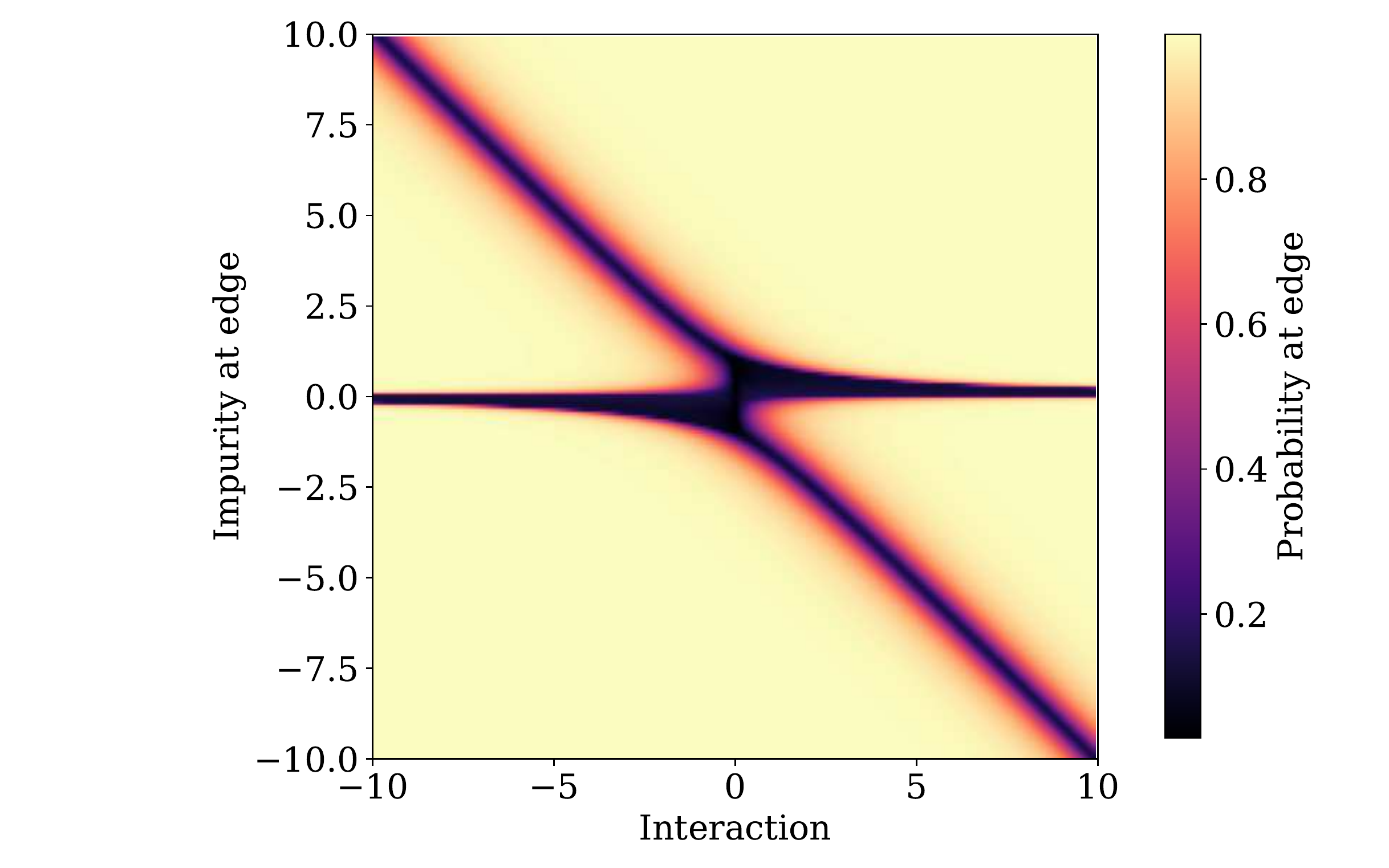}}
\hspace*{-.6cm}\subfigure[]{\includegraphics[width=0.52\textwidth]{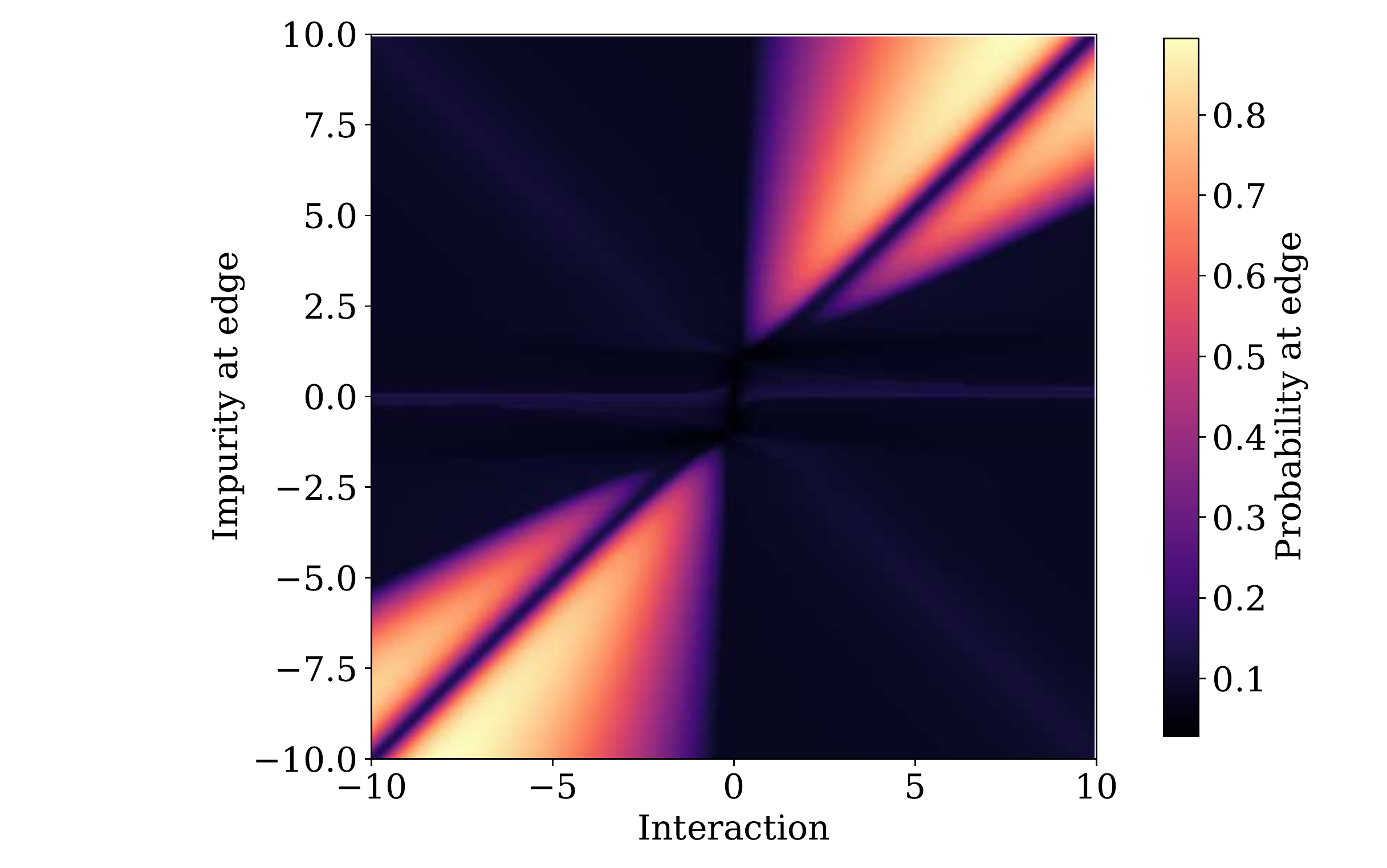}}
\caption{Probabilities $|\psi (0,0)|^2$ of the two highest IPR states at the 
left edge of a 25-site system plotted versus $(u,A)$. The probability is 
higher in the lighter colored regions. Plot (a) shows that one edge state 
exists for a large range of parameter values, while plot (b) shows that a 
second edge state exists in some smaller regions of parameters. We have set 
$g=1$.} \label{ss1fig23} \end{figure}

For the Bose-Hubbard model with an edge potential as described in 
Eq.~\eqref{c2}, we consider a 25-site system with two bosons and 
numerically find the probability of the two particles
to be at the edge, $|\psi (0,0)|^2$, as a function of the parameters 
$(u,A)$. The results are shown in Figs.~\ref{ss1fig23} (a) and (b) for
the states with the largest two IPR values. Just as in Fig.~\ref{ss1fig21},
there can be zero, one or two two-particle bound states which are localized 
near the leftmost site of the system.


We can understand the existence of a two-particle bound state localized
near $n=0$ using a perturbative argument in the limit $|u| \gg |g|, |A|$.
To see this, we write the Hamiltonian in Eq.~\eqref{c2} in the form
$H = H_0 + V$, where
\bea H_0 &=& \frac{u}{2} ~\sum_{n=0}^{\infty} \rho_{n} (\rho_{n}-1) ~+~
A ~b_{0}^{\dg} b_{0}, \non \\
V &=& - ~g ~\sum_{n=0}^{\infty} ~(b_{n}^{\dg}b_{n+1} ~+~ b_{n+1}^{\dg}b_{n}).
\label{bh1} \eea
We now consider a system with two particles. The eigenstates of $H_0$ are of 
the following kinds. States where there is no particle at $n=0$ and no site 
has more than one particle (all these states have zero energy), there is one
particle at $n=0$ and the other particle is at some other site (these states
have energy $A$), the two particles are at the same site which is not at
$n=0$ (these have energy $u$), and both the particles are at $n=0$ (this state
has energy $u+2A$). We will now work within the space of states where
the two particles are at the same site (which may or may not by $n=0$) and
derive an effective Hamiltonian within this space to second order in $g$
(i.e., the perturbation $V$ in Eq.~\eqref{bh1}).

Starting with an initial state $|n,n\ra$ (where $n=0,1,2,\cdots$), the 
hopping can take us to a final state $|n+1,n+1\ra$ (or $|n-1,n-1\ra$) through 
the intermediate state $|n,n+1 \ra$ (or $|n-1,n \ra$ respectively). The matrix 
element connecting the initial or final state to the intermediate state
is $- g \sqrt{2}$. The energy denominator, given by the difference of
the initial and intermediate state energies, is given by $u$ (if $n=0$,
the energy denominator is $u+2A$, but we can approximate this by $u$
since we are assuming that $|u| \gg |A|$). The hopping can also take us from 
a state $|n,n\ra$ back to the same state $|n,n\ra$ in two ways (through the 
intermediate states ($|n,n+1 \ra$ and $|n-1,n \ra$) if $n \ge 1$ but in only 
one way (through the intermediate state $|0,1\ra$) if $n=0$. Putting all this
together and using the notation $|n \ra$ to denote the state in which both 
particles are at site $n$ and $d_n$ and $d_n^\dg$ as the annihilation
and creation operator for two particles at site $n$, we see that the 
effective Hamiltonian in this space is given by
\bea H_{eff} &=& ~\frac{2 g^2}{u} ~\sum_{n=0}^\infty ~(d_n^\dg d_{n+1} 
~+~ d_{n+1}^\dg d_n) \non \\
&& + (u + 2A + \frac{2 g^2}{u}) ~d_0^\dg d_0 + (u + \frac{4 g^2}{u})~
\sum_{n=1}^\infty ~d_n^\dg d_n. \non \\
&& \label{bh2} \eea
We see that this Hamiltonian has a single particle (which is actually a pair 
of bosons) hopping amplitude given by $2 g^2/u$, a chemical potential 
$u + 4g^2/u$ at {\it all} sites, and a potential $2A-2g^2/u$ at the site $n=0$.
We now see that if $u > 0$, the result for a single particle with an edge 
potential discussed at the end of the previous section implies that there will 
be a bound state localized near $n=0$ if either $2A-2g^2/u < - 2g^2/u$ or
$2A-2g^2/u > 2g^2/u$, i.e., if 
\beq {\rm either}~~~A ~<~ 0 ~~~{\rm or}~~~ A ~>~ \frac{2 g^2}{u}. 
\label{bh3} \eeq
If $u < 0$, these conditions change to $2A-2g^2/u < 2g^2/u$ or $2A-2g^2/u 
> - 2g^2/u$, i.e., 
\beq {\rm either}~~~A ~<~ \frac{2g^2}{u} ~~~{\rm or}~~~ A ~>~ 0. 
\label{bh4} \eeq
We see that Eqs.~\eqref{bh3} and \eqref{bh4}) correctly describe the regions 
of bound states in Fig.~\ref{ss1fig23} (a) for $u > 0$ and $u < 0$
respectively, when $|A| \ll |u|$.

When $|u|, |A| \gg |g|$, but $|u|$ and $|A|$ are of the same order, the 
perturbation theory described above breaks down. However, we can understand 
why there are no two-particle bound states localized near $n=0$ close to the 
line $u + A = 0$ as we see in Fig.~\ref{ss1fig23} (a). Ignoring the
hopping $g$ entirely, we know that the state where both particles are at $n=0$
has energy $u + 2A$ while all the states where one particle is at $n=0$ and
the other particle is at any other state have energy $A$. If these two states
have the same energy, and a small hopping $g$ is turned on, the state with
two particles at $n=0$ will mix with the states where one particle remains at 
$n=0$ and the other particle escapes far away from there. Hence we no longer
have a two-particle bound state localized near $n=0$ as an eigenstate of the 
Hamiltonian.

The above arguments do not explain the existence of a second bound state
that we see in Fig.~\ref{ss1fig23} (b) in some small regions in the 
parameter space.

\end{document}